\PassOptionsToPackage{pdfencoding=auto,psdextra}{hyperref}

\documentclass[longauth,
  kazakh,
  russian,
  english]{aa}
\usepackage[LGR,T2A,T1]{fontenc}
\usepackage{CJKutf8}
\newcommand{\tinier}{\fontsize{7.5pt}{7.5pt}\selectfont}
\usepackage{setspace}

\usepackage{graphicx}
\usepackage{natbib}
\usepackage{scalerel}

\usepackage[table]{xcolor}
\usepackage{booktabs}
\usepackage{orcidlink}
\definecolor{color1}{HTML}{2D6A4F}
\definecolor{color2}{HTML}{52b788}
\definecolor{color3}{HTML}{74C69D}
\definecolor{color4}{HTML}{95D5B2}
\definecolor{color5}{HTML}{B7E4C7}
\definecolor{color6}{HTML}{D8F3DC}

\usepackage{txfonts}

\usepackage{hyperref}
\hypersetup{
    colorlinks=true,
    linkcolor=blue,
    filecolor=magenta,      
    urlcolor=blue,
    citecolor=blue
}
\usepackage{tikz}
\usetikzlibrary{shapes.geometric, arrows}
\usepackage{adjustbox}

\tikzset{
   flowbox/.style 2 args={
    rectangle,
    rounded corners,
    minimum width=3.5cm,
    minimum height=0.1cm,
    text width=3.5cm,
    align=center,
    draw=black,
    fill=#1,
    text=#2,
    font=\tiny
  }
}

\tikzset{
   wideflowbox/.style 2 args={
    rectangle,
    rounded corners,
    minimum width=4.5cm, 
    minimum height=0.1cm,
    text centered, 
    draw=black, 
    text width=4.5cm, 
    fill=#1,
    text=#2,
    font=\tiny
  }
}

\tikzset{
    narrowflowbox/.style 2 args={
    rectangle,
    rounded corners,
    minimum width=2.5cm, 
    minimum height=0.1cm,
    text centered, 
    draw=black, 
    text width=2cm, 
    fill=#1,
    text=#2,
    font=\tiny
  }
}

\tikzstyle{arrow} = [thick,->,>=stealth, font=\tiny]

\makeatletter
\renewcommand*\aa@pageof{, page \thepage{} of \pageref*{LastPage}}
\makeatother

\usepackage[switch, modulo]{lineno}
              
\linenumbers

\renewcommand{\linenumbers}[0]{}

\usepackage{euclid}

\newcommand*{\rt}{\textcolor{black}}

\newcommand*{\SWCt}{SW-Classify}
\newcommand*{\SWRt}{SW-Refine}
\newcommand*{\Lines}{DR1-SL01}
\newcommand*{\Nightingale}{DR1-SL03}
\newcommand*{\Nersesian}{DR1-SL04}

\newcommand*{\Fogliardi}{DR1-SL06}

\newcommand*{\Vincken}{Euclid Collaboration: Vincken et al. (in prep.)}

\newcommand*{\Vinckensegmentation}{DR1-SL07}

\newcommand*{\Li}{TianDR1dspl}
\newcommand*{\Ecker}{Euclid Collaboration: Ecker et al. (2026, in prep.)}

\newcommand*{\Refine}{\citet{Holloway_Refine}}
\newcommand{\ignoreme}[1]{}
\newcommand*{\NCGij}{\ensuremath{N_{\rm{C_{\it{i}}|G_{\it{j}}}}}}
\newcommand*{\NGj}{\ensuremath{N_{\rm{G_{\it{j}}}}}}
\newcommand*{\Lens}{\ensuremath{{\rm{Lens}}}}
\newcommand*{\Nonlens}{{\rm{Non}}$-${\rm{Lens}}}

\begin{document}
%
%
\title{Euclid Data Release 1 (DR1)}
\subtitle{Euclid Lens Identification Pipeline and Scientific Exploration (ECLIPSE) B: Discovery pipeline overview
\thanks{This publication has been made possible by the participation of volunteers in the Space Warps project. 
Their contributions are individually acknowledged in Appendix \ref{Ap: SW Volunteers} and on the Space Warps \href{https://www.zooniverse.org/projects/aprajita/space-warps-esa-euclid/about/team}{`The Team'} webpage.}}

\newcommand{\orcid}[1]{\,\orcidlink{#1}} 
\author{Euclid Collaboration: P.~Holloway\orcid{0009-0002-8896-6100}\thanks{\email{philip.holloway@port.ac.uk}}\inst{\ref{aff1}}
\and A.~Verma\orcid{0000-0002-0730-0781}\inst{\ref{aff2}}
\and N.~E.~P.~Lines\orcid{0009-0004-7751-1914}\inst{\ref{aff1}}
\and T.~Li\orcid{0009-0005-5008-0381}\inst{\ref{aff1}}
\and S.~H.~Vincken\orcid{0009-0005-7305-2359}\inst{\ref{aff3}}
\and J.~A.~Acevedo~Barroso\orcid{0000-0002-9654-1711}\inst{\ref{aff4}}
\and T.~E.~Collett\orcid{0000-0001-5564-3140}\inst{\ref{aff1}}
\and S.~Schuldt\orcid{0000-0003-2497-6334}\inst{\ref{aff5},\ref{aff6},\ref{aff7}}
\and M.~Fogliardi\orcid{0009-0006-4964-5311}\inst{\ref{aff8}}
\and A.~Manj\'on-Garc\'ia\orcid{0000-0002-7413-8825}\inst{\ref{aff9}}
\and J.~Wilde\orcid{0000-0002-4460-7379}\inst{\ref{aff10}}
\and K.~Rojas\orcid{0000-0003-1391-6854}\inst{\ref{aff11}}
\and R.~B.~Metcalf\orcid{0000-0003-3167-2574}\inst{\ref{aff12},\ref{aff13}}
\and G.~Despali\orcid{0000-0001-6150-4112}\inst{\ref{aff12},\ref{aff13},\ref{aff14}}
\and F.~Courbin\orcid{0000-0003-0758-6510}\inst{\ref{aff10},\ref{aff15},\ref{aff16}}
\and J.~W.~Nightingale\orcid{0000-0002-8987-7401}\inst{\ref{aff17}}
\and R.~Gavazzi\orcid{0000-0002-5540-6935}\inst{\ref{aff18},\ref{aff19}}
\and H.~Degaudenzi\orcid{0000-0002-5887-6799}\inst{\ref{aff20}}
\and P.~Fleury\orcid{0000-0001-9292-3651}\inst{\ref{aff21}}
\and A.~Sonnenfeld\orcid{0000-0002-6061-5977}\inst{\ref{aff22},\ref{aff23},\ref{aff24}}
\and P.~J.~Marshall\orcid{0000-0002-0113-5770}\inst{\ref{aff25},\ref{aff26}}
\and A.~More\orcid{0000-0001-7714-7076}\inst{\ref{aff27},\ref{aff28}}
\and E.~Baeten\inst{\ref{aff29}}
\and C.~Cornen\orcid{0000-0002-7786-2798}\inst{\ref{aff29}}
\and C.~Macmillan\inst{\ref{aff29}}
\and D.~J.~Bowden\orcid{0009-0008-6114-1401}\inst{\ref{aff2}}
\and R.~Ca\~nameras\orcid{0000-0002-2468-5169}\inst{\ref{aff18}}
\and J.~H.-H.~Chan\orcid{0000-0001-8797-725X}\inst{\ref{aff30},\ref{aff31}}
\and M.~Grespan\orcid{0000-0002-8213-0217}\inst{\ref{aff2},\ref{aff32}}
\and Q.~He\orcid{0000-0003-3672-9365}\inst{\ref{aff33}}
\and D.~P.~Johnson\orcid{0000-0002-0311-2513}\inst{\ref{aff21}}
\and N.~Massari\orcid{0009-0006-8421-930X}\inst{\ref{aff18}}
\and A.~Nersesian\orcid{0000-0001-6843-409X}\inst{\ref{aff34},\ref{aff35}}
\and C.~Olave\inst{\ref{aff36}}
\and A.~Paulino-Afonso\orcid{0000-0002-0943-0694}\inst{\ref{aff37},\ref{aff38}}
\and C.~Saulder\orcid{0000-0002-0408-5633}\inst{\ref{aff39},\ref{aff40}}
\and D.~Sluse\orcid{0000-0001-6116-2095}\inst{\ref{aff34}}
\and G.~Smadja\inst{\ref{aff41}}
\and D.~Scott\orcid{0000-0002-6878-9840}\inst{\ref{aff42}}
\and H.~Srivastava\orcid{0009-0002-3254-0952}\inst{\ref{aff42}}
\and D.~Stern\orcid{0000-0003-2686-9241}\inst{\ref{aff4}}
\and T.~T.~Thai\orcid{0000-0002-8408-4816}\inst{\ref{aff43}}
\and C.~Tortora\orcid{0000-0001-7958-6531}\inst{\ref{aff44}}
\and N.~Wielders\orcid{0009-0001-8110-819X}\inst{\ref{aff17}}
\and I.~T.~Andika\orcid{0000-0001-6102-9526}\inst{\ref{aff40}}
\and A.~Angress\orcid{0009-0005-5250-9487}\inst{\ref{aff45},\ref{aff31}}
\and M.~Bradac\orcid{0000-0001-5984-0395}\inst{\ref{aff46}}
\and B.~C.~Nagam\orcid{0000-0002-3724-7694}\inst{\ref{aff47}}
\and D.~Delley\orcid{0000-0002-4958-7469}\inst{\ref{aff39}}
\and J.~M.~Diego\orcid{0000-0001-9065-3926}\inst{\ref{aff48}}
\and L.~R.~Ecker\orcid{0009-0005-3508-2469}\inst{\ref{aff40},\ref{aff39},\ref{aff49}}
\and W.~J.~R.~Enzi\orcid{0009-0004-2992-3148}\inst{\ref{aff1},\ref{aff50}}
\and A.~Galan\orcid{0000-0003-2547-9815}\inst{\ref{aff20}}
\and G.~Granata\orcid{0000-0002-9512-3788}\inst{\ref{aff1}}
\and T.~Harvey\orcid{0000-0002-4130-636X}\inst{\ref{aff51},\ref{aff52}}
\and N.~B.~Hogg\orcid{0000-0001-9346-4477}\inst{\ref{aff53},\ref{aff54}}
\and K.~Jahnke\orcid{0000-0003-3804-2137}\inst{\ref{aff52}}
\and K.~Kuijken\orcid{0000-0002-3827-0175}\inst{\ref{aff55}}
\and T.~Li\orcid{0009-0003-5064-1914}\inst{\ref{aff56}}
\and A.~Melo\orcid{0000-0002-6449-3970}\inst{\ref{aff57},\ref{aff58}}
\and M.~Meneghetti\orcid{0000-0003-1225-7084}\inst{\ref{aff13},\ref{aff14}}
\and L.~A.~Moustakas\orcid{0000-0003-3030-2360}\inst{\ref{aff4}}
\and D.~O'Ryan\orcid{0000-0003-1217-4617}\inst{\ref{aff59}}
\and J.~Pearson\orcid{0000-0001-8555-8561}\inst{\ref{aff60}}
\and H.~Qu\orcid{0009-0006-0299-0265}\inst{\ref{aff61}}
\and D.~Ryczanowski\orcid{0000-0002-4429-3429}\inst{\ref{aff1},\ref{aff62}}
\and T.~Saifollahi\orcid{0000-0002-9554-7660}\inst{\ref{aff63},\ref{aff64}}
\and S.~Serjeant\orcid{0000-0002-0517-7943}\inst{\ref{aff60}}
\and C.~Spiniello\orcid{0000-0002-3909-6359}\inst{\ref{aff65},\ref{aff2}}
\and C.~J.~Storfer\orcid{0000-0002-0385-0014}\inst{\ref{aff66}}
\and L.~Trobbiani\orcid{0009-0000-8058-834X}\inst{\ref{aff36},\ref{aff50}}
\and X.~Xu\orcid{0009-0002-0985-2309}\inst{\ref{aff56},\ref{aff67}}
\and M.~Zumalacarregui\orcid{0000-0002-9943-6490}\inst{\ref{aff68},\ref{aff69}}
\and N.~Aghanim\orcid{0000-0002-6688-8992}\inst{\ref{aff70}}
\and B.~Altieri\orcid{0000-0003-3936-0284}\inst{\ref{aff59}}
\and S.~Andreon\orcid{0000-0002-2041-8784}\inst{\ref{aff71}}
\and N.~Auricchio\orcid{0000-0003-4444-8651}\inst{\ref{aff13}}
\and C.~Baccigalupi\orcid{0000-0002-8211-1630}\inst{\ref{aff72},\ref{aff73},\ref{aff74},\ref{aff75}}
\and M.~Baldi\orcid{0000-0003-4145-1943}\inst{\ref{aff36},\ref{aff13},\ref{aff14}}
\and S.~Bardelli\orcid{0000-0002-8900-0298}\inst{\ref{aff13}}
\and P.~Battaglia\orcid{0000-0002-7337-5909}\inst{\ref{aff13}}
\and R.~Bender\orcid{0000-0001-7179-0626}\inst{\ref{aff39},\ref{aff40}}
\and A.~Biviano\orcid{0000-0002-0857-0732}\inst{\ref{aff73},\ref{aff72}}
\and M.~Bolzonella\orcid{0000-0003-3278-4607}\inst{\ref{aff13}}
\and E.~Branchini\orcid{0000-0002-0808-6908}\inst{\ref{aff76},\ref{aff77},\ref{aff71}}
\and M.~Brescia\orcid{0000-0001-9506-5680}\inst{\ref{aff78},\ref{aff44}}
\and S.~Camera\orcid{0000-0003-3399-3574}\inst{\ref{aff79},\ref{aff80},\ref{aff81}}
\and G.~Ca\~nas-Herrera\orcid{0000-0003-2796-2149}\inst{\ref{aff55}}
\and V.~Capobianco\orcid{0000-0002-3309-7692}\inst{\ref{aff81}}
\and C.~Carbone\orcid{0000-0003-0125-3563}\inst{\ref{aff7}}
\and V.~F.~Cardone\inst{\ref{aff82},\ref{aff83}}
\and J.~Carretero\orcid{0000-0002-3130-0204}\inst{\ref{aff84},\ref{aff85}}
\and M.~Castellano\orcid{0000-0001-9875-8263}\inst{\ref{aff82}}
\and G.~Castignani\orcid{0000-0001-6831-0687}\inst{\ref{aff13}}
\and S.~Cavuoti\orcid{0000-0002-3787-4196}\inst{\ref{aff44},\ref{aff86}}
\and K.~C.~Chambers\orcid{0000-0001-6965-7789}\inst{\ref{aff66}}
\and A.~Cimatti\inst{\ref{aff87}}
\and C.~Colodro-Conde\inst{\ref{aff88}}
\and G.~Congedo\orcid{0000-0003-2508-0046}\inst{\ref{aff89}}
\and C.~J.~Conselice\orcid{0000-0003-1949-7638}\inst{\ref{aff51}}
\and L.~Conversi\orcid{0000-0002-6710-8476}\inst{\ref{aff90},\ref{aff59}}
\and Y.~Copin\orcid{0000-0002-5317-7518}\inst{\ref{aff41}}
\and H.~M.~Courtois\orcid{0000-0003-0509-1776}\inst{\ref{aff91}}
\and M.~Cropper\orcid{0000-0003-4571-9468}\inst{\ref{aff92}}
\and G.~De~Lucia\orcid{0000-0002-6220-9104}\inst{\ref{aff73}}
\and C.~Dolding\orcid{0009-0003-7199-6108}\inst{\ref{aff92}}
\and H.~Dole\orcid{0000-0002-9767-3839}\inst{\ref{aff70}}
\and F.~Dubath\orcid{0000-0002-6533-2810}\inst{\ref{aff20}}
\and X.~Dupac\inst{\ref{aff59}}
\and S.~Dusini\orcid{0000-0002-1128-0664}\inst{\ref{aff93}}
\and S.~Escoffier\orcid{0000-0002-2847-7498}\inst{\ref{aff94}}
\and M.~Farina\orcid{0000-0002-3089-7846}\inst{\ref{aff95}}
\and R.~Farinelli\inst{\ref{aff13}}
\and F.~Faustini\orcid{0000-0001-6274-5145}\inst{\ref{aff82}}
\and S.~Ferriol\inst{\ref{aff41}}
\and F.~Finelli\orcid{0000-0002-6694-3269}\inst{\ref{aff13},\ref{aff96}}
\and S.~Fotopoulou\orcid{0000-0002-9686-254X}\inst{\ref{aff97}}
\and M.~Frailis\orcid{0000-0002-7400-2135}\inst{\ref{aff73}}
\and E.~Franceschi\orcid{0000-0002-0585-6591}\inst{\ref{aff13}}
\and M.~Fumana\orcid{0000-0001-6787-5950}\inst{\ref{aff7}}
\and L.~Gabarra\orcid{0000-0002-8486-8856}\inst{\ref{aff2}}
\and S.~Galeotta\orcid{0000-0002-3748-5115}\inst{\ref{aff73}}
\and K.~George\orcid{0000-0002-1734-8455}\inst{\ref{aff98}}
\and B.~Gillis\orcid{0000-0002-4478-1270}\inst{\ref{aff89}}
\and C.~Giocoli\orcid{0000-0002-9590-7961}\inst{\ref{aff13},\ref{aff14}}
\and J.~Gracia-Carpio\orcid{0000-0003-4689-3134}\inst{\ref{aff39}}
\and A.~Grazian\orcid{0000-0002-5688-0663}\inst{\ref{aff99}}
\and F.~Grupp\inst{\ref{aff39},\ref{aff40}}
\and W.~G.~Hartley\orcid{0000-0002-4563-5101}\inst{\ref{aff20}}
\and S.~V.~H.~Haugan\orcid{0000-0001-9648-7260}\inst{\ref{aff100}}
\and S.~Hemmati\orcid{0000-0003-2226-5395}\inst{\ref{aff61}}
\and H.~Hoekstra\orcid{0000-0002-0641-3231}\inst{\ref{aff55}}
\and W.~Holmes\orcid{0009-0007-8554-4646}\inst{\ref{aff4}}
\and F.~Hormuth\inst{\ref{aff101}}
\and A.~Hornstrup\orcid{0000-0002-3363-0936}\inst{\ref{aff102},\ref{aff103}}
\and M.~Huertas-Company\orcid{0000-0002-1416-8483}\inst{\ref{aff88},\ref{aff104},\ref{aff105}}
\and M.~Jhabvala\inst{\ref{aff106}}
\and B.~Joachimi\orcid{0000-0001-7494-1303}\inst{\ref{aff107}}
\and S.~Kermiche\orcid{0000-0002-0302-5735}\inst{\ref{aff94}}
\and A.~Kiessling\orcid{0000-0002-2590-1273}\inst{\ref{aff4}}
\and B.~Kubik\orcid{0009-0006-5823-4880}\inst{\ref{aff41}}
\and M.~K\"ummel\orcid{0000-0003-2791-2117}\inst{\ref{aff40}}
\and M.~Kunz\orcid{0000-0002-3052-7394}\inst{\ref{aff108}}
\and H.~Kurki-Suonio\orcid{0000-0002-4618-3063}\inst{\ref{aff6},\ref{aff109}}
\and A.~M.~C.~Le~Brun\orcid{0000-0002-0936-4594}\inst{\ref{aff110}}
\and S.~Ligori\orcid{0000-0003-4172-4606}\inst{\ref{aff81}}
\and P.~B.~Lilje\orcid{0000-0003-4324-7794}\inst{\ref{aff100}}
\and V.~Lindholm\orcid{0000-0003-2317-5471}\inst{\ref{aff6},\ref{aff109}}
\and I.~Lloro\orcid{0000-0001-5966-1434}\inst{\ref{aff111}}
\and M.~Magliocchetti\orcid{0000-0001-9158-4838}\inst{\ref{aff95}}
\and G.~Mainetti\orcid{0000-0003-2384-2377}\inst{\ref{aff112}}
\and O.~Mansutti\orcid{0000-0001-5758-4658}\inst{\ref{aff73}}
\and O.~Marggraf\orcid{0000-0001-7242-3852}\inst{\ref{aff113}}
\and M.~Martinelli\orcid{0000-0002-6943-7732}\inst{\ref{aff82},\ref{aff83}}
\and N.~Martinet\orcid{0000-0003-2786-7790}\inst{\ref{aff18}}
\and F.~Marulli\orcid{0000-0002-8850-0303}\inst{\ref{aff12},\ref{aff13},\ref{aff14}}
\and R.~J.~Massey\orcid{0000-0002-6085-3780}\inst{\ref{aff114}}
\and N.~Mauri\orcid{0000-0001-8196-1548}\inst{\ref{aff87},\ref{aff14}}
\and E.~Medinaceli\orcid{0000-0002-4040-7783}\inst{\ref{aff13}}
\and S.~Mei\orcid{0000-0002-2849-559X}\inst{\ref{aff115},\ref{aff116}}
\and M.~Melchior\inst{\ref{aff3}}
\and E.~Merlin\orcid{0000-0001-6870-8900}\inst{\ref{aff99}}
\and G.~Meylan\orcid{0000-0001-6503-0209}\inst{\ref{aff117}}
\and P.~Monaco\orcid{0000-0003-2083-7564}\inst{\ref{aff118},\ref{aff73},\ref{aff74},\ref{aff72}}
\and A.~Mora\orcid{0000-0002-1922-8529}\inst{\ref{aff119}}
\and M.~Moresco\orcid{0000-0002-7616-7136}\inst{\ref{aff12},\ref{aff13}}
\and C.~Moretti\orcid{0000-0003-3314-8936}\inst{\ref{aff73},\ref{aff72},\ref{aff74}}
\and L.~Moscardini\orcid{0000-0002-3473-6716}\inst{\ref{aff12},\ref{aff13},\ref{aff14}}
\and C.~Neissner\orcid{0000-0001-8524-4968}\inst{\ref{aff120},\ref{aff85}}
\and R.~C.~Nichol\orcid{0000-0003-0939-6518}\inst{\ref{aff121}}
\and S.-M.~Niemi\orcid{0009-0005-0247-0086}\inst{\ref{aff122}}
\and C.~Padilla\orcid{0000-0001-7951-0166}\inst{\ref{aff120}}
\and S.~Paltani\orcid{0000-0002-8108-9179}\inst{\ref{aff20}}
\and F.~Pasian\orcid{0000-0002-4869-3227}\inst{\ref{aff73}}
\and W.~J.~Percival\orcid{0000-0002-0644-5727}\inst{\ref{aff123},\ref{aff124},\ref{aff125}}
\and V.~Pettorino\orcid{0000-0002-4203-9320}\inst{\ref{aff122}}
\and A.~Pezzotta\orcid{0000-0003-0726-2268}\inst{\ref{aff71}}
\and G.~Polenta\orcid{0000-0003-4067-9196}\inst{\ref{aff126}}
\and M.~Poncet\inst{\ref{aff64}}
\and L.~A.~Popa\inst{\ref{aff127}}
\and F.~Raison\orcid{0000-0002-7819-6918}\inst{\ref{aff39}}
\and R.~Rebolo\orcid{0000-0003-3767-7085}\inst{\ref{aff88},\ref{aff128},\ref{aff129}}
\and A.~Renzi\orcid{0000-0001-9856-1970}\inst{\ref{aff130},\ref{aff93},\ref{aff13}}
\and J.~Rhodes\orcid{0000-0002-4485-8549}\inst{\ref{aff4}}
\and G.~Riccio\inst{\ref{aff44}}
\and I.~Risso\orcid{0000-0003-2525-7761}\inst{\ref{aff76},\ref{aff77},\ref{aff71}}
\and F.~Rizzo\orcid{0000-0002-9407-585X}\inst{\ref{aff73}}
\and E.~Romelli\orcid{0000-0003-3069-9222}\inst{\ref{aff73}}
\and M.~Roncarelli\orcid{0000-0001-9587-7822}\inst{\ref{aff13}}
\and B.~Rusholme\orcid{0000-0001-7648-4142}\inst{\ref{aff61}}
\and R.~Saglia\orcid{0000-0003-0378-7032}\inst{\ref{aff40},\ref{aff39}}
\and Z.~Sakr\orcid{0000-0002-4823-3757}\inst{\ref{aff69},\ref{aff131},\ref{aff132}}
\and D.~Sapone\orcid{0000-0001-7089-4503}\inst{\ref{aff133}}
\and B.~Sartoris\orcid{0000-0003-1337-5269}\inst{\ref{aff39},\ref{aff73}}
\and P.~Schneider\orcid{0000-0001-8561-2679}\inst{\ref{aff113}}
\and T.~Schrabback\orcid{0000-0002-6987-7834}\inst{\ref{aff134}}
\and A.~Secroun\orcid{0000-0003-0505-3710}\inst{\ref{aff94}}
\and E.~Sihvola\orcid{0000-0003-1804-7715}\inst{\ref{aff135}}
\and P.~Simon\inst{\ref{aff113}}
\and C.~Sirignano\orcid{0000-0002-0995-7146}\inst{\ref{aff130},\ref{aff93}}
\and G.~Sirri\orcid{0000-0003-2626-2853}\inst{\ref{aff14}}
\and L.~Stanco\orcid{0000-0002-9706-5104}\inst{\ref{aff93}}
\and J.-L.~Starck\orcid{0000-0003-2177-7794}\inst{\ref{aff136}}
\and P.~Tallada-Cresp\'{i}\orcid{0000-0002-1336-8328}\inst{\ref{aff84},\ref{aff85}}
\and A.~N.~Taylor\inst{\ref{aff89}}
\and H.~I.~Teplitz\orcid{0000-0002-7064-5424}\inst{\ref{aff137}}
\and I.~Tereno\orcid{0000-0002-4537-6218}\inst{\ref{aff138},\ref{aff139}}
\and N.~Tessore\orcid{0000-0002-9696-7931}\inst{\ref{aff92}}
\and S.~Toft\orcid{0000-0003-3631-7176}\inst{\ref{aff140},\ref{aff141}}
\and R.~Toledo-Moreo\orcid{0000-0002-2997-4859}\inst{\ref{aff142},\ref{aff143}}
\and F.~Torradeflot\orcid{0000-0003-1160-1517}\inst{\ref{aff85},\ref{aff84}}
\and I.~Tutusaus\orcid{0000-0002-3199-0399}\inst{\ref{aff144},\ref{aff145},\ref{aff131}}
\and J.~Valiviita\orcid{0000-0001-6225-3693}\inst{\ref{aff6},\ref{aff109}}
\and T.~Vassallo\orcid{0000-0001-6512-6358}\inst{\ref{aff73},\ref{aff98}}
\and Y.~Wang\orcid{0000-0002-4749-2984}\inst{\ref{aff61}}
\and J.~Weller\orcid{0000-0002-8282-2010}\inst{\ref{aff40},\ref{aff39}}
\and A.~Zacchei\orcid{0000-0003-0396-1192}\inst{\ref{aff73},\ref{aff72}}
\and G.~Zamorani\orcid{0000-0002-2318-301X}\inst{\ref{aff13}}
\and F.~M.~Zerbi\orcid{0000-0002-9996-973X}\inst{\ref{aff71}}
\and E.~Zucca\orcid{0000-0002-5845-8132}\inst{\ref{aff13}}
\and M.~Ballardini\orcid{0000-0003-4481-3559}\inst{\ref{aff8},\ref{aff146},\ref{aff13}}
\and P.~Bergamini\orcid{0000-0003-1383-9414}\inst{\ref{aff13}}
\and E.~Bozzo\orcid{0000-0002-8201-1525}\inst{\ref{aff20}}
\and C.~Burigana\orcid{0000-0002-3005-5796}\inst{\ref{aff50},\ref{aff96}}
\and R.~Cabanac\orcid{0000-0001-6679-2600}\inst{\ref{aff131}}
\and M.~Calabrese\orcid{0000-0002-2637-2422}\inst{\ref{aff147},\ref{aff7}}
\and A.~Cappi\inst{\ref{aff148},\ref{aff13}}
\and F.~Caro\orcid{0009-0003-1053-0507}\inst{\ref{aff82}}
\and T.~Castro\orcid{0000-0002-6292-3228}\inst{\ref{aff149},\ref{aff73},\ref{aff72}}
\and J.~A.~Escartin~Vigo\inst{\ref{aff39}}
\and J.~Garc\'ia-Bellido\orcid{0000-0002-9370-8360}\inst{\ref{aff69}}
\and T.~Gasparetto\orcid{0000-0002-7913-4866}\inst{\ref{aff82}}
\and A.~Loureiro\orcid{0000-0002-4371-0876}\inst{\ref{aff150},\ref{aff151}}
\and J.~Macias-Perez\orcid{0000-0002-5385-2763}\inst{\ref{aff152}}
\and R.~Maoli\orcid{0000-0002-6065-3025}\inst{\ref{aff153},\ref{aff82}}
\and A.~A.~Nucita\inst{\ref{aff154},\ref{aff155},\ref{aff156}}
\and M.~P\"ontinen\orcid{0000-0001-5442-2530}\inst{\ref{aff6}}
\and E.~Sarpa\orcid{0000-0002-1256-655X}\inst{\ref{aff73}}
\and V.~Scottez\orcid{0009-0008-3864-940X}\inst{\ref{aff157},\ref{aff158}}
\and M.~Sereno\orcid{0000-0003-0302-0325}\inst{\ref{aff13},\ref{aff14}}
\and M.~Tenti\orcid{0000-0002-4254-5901}\inst{\ref{aff14}}
\and M.~Tucci\inst{\ref{aff20}}
\and M.~Viel\orcid{0000-0002-2642-5707}\inst{\ref{aff72},\ref{aff73},\ref{aff75},\ref{aff74},\ref{aff159}}
\and M.~Wiesmann\orcid{0009-0000-8199-5860}\inst{\ref{aff100}}
\and Y.~Akrami\orcid{0000-0002-2407-7956}\inst{\ref{aff69},\ref{aff160}}
\and G.~Angora\orcid{0000-0002-0316-6562}\inst{\ref{aff44},\ref{aff8}}
\and S.~Anselmi\orcid{0000-0002-3579-9583}\inst{\ref{aff93},\ref{aff130},\ref{aff161}}
\and M.~Archidiacono\orcid{0000-0003-4952-9012}\inst{\ref{aff162},\ref{aff163}}
\and G.~Aric\`o\orcid{0000-0002-2802-2928}\inst{\ref{aff14}}
\and F.~Atrio-Barandela\orcid{0000-0002-2130-2513}\inst{\ref{aff164}}
\and M.~Baes\orcid{0000-0002-3930-2757}\inst{\ref{aff35}}
\and L.~Bazzanini\orcid{0000-0003-0727-0137}\inst{\ref{aff8},\ref{aff13}}
\and D.~Bertacca\orcid{0000-0002-2490-7139}\inst{\ref{aff130},\ref{aff99},\ref{aff93}}
\and M.~Bethermin\orcid{0000-0002-3915-2015}\inst{\ref{aff63}}
\and F.~Beutler\orcid{0000-0003-0467-5438}\inst{\ref{aff89}}
\and A.~Blanchard\orcid{0000-0001-8555-9003}\inst{\ref{aff131}}
\and L.~Blot\orcid{0000-0002-9622-7167}\inst{\ref{aff165},\ref{aff110}}
\and M.~L.~Brown\orcid{0000-0002-0370-8077}\inst{\ref{aff51}}
\and S.~Bruton\orcid{0000-0002-6503-5218}\inst{\ref{aff166}}
\and A.~Calabro\orcid{0000-0003-2536-1614}\inst{\ref{aff82}}
\and B.~Camacho~Quevedo\orcid{0000-0002-8789-4232}\inst{\ref{aff72},\ref{aff75},\ref{aff73}}
\and C.~S.~Carvalho\inst{\ref{aff139}}
\and F.~Cogato\orcid{0000-0003-4632-6113}\inst{\ref{aff12},\ref{aff13}}
\and S.~Conseil\orcid{0000-0002-3657-4191}\inst{\ref{aff41}}
\and A.~R.~Cooray\orcid{0000-0002-3892-0190}\inst{\ref{aff56}}
\and P.~Corcho-Caballero\orcid{0000-0001-6327-7080}\inst{\ref{aff33}}
\and B.~Csizi\orcid{0000-0003-3227-6581}\inst{\ref{aff134}}
\and O.~Cucciati\orcid{0000-0002-9336-7551}\inst{\ref{aff13}}
\and S.~Davini\orcid{0000-0003-3269-1718}\inst{\ref{aff77}}
\and T.~de~Boer\orcid{0000-0001-5486-2747}\inst{\ref{aff66}}
\and F.~De~Paolis\orcid{0000-0001-6460-7563}\inst{\ref{aff154},\ref{aff155},\ref{aff156}}
\and G.~Desprez\orcid{0000-0001-8325-1742}\inst{\ref{aff33}}
\and A.~D\'iaz-S\'anchez\orcid{0000-0003-0748-4768}\inst{\ref{aff9}}
\and S.~Di~Domizio\orcid{0000-0003-2863-5895}\inst{\ref{aff76},\ref{aff77}}
\and V.~Duret\orcid{0009-0009-0383-4960}\inst{\ref{aff94}}
\and M.~Y.~Elkhashab\orcid{0000-0001-9306-2603}\inst{\ref{aff73},\ref{aff74},\ref{aff118},\ref{aff72}}
\and Y.~Fang\orcid{0000-0002-0334-6950}\inst{\ref{aff40}}
\and A.~Farina\orcid{0009-0000-3420-929X}\inst{\ref{aff71},\ref{aff77}}
\and A.~Finoguenov\orcid{0000-0002-4606-5403}\inst{\ref{aff6}}
\and A.~Franco\orcid{0000-0002-4761-366X}\inst{\ref{aff154},\ref{aff155},\ref{aff156}}
\and Y.~Fu\orcid{0000-0002-0759-0504}\inst{\ref{aff55},\ref{aff33}}
\and K.~Ganga\orcid{0000-0001-8159-8208}\inst{\ref{aff115}}
\and E.~Gaztanaga\orcid{0000-0001-9632-0815}\inst{\ref{aff144},\ref{aff145},\ref{aff1}}
\and Z.~Ghaffari\orcid{0000-0002-6467-8078}\inst{\ref{aff73},\ref{aff72}}
\and F.~Giacomini\orcid{0000-0002-3129-2814}\inst{\ref{aff14}}
\and F.~Gianotti\orcid{0000-0003-4666-119X}\inst{\ref{aff13}}
\and G.~Gozaliasl\orcid{0000-0002-0236-919X}\inst{\ref{aff167},\ref{aff6}}
\and A.~Gruppuso\orcid{0000-0001-9272-5292}\inst{\ref{aff13},\ref{aff14}}
\and C.~M.~Gutierrez\orcid{0000-0001-7854-783X}\inst{\ref{aff88},\ref{aff129}}
\and A.~Hall\orcid{0000-0002-3139-8651}\inst{\ref{aff89}}
\and C.~Hern\'andez-Monteagudo\orcid{0000-0001-5471-9166}\inst{\ref{aff129},\ref{aff88}}
\and H.~Hildebrandt\orcid{0000-0002-9814-3338}\inst{\ref{aff168}}
\and J.~Hjorth\orcid{0000-0002-4571-2306}\inst{\ref{aff169}}
\and M.~Jauzac\orcid{0000-0003-1974-8732}\inst{\ref{aff170},\ref{aff114},\ref{aff171},\ref{aff172},\ref{aff131}}
\and J.~J.~E.~Kajava\orcid{0000-0002-3010-8333}\inst{\ref{aff173},\ref{aff5},\ref{aff174}}
\and Y.~Kang\orcid{0009-0000-8588-7250}\inst{\ref{aff134}}
\and V.~Kansal\orcid{0000-0002-4008-6078}\inst{\ref{aff175},\ref{aff176}}
\and D.~Karagiannis\orcid{0000-0002-4927-0816}\inst{\ref{aff8},\ref{aff177}}
\and J.~Kim\orcid{0000-0003-2776-2761}\inst{\ref{aff2}}
\and C.~C.~Kirkpatrick\inst{\ref{aff135}}
\and A.~Kov\'acs\orcid{0000-0002-5825-579X}\inst{\ref{aff178},\ref{aff179}}
\and I.~Kova{\v{c}}i{\'{c}}\orcid{0000-0001-6751-3263}\inst{\ref{aff35}}
\and K.~Koyama\orcid{0000-0001-6727-6915}\inst{\ref{aff1}}
\and S.~Kruk\orcid{0000-0001-8010-8879}\inst{\ref{aff59}}
\and M.~C.~Lam\orcid{0000-0002-9347-2298}\inst{\ref{aff89}}
\and M.~Lattanzi\orcid{0000-0003-1059-2532}\inst{\ref{aff146}}
\and L.~Legrand\orcid{0000-0003-0610-5252}\inst{\ref{aff180},\ref{aff70}}
\and G.~Leroy\orcid{0009-0004-2523-4425}\inst{\ref{aff170},\ref{aff114}}
\and G.~F.~Lesci\orcid{0000-0002-4607-2830}\inst{\ref{aff12},\ref{aff13}}
\and J.~Lesgourgues\orcid{0000-0001-7627-353X}\inst{\ref{aff181}}
\and T.~I.~Liaudat\orcid{0000-0002-9104-314X}\inst{\ref{aff182}}
\and L.~Linke\orcid{0000-0002-2622-8113}\inst{\ref{aff134}}
\and S.~J.~Liu\orcid{0000-0001-7680-2139}\inst{\ref{aff95}}
\and G.~Maggio\orcid{0000-0003-4020-4836}\inst{\ref{aff73}}
\and F.~Mannucci\orcid{0000-0002-4803-2381}\inst{\ref{aff183}}
\and F.~R.~Marleau\orcid{0000-0002-1442-2947}\inst{\ref{aff134}}
\and C.~J.~A.~P.~Martins\orcid{0000-0002-4886-9261}\inst{\ref{aff37},\ref{aff38}}
\and M.~Miluzio\inst{\ref{aff59},\ref{aff184}}
\and G.~Morgante\inst{\ref{aff13}}
\and S.~Nadathur\orcid{0000-0001-9070-3102}\inst{\ref{aff1}}
\and K.~Naidoo\orcid{0000-0002-9182-1802}\inst{\ref{aff1},\ref{aff52}}
\and A.~Navarro-Alsina\orcid{0000-0002-3173-2592}\inst{\ref{aff113}}
\and S.~Nesseris\orcid{0000-0002-0567-0324}\inst{\ref{aff69}}
\and M.~Oguri\orcid{0000-0003-3484-399X}\inst{\ref{aff185},\ref{aff186}}
\and F.~Oppizzi\orcid{0000-0003-3904-8370}\inst{\ref{aff77},\ref{aff93},\ref{aff130}}
\and F.~Pace\orcid{0000-0001-8039-0480}\inst{\ref{aff79},\ref{aff80},\ref{aff81},\ref{aff187}}
\and D.~Paoletti\orcid{0000-0003-4761-6147}\inst{\ref{aff13},\ref{aff96}}
\and F.~Passalacqua\orcid{0000-0002-8606-4093}\inst{\ref{aff93}}
\and K.~Paterson\orcid{0000-0001-8340-3486}\inst{\ref{aff52}}
\and L.~Patrizii\inst{\ref{aff14}}
\and C.~Pattison\orcid{0000-0003-3272-2617}\inst{\ref{aff1}}
\and R.~Paviot\orcid{0009-0002-8108-3460}\inst{\ref{aff136},\ref{aff64}}
\and A.~Pisani\orcid{0000-0002-6146-4437}\inst{\ref{aff94}}
\and D.~Potter\orcid{0000-0002-0757-5195}\inst{\ref{aff188}}
\and G.~W.~Pratt\inst{\ref{aff136}}
\and S.~Quai\orcid{0000-0002-0449-8163}\inst{\ref{aff12},\ref{aff13}}
\and M.~Radovich\orcid{0000-0002-3585-866X}\inst{\ref{aff99}}
\and G.~Rodighiero\orcid{0000-0002-9415-2296}\inst{\ref{aff130},\ref{aff99}}
\and W.~Roster\orcid{0000-0002-9149-6528}\inst{\ref{aff39}}
\and S.~Sacquegna\orcid{0000-0002-8433-6630}\inst{\ref{aff189}}
\and M.~Sahl\'en\orcid{0000-0003-0973-4804}\inst{\ref{aff190}}
\and D.~B.~Sanders\orcid{0000-0002-1233-9998}\inst{\ref{aff66}}
\and A.~Schneider\orcid{0000-0001-7055-8104}\inst{\ref{aff188}}
\and D.~Sciotti\orcid{0009-0008-4519-2620}\inst{\ref{aff82},\ref{aff83}}
\and E.~Sellentin\orcid{0009-0002-2655-3458}\inst{\ref{aff191},\ref{aff55}}
\and F.~Shankar\orcid{0000-0001-8973-5051}\inst{\ref{aff192}}
\and L.~C.~Smith\orcid{0000-0002-3259-2771}\inst{\ref{aff53}}
\and J.~G.~Sorce\orcid{0000-0002-2307-2432}\inst{\ref{aff193},\ref{aff70}}
\and M.~Talia\orcid{0000-0003-4352-2063}\inst{\ref{aff12},\ref{aff13}}
\and K.~Tanidis\orcid{0000-0001-9843-5130}\inst{\ref{aff194}}
\and C.~Tao\orcid{0000-0001-7961-8177}\inst{\ref{aff94}}
\and F.~Tarsitano\orcid{0000-0002-5919-0238}\inst{\ref{aff195},\ref{aff196},\ref{aff20}}
\and G.~Testera\orcid{0000-0003-2970-766X}\inst{\ref{aff77}}
\and R.~Teyssier\orcid{0000-0001-7689-0933}\inst{\ref{aff197}}
\and S.~Tosi\orcid{0000-0002-7275-9193}\inst{\ref{aff76},\ref{aff71},\ref{aff77}}
\and A.~Troja\orcid{0000-0003-0239-4595}\inst{\ref{aff73}}
\and C.~Uhlemann\orcid{0000-0001-7831-1579}\inst{\ref{aff198},\ref{aff17}}
\and C.~Valieri\inst{\ref{aff14}}
\and A.~Venhola\orcid{0000-0001-6071-4564}\inst{\ref{aff199}}
\and D.~Vergani\orcid{0000-0003-0898-2216}\inst{\ref{aff13}}
\and G.~Vernardos\orcid{0000-0001-8554-7248}\inst{\ref{aff31},\ref{aff30}}
\and G.~Verza\orcid{0000-0002-1886-8348}\thanks{Deceased}\inst{\ref{aff200},\ref{aff201}}
\and P.~Vielzeuf\orcid{0000-0003-2035-9339}\inst{\ref{aff94}}
\and E.~Vilenius\orcid{0000-0002-6184-7681}\inst{\ref{aff92}}
\and S.~Vinciguerra\orcid{0009-0005-4018-3184}\inst{\ref{aff18}}
\and M.~von~Wietersheim-Kramsta\orcid{0000-0003-4986-5091}\inst{\ref{aff114},\ref{aff170}}
\and M.~Walmsley\orcid{0000-0002-6408-4181}\inst{\ref{aff202},\ref{aff51}}
\and N.~A.~Walton\orcid{0000-0003-3983-8778}\inst{\ref{aff53}}
\and L.~Wang\orcid{0000-0002-6736-9158}\inst{\ref{aff203},\ref{aff33}}
\and A.~H.~Wright\orcid{0000-0001-7363-7932}\inst{\ref{aff168}}}
										   
\institute{Institute of Cosmology and Gravitation, University of Portsmouth, Portsmouth PO1 3FX, UK\label{aff1}
\and
Department of Physics, University of Oxford, Keble Road, Oxford OX1 3RH, UK\label{aff2}
\and
University of Applied Sciences and Arts of Northwestern Switzerland, School of Engineering, 5210 Windisch, Switzerland\label{aff3}
\and
Jet Propulsion Laboratory, California Institute of Technology, 4800 Oak Grove Drive, Pasadena, CA, 91109, USA\label{aff4}
\and
Finnish Centre for Astronomy with ESO (FINCA), Quantum, Vesilinnantie 5, University of Turku, 20014 Turku, Finland\label{aff5}
\and
Department of Physics, P.O. Box 64, University of Helsinki, 00014 Helsinki, Finland\label{aff6}
\and
INAF-IASF Milano, Via Alfonso Corti 12, 20133 Milano, Italy\label{aff7}
\and
Dipartimento di Fisica e Scienze della Terra, Universit\`a degli Studi di Ferrara, Via Giuseppe Saragat 1, 44122 Ferrara, Italy\label{aff8}
\and
Departamento F\'isica Aplicada, Universidad Polit\'ecnica de Cartagena, Campus Muralla del Mar, 30202 Cartagena, Murcia, Spain\label{aff9}
\and
Institut de Ci\`{e}ncies del Cosmos (ICCUB), Universitat de Barcelona (IEEC-UB), Mart\'{i} i Franqu\`{e}s 1, 08028 Barcelona, Spain\label{aff10}
\and
University of Applied Sciences and Arts of Northwestern Switzerland, School of Computer Science, 5210 Windisch, Switzerland\label{aff11}
\and
Dipartimento di Fisica e Astronomia "Augusto Righi" - Alma Mater Studiorum Universit\`a di Bologna, via Piero Gobetti 93/2, 40129 Bologna, Italy\label{aff12}
\and
INAF-Osservatorio di Astrofisica e Scienza dello Spazio di Bologna, Via Piero Gobetti 93/3, 40129 Bologna, Italy\label{aff13}
\and
INFN-Sezione di Bologna, Viale Berti Pichat 6/2, 40127 Bologna, Italy\label{aff14}
\and
Instituci\'o Catalana de Recerca i Estudis Avan\c{c}ats (ICREA), Passeig de Llu\'{\i}s Companys 23, 08010 Barcelona, Spain\label{aff15}
\and
Institut de Ciencies de l'Espai (IEEC-CSIC), Campus UAB, Carrer de Can Magrans, s/n Cerdanyola del Vall\'es, 08193 Barcelona, Spain\label{aff16}
\and
School of Mathematics, Statistics and Physics, Newcastle University, Herschel Building, Newcastle-upon-Tyne, NE1 7RU, UK\label{aff17}
\and
Aix-Marseille Universit\'e, CNRS, CNES, LAM, Marseille, France\label{aff18}
\and
Institut d'Astrophysique de Paris, UMR 7095, CNRS, and Sorbonne Universit\'e, 98 bis boulevard Arago, 75014 Paris, France\label{aff19}
\and
Department of Astronomy, University of Geneva, ch. d'Ecogia 16, 1290 Versoix, Switzerland\label{aff20}
\and
Laboratoire univers et particules de Montpellier, Universit\'e de Montpellier, CNRS, 34090 Montpellier, France\label{aff21}
\and
Department of Astronomy, School of Physics and Astronomy, Shanghai Jiao Tong University, Shanghai 200240, China\label{aff22}
\and
State Key Laboratory of Dark Matter Physics, School of Physics and Astronomy, Shanghai Jiao Tong University, Shanghai 200240, China\label{aff23}
\and
Shanghai Key Laboratory for Particle Physics and Cosmology, Shanghai 200240, China\label{aff24}
\and
Kavli Institute for Particle Astrophysics \& Cosmology (KIPAC), Stanford University, Stanford, CA 94305, USA\label{aff25}
\and
SLAC National Accelerator Laboratory, 2575 Sand Hill Road, Menlo Park, CA 94025, USA\label{aff26}
\and
The Inter-University Centre for Astronomy and Astrophysics, Post Bag 4, Ganeshkhind, Pune 411007, India\label{aff27}
\and
Kavli Institute for the Physics and Mathematics of the Universe (WPI), University of Tokyo, Kashiwa, Chiba 277-8583, Japan\label{aff28}
\and
Citizen Scientist, Zooniverse c/o University of Oxford,  Keble Road, Oxford OX1 3RH, UK\label{aff29}
\and
American Museum of Natural History, Department of Astrophysics, New York, NY 10024, USA\label{aff30}
\and
Department of Physics and Astronomy, Lehman College of the CUNY, Bronx, NY 10468, USA\label{aff31}
\and
National Centre for Nuclear Research, ul. Pasteura 7, 02-093, Warsaw, Poland\label{aff32}
\and
Kapteyn Astronomical Institute, University of Groningen, PO Box 800, 9700 AV Groningen, The Netherlands\label{aff33}
\and
STAR Institute, University of Li{\`e}ge, Quartier Agora, All\'ee du six Ao\^ut 19c, 4000 Li\`ege, Belgium\label{aff34}
\and
Universiteit Gent, Department of Physics and Astronomy, Proeftuinstraat 86 N3, 9000 Ghent, Belgium
\label{aff35}
\and
Dipartimento di Fisica e Astronomia, Universit\`a di Bologna, Via Gobetti 93/2, 40129 Bologna, Italy\label{aff36}
\and
Centro de Astrof\'{\i}sica da Universidade do Porto, Rua das Estrelas, 4150-762 Porto, Portugal\label{aff37}
\and
Instituto de Astrof\'isica e Ci\^encias do Espa\c{c}o, Universidade do Porto, CAUP, Rua das Estrelas, PT4150-762 Porto, Portugal\label{aff38}
\and
Max Planck Institute for Extraterrestrial Physics, Giessenbachstr. 1, 85748 Garching, Germany\label{aff39}
\and
Universit\"ats-Sternwarte M\"unchen, Fakult\"at f\"ur Physik, Ludwig-Maximilians-Universit\"at M\"unchen, Scheinerstr.~1, 81679 M\"unchen, Germany\label{aff40}
\and
Universit\'e Lyon 1, CNRS/IN2P3, IP2I Lyon, Villeurbanne, France\label{aff41}
\and
Department of Physics and Astronomy, University of British Columbia, Vancouver, BC V6T 1Z1, Canada\label{aff42}
\and
National Astronomical Observatory of Japan, 2-21-1 Osawa, Mitaka, Tokyo 181-8588, Japan\label{aff43}
\and
INAF-Osservatorio Astronomico di Capodimonte, Via Moiariello 16, 80131 Napoli, Italy\label{aff44}
\and
CUNY Graduate Center, City University of New York, 365 5th Ave., New York, NY 10016, USA\label{aff45}
\and
University of Ljubljana, Faculty of mathematics and physics, Jadranska ulica 19, 1000 Ljubljana, Slovenia\label{aff46}
\and
Minnesota Institute for Astrophysics, University of Minnesota, 116 Church St SE, Minneapolis, MN 55455, USA\label{aff47}
\and
Instituto de F\'isica de Cantabria, Edificio Juan Jord\'a, Avenida de los Castros, 39005 Santander, Spain\label{aff48}
\and
Max-Planck-Institut f\"ur Astrophysik, Karl-Schwarzschild-Str.~1, 85748 Garching, Germany\label{aff49}
\and
INAF, Istituto di Radioastronomia, Via Piero Gobetti 101, 40129 Bologna, Italy\label{aff50}
\and
Jodrell Bank Centre for Astrophysics, Department of Physics and Astronomy, University of Manchester, Oxford Road, Manchester M13 9PL, UK\label{aff51}
\and
Max-Planck-Institut f\"ur Astronomie, K\"onigstuhl 17, 69117 Heidelberg, Germany\label{aff52}
\and
Institute of Astronomy, University of Cambridge, Madingley Road, Cambridge CB3 0HA, UK\label{aff53}
\and
Kavli Institute for Cosmology Cambridge, Madingley Road, Cambridge, CB3 0HA, UK\label{aff54}
\and
Leiden Observatory, Leiden University, Einsteinweg 55, 2333 CC Leiden, The Netherlands\label{aff55}
\and
Department of Physics \& Astronomy, University of California Irvine, Irvine CA 92697, USA\label{aff56}
\and
Departamento de F\'isica, Universidad de Santiago de Chile, Av. Victor Jara 3493, Santiago, Chile\label{aff57}
\and
Center for Interdisciplinary Research in Astrophysics Space Science (CIRAS), Universidad de Santiago, Chile\label{aff58}
\and
ESAC/ESA, Camino Bajo del Castillo, s/n., Urb. Villafranca del Castillo, 28692 Villanueva de la Ca\~nada, Madrid, Spain\label{aff59}
\and
School of Physical Sciences, The Open University, Milton Keynes, MK7 6AA, UK\label{aff60}
\and
Caltech/IPAC, 1200 E. California Blvd., Pasadena, CA 91125, USA\label{aff61}
\and
School of Physics and Astronomy, University of Birmingham, Birmingham, B15 2TT, UK\label{aff62}
\and
Universit\'e de Strasbourg, CNRS, Observatoire astronomique de Strasbourg, UMR 7550, 67000 Strasbourg, France\label{aff63}
\and
Centre National d'Etudes Spatiales -- Centre spatial de Toulouse, 18 avenue Edouard Belin, 31401 Toulouse Cedex 9, France\label{aff64}
\and
European Southern Observatory, Karl-Schwarzschild-Str.~2, 85748 Garching, Germany\label{aff65}
\and
Institute for Astronomy, University of Hawaii, 2680 Woodlawn Drive, Honolulu, HI 96822, USA\label{aff66}
\and
University of Southern California, 3551 Trousdale Parkway, Los Angeles, CA 90089, USA\label{aff67}
\and
Max Planck Institute for Gravitational Physics (Albert Einstein Institute), Am Muhlenberg 1, D-14476 Potsdam-Golm, Germany\label{aff68}
\and
Instituto de F\'isica Te\'orica UAM-CSIC, Campus de Cantoblanco, 28049 Madrid, Spain\label{aff69}
\and
Universit\'e Paris-Saclay, CNRS, Institut d'astrophysique spatiale, 91405, Orsay, France\label{aff70}
\and
INAF-Osservatorio Astronomico di Brera, Via Brera 28, 20122 Milano, Italy\label{aff71}
\and
IFPU, Institute for Fundamental Physics of the Universe, via Beirut 2, 34151 Trieste, Italy\label{aff72}
\and
INAF-Osservatorio Astronomico di Trieste, Via G. B. Tiepolo 11, 34143 Trieste, Italy\label{aff73}
\and
INFN, Sezione di Trieste, Via Valerio 2, 34127 Trieste TS, Italy\label{aff74}
\and
SISSA, International School for Advanced Studies, Via Bonomea 265, 34136 Trieste TS, Italy\label{aff75}
\and
Dipartimento di Fisica, Universit\`a di Genova, Via Dodecaneso 33, 16146, Genova, Italy\label{aff76}
\and
INFN-Sezione di Genova, Via Dodecaneso 33, 16146, Genova, Italy\label{aff77}
\and
Department of Physics "E. Pancini", University Federico II, Via Cinthia 6, 80126, Napoli, Italy\label{aff78}
\and
Dipartimento di Fisica, Universit\`a degli Studi di Torino, Via P. Giuria 1, 10125 Torino, Italy\label{aff79}
\and
INFN-Sezione di Torino, Via P. Giuria 1, 10125 Torino, Italy\label{aff80}
\and
INAF-Osservatorio Astrofisico di Torino, Via Osservatorio 20, 10025 Pino Torinese (TO), Italy\label{aff81}
\and
INAF-Osservatorio Astronomico di Roma, Via Frascati 33, 00078 Monteporzio Catone, Italy\label{aff82}
\and
INFN-Sezione di Roma, Piazzale Aldo Moro, 2 - c/o Dipartimento di Fisica, Edificio G. Marconi, 00185 Roma, Italy\label{aff83}
\and
Centro de Investigaciones Energ\'eticas, Medioambientales y Tecnol\'ogicas (CIEMAT), Avenida Complutense 40, 28040 Madrid, Spain\label{aff84}
\and
Port d'Informaci\'{o} Cient\'{i}fica, Campus UAB, C. Albareda s/n, 08193 Bellaterra (Barcelona), Spain\label{aff85}
\and
INFN -- Sezione di Napoli, Via Cinthia 6, 80126, Napoli, Italy\label{aff86}
\and
Dipartimento di Fisica e Astronomia "Augusto Righi" - Alma Mater Studiorum Universit\`a di Bologna, Viale Berti Pichat 6/2, 40127 Bologna, Italy\label{aff87}
\and
Instituto de Astrof\'{\i}sica de Canarias, E-38205 La Laguna, Tenerife, Spain\label{aff88}
\and
Institute for Astronomy, University of Edinburgh, Royal Observatory, Blackford Hill, Edinburgh EH9 3HJ, UK\label{aff89}
\and
European Space Agency/ESRIN, Largo Galileo Galilei 1, 00044 Frascati, Roma, Italy\label{aff90}
\and
Universit\'e Lyon 1, CNRS/IN2P3, IUF, IP2I Lyon, Villeurbanne, France \label{aff91}
\and
Mullard Space Science Laboratory, University College London, Holmbury St Mary, Dorking, Surrey RH5 6NT, UK\label{aff92}
\and
INFN-Padova, Via Marzolo 8, 35131 Padova, Italy\label{aff93}
\and
Aix-Marseille Universit\'e, CNRS/IN2P3, CPPM, Marseille, France\label{aff94}
\and
INAF-Istituto di Astrofisica e Planetologia Spaziali, via del Fosso del Cavaliere, 100, 00100 Roma, Italy\label{aff95}
\and
INFN-Bologna, Via Irnerio 46, 40126 Bologna, Italy\label{aff96}
\and
School of Physics, HH Wills Physics Laboratory, University of Bristol, Tyndall Avenue, Bristol, BS8 1TL, UK\label{aff97}
\and
University Observatory, LMU Faculty of Physics, Scheinerstr.~1, 81679 Munich, Germany\label{aff98}
\and
INAF-Osservatorio Astronomico di Padova, Via dell'Osservatorio 5, 35122 Padova, Italy\label{aff99}
\and
Institute of Theoretical Astrophysics, University of Oslo, P.O. Box 1029 Blindern, 0315 Oslo, Norway\label{aff100}
\and
Felix Hormuth Engineering, Goethestr. 17, 69181 Leimen, Germany\label{aff101}
\and
Technical University of Denmark, Elektrovej 327, 2800 Kgs. Lyngby, Denmark\label{aff102}
\and
Cosmic Dawn Center (DAWN), Denmark\label{aff103}
\and
Universit\'e PSL, Observatoire de Paris, Sorbonne Universit\'e, CNRS, LERMA, 75014, Paris, France\label{aff104}
\and
Universit\'e Paris-Cit\'e, 5 Rue Thomas Mann, 75013, Paris, France\label{aff105}
\and
NASA Goddard Space Flight Center, Greenbelt, MD 20771, USA\label{aff106}
\and
Department of Physics and Astronomy, University College London, Gower Street, London WC1E 6BT, UK\label{aff107}
\and
Universit\'e de Gen\`eve, D\'epartement de Physique Th\'eorique and Centre for Astroparticle Physics, 24 quai Ernest-Ansermet, CH-1211 Gen\`eve 4, Switzerland\label{aff108}
\and
Helsinki Institute of Physics, Gustaf H{\"a}llstr{\"o}min katu 2, University of Helsinki, 00014 Helsinki, Finland\label{aff109}
\and
Laboratoire d'etude de l'Univers et des phenomenes eXtremes, Observatoire de Paris, Universit\'e PSL, Sorbonne Universit\'e, CNRS, 92190 Meudon, France\label{aff110}
\and
SKAO, Jodrell Bank, Lower Withington, Macclesfield SK11 9FT, UK\label{aff111}
\and
Centre de Calcul de l'IN2P3/CNRS, 21 avenue Pierre de Coubertin 69627 Villeurbanne Cedex, France\label{aff112}
\and
Universit\"at Bonn, Argelander-Institut f\"ur Astronomie, Auf dem H\"ugel 71, 53121 Bonn, Germany\label{aff113}
\and
Department of Physics, Institute for Computational Cosmology, Durham University, South Road, Durham, DH1 3LE, UK\label{aff114}
\and
Universit\'e Paris Cit\'e, CNRS, Astroparticule et Cosmologie, 75013 Paris, France\label{aff115}
\and
CNRS-UCB International Research Laboratory, Centre Pierre Bin\'etruy, IRL2007, CPB-IN2P3, Berkeley, USA\label{aff116}
\and
Institute of Physics, Laboratory of Astrophysics, Ecole Polytechnique F\'ed\'erale de Lausanne (EPFL), Observatoire de Sauverny, 1290 Versoix, Switzerland\label{aff117}
\and
Dipartimento di Fisica - Sezione di Astronomia, Universit\`a di Trieste, Via Tiepolo 11, 34131 Trieste, Italy\label{aff118}
\and
Telespazio UK S.L. for European Space Agency (ESA), Camino bajo del Castillo, s/n, Urbanizacion Villafranca del Castillo, Villanueva de la Ca\~nada, 28692 Madrid, Spain\label{aff119}
\and
Institut de F\'{i}sica d'Altes Energies (IFAE), The Barcelona Institute of Science and Technology, Campus UAB, 08193 Bellaterra (Barcelona), Spain\label{aff120}
\and
School of Mathematics and Physics, University of Surrey, Guildford, Surrey, GU2 7XH, UK\label{aff121}
\and
European Space Agency/ESTEC, Keplerlaan 1, 2201 AZ Noordwijk, The Netherlands\label{aff122}
\and
Waterloo Centre for Astrophysics, University of Waterloo, Waterloo, Ontario N2L 3G1, Canada\label{aff123}
\and
Department of Physics and Astronomy, University of Waterloo, Waterloo, Ontario N2L 3G1, Canada\label{aff124}
\and
Perimeter Institute for Theoretical Physics, Waterloo, Ontario N2L 2Y5, Canada\label{aff125}
\and
Space Science Data Center, Italian Space Agency, via del Politecnico snc, 00133 Roma, Italy\label{aff126}
\and
Institute of Space Science, Str. Atomistilor, nr. 409 M\u{a}gurele, Ilfov, 077125, Romania\label{aff127}
\and
Consejo Superior de Investigaciones Cientificas, Calle Serrano 117, 28006 Madrid, Spain\label{aff128}
\and
Universidad de La Laguna, Dpto. Astrof\'\i sica, E-38206 La Laguna, Tenerife, Spain\label{aff129}
\and
Dipartimento di Fisica e Astronomia "G. Galilei", Universit\`a di Padova, Via Marzolo 8, 35131 Padova, Italy\label{aff130}
\and
Institut de Recherche en Astrophysique et Plan\'etologie (IRAP), Universit\'e de Toulouse, CNRS, UPS, CNES, 14 Av. Edouard Belin, 31400 Toulouse, France\label{aff131}
\and
Universit\'e St Joseph; Faculty of Sciences, Beirut, Lebanon\label{aff132}
\and
Departamento de F\'isica, FCFM, Universidad de Chile, Blanco Encalada 2008, Santiago, Chile\label{aff133}
\and
Universit\"at Innsbruck, Institut f\"ur Astro- und Teilchenphysik, Technikerstr. 25/8, 6020 Innsbruck, Austria\label{aff134}
\and
Department of Physics and Helsinki Institute of Physics, Gustaf H\"allstr\"omin katu 2, University of Helsinki, 00014 Helsinki, Finland\label{aff135}
\and
Universit\'e Paris-Saclay, Universit\'e Paris Cit\'e, CEA, CNRS, AIM, 91191, Gif-sur-Yvette, France\label{aff136}
\and
Infrared Processing and Analysis Center, California Institute of Technology, Pasadena, CA 91125, USA\label{aff137}
\and
Departamento de F\'isica, Faculdade de Ci\^encias, Universidade de Lisboa, Edif\'icio C8, Campo Grande, PT1749-016 Lisboa, Portugal\label{aff138}
\and
Instituto de Astrof\'isica e Ci\^encias do Espa\c{c}o, Faculdade de Ci\^encias, Universidade de Lisboa, Tapada da Ajuda, 1349-018 Lisboa, Portugal\label{aff139}
\and
Cosmic Dawn Center (DAWN)\label{aff140}
\and
Niels Bohr Institute, University of Copenhagen, Jagtvej 128, 2200 Copenhagen, Denmark\label{aff141}
\and
Universidad Polit\'ecnica de Cartagena, Departamento de Electr\'onica y Tecnolog\'ia de Computadoras,  Plaza del Hospital 1, 30202 Cartagena, Spain\label{aff142}
\and
European University of Technology EUt+, European Union\label{aff143}
\and
Institute of Space Sciences (ICE, CSIC), Campus UAB, Carrer de Can Magrans, s/n, 08193 Barcelona, Spain\label{aff144}
\and
Institut d'Estudis Espacials de Catalunya (IEEC),  Edifici RDIT, Campus UPC, 08860 Castelldefels, Barcelona, Spain\label{aff145}
\and
Istituto Nazionale di Fisica Nucleare, Sezione di Ferrara, Via Giuseppe Saragat 1, 44122 Ferrara, Italy\label{aff146}
\and
Astronomical Observatory of the Autonomous Region of the Aosta Valley (OAVdA), Loc. Lignan 39, I-11020, Nus (Aosta Valley), Italy\label{aff147}
\and
Universit\'e C\^{o}te d'Azur, Observatoire de la C\^{o}te d'Azur, CNRS, Laboratoire Lagrange, Bd de l'Observatoire, CS 34229, 06304 Nice cedex 4, France\label{aff148}
\and
Department of Mathematical Physics, Institute of Physics, University of S\~ao Paulo, R. do Mat\~ao 1371, 05508-090, S\~ao Paulo, SP, Brazil\label{aff149}
\and
Oskar Klein Centre for Cosmoparticle Physics, Department of Physics, Stockholm University, Stockholm, SE-106 91, Sweden\label{aff150}
\and
Astrophysics Group, Blackett Laboratory, Imperial College London, London SW7 2AZ, UK\label{aff151}
\and
Univ. Grenoble Alpes, CNRS, Grenoble INP, LPSC-IN2P3, 53, Avenue des Martyrs, 38000 Grenoble, France\label{aff152}
\and
Dipartimento di Fisica, Sapienza Universit\`a di Roma, Piazzale Aldo Moro 2, 00185 Roma, Italy\label{aff153}
\and
Department of Mathematics and Physics E. De Giorgi, University of Salento, Via per Arnesano, CP-I93, 73100, Lecce, Italy\label{aff154}
\and
INFN, Sezione di Lecce, Via per Arnesano, CP-193, 73100, Lecce, Italy\label{aff155}
\and
INAF-Sezione di Lecce, c/o Dipartimento Matematica e Fisica, Via per Arnesano, 73100, Lecce, Italy\label{aff156}
\and
Institut d'Astrophysique de Paris, 98bis Boulevard Arago, 75014, Paris, France\label{aff157}
\and
ICL, Junia, Universit\'e Catholique de Lille, LITL, 59000 Lille, France\label{aff158}
\and
ICSC - Centro Nazionale di Ricerca in High Performance Computing, Big Data e Quantum Computing, Via Magnanelli 2, Bologna, Italy\label{aff159}
\and
CERCA/ISO, Department of Physics, Case Western Reserve University, 10900 Euclid Avenue, Cleveland, OH 44106, USA\label{aff160}
\and
Laboratoire Univers et Th\'eorie, Observatoire de Paris, Universit\'e PSL, Universit\'e Paris Cit\'e, CNRS, 92190 Meudon, France\label{aff161}
\and
Dipartimento di Fisica "Aldo Pontremoli", Universit\`a degli Studi di Milano, Via Celoria 16, 20133 Milano, Italy\label{aff162}
\and
INFN-Sezione di Milano, Via Celoria 16, 20133 Milano, Italy\label{aff163}
\and
Departamento de F{\'\i}sica Fundamental. Universidad de Salamanca. Plaza de la Merced s/n. 37008 Salamanca, Spain\label{aff164}
\and
Center for Data-Driven Discovery, Kavli IPMU (WPI), UTIAS, The University of Tokyo, Kashiwa, Chiba 277-8583, Japan\label{aff165}
\and
California Institute of Technology, 1200 E California Blvd, Pasadena, CA 91125, USA\label{aff166}
\and
Department of Computer Science, Aalto University, PO Box 15400, Espoo, FI-00 076, Finland\label{aff167}
\and
Ruhr University Bochum, Faculty of Physics and Astronomy, Astronomical Institute (AIRUB), German Centre for Cosmological Lensing (GCCL), 44780 Bochum, Germany\label{aff168}
\and
DARK, Niels Bohr Institute, University of Copenhagen, Jagtvej 155, 2200 Copenhagen, Denmark\label{aff169}
\and
Department of Physics, Centre for Extragalactic Astronomy, Durham University, South Road, Durham, DH1 3LE, UK\label{aff170}
\and
Astrophysics Research Centre, University of KwaZulu-Natal, Westville Campus, Durban 4041, South Africa\label{aff171}
\and
School of Mathematics, Statistics \& Computer Science, University of KwaZulu-Natal, Westville Campus, Durban 4041, South Africa\label{aff172}
\and
Department of Physics and Astronomy, Vesilinnantie 5, University of Turku, 20014 Turku, Finland\label{aff173}
\and
Serco for European Space Agency (ESA), Camino bajo del Castillo, s/n, Urbanizacion Villafranca del Castillo, Villanueva de la Ca\~nada, 28692 Madrid, Spain\label{aff174}
\and
ARC Centre of Excellence for Dark Matter Particle Physics, Melbourne, Australia\label{aff175}
\and
Centre for Astrophysics \& Supercomputing, Swinburne University of Technology,  Hawthorn, Victoria 3122, Australia\label{aff176}
\and
Department of Physics and Astronomy, University of the Western Cape, Bellville, Cape Town, 7535, South Africa\label{aff177}
\and
MTA-CSFK Lend\"ulet Large-Scale Structure Research Group, Konkoly-Thege Mikl\'os \'ut 15-17, H-1121 Budapest, Hungary\label{aff178}
\and
Konkoly Observatory, HUN-REN CSFK, MTA Centre of Excellence, Budapest, Konkoly Thege Mikl\'os {\'u}t 15-17. H-1121, Hungary\label{aff179}
\and
Brazilian Center for Research in Physics (CBPF), Dr. Xavier Sigaud st. 150, zip 22290-180, Rio de Janeiro, RJ, Brazil\label{aff180}
\and
Institute for Theoretical Particle Physics and Cosmology (TTK), RWTH Aachen University, 52056 Aachen, Germany\label{aff181}
\and
IRFU, CEA, Universit\'e Paris-Saclay 91191 Gif-sur-Yvette Cedex, France\label{aff182}
\and
INAF-Osservatorio Astrofisico di Arcetri, Largo E. Fermi 5, 50125, Firenze, Italy\label{aff183}
\and
HE Space for European Space Agency (ESA), Camino bajo del Castillo, s/n, Urbanizacion Villafranca del Castillo, Villanueva de la Ca\~nada, 28692 Madrid, Spain\label{aff184}
\and
Center for Frontier Science, Chiba University, 1-33 Yayoi-cho, Inage-ku, Chiba 263-8522, Japan\label{aff185}
\and
Department of Physics, Graduate School of Science, Chiba University, 1-33 Yayoi-Cho, Inage-Ku, Chiba 263-8522, Japan\label{aff186}
\and
Instituto de Astrof\'isica e Ci\^encias do Espa\c{c}o, Faculdade de Ci\^encias, Universidade de Lisboa, Campo Grande, 1749-016 Lisboa, Portugal\label{aff187}
\and
Department of Astrophysics, University of Zurich, Winterthurerstrasse 190, 8057 Zurich, Switzerland\label{aff188}
\and
INAF - Osservatorio Astronomico d'Abruzzo, Via Maggini, 64100, Teramo, Italy\label{aff189}
\and
Theoretical astrophysics, Department of Physics and Astronomy, Uppsala University, Box 524, 751 20 Uppsala, Sweden\label{aff190}
\and
Mathematical Institute, University of Leiden, Einsteinweg 55, 2333 CA Leiden, The Netherlands\label{aff191}
\and
School of Physics \& Astronomy, University of Southampton, Highfield Campus, Southampton SO17 1BJ, UK\label{aff192}
\and
Univ. Lille, CNRS, Centrale Lille, UMR 9189 CRIStAL, 59000 Lille, France\label{aff193}
\and
Center for Astrophysics and Cosmology, University of Nova Gorica, Nova Gorica, Slovenia\label{aff194}
\and
Kobayashi-Maskawa Institute for the Origin of Particles and the Universe, Nagoya University, Chikusa-ku, Nagoya, 464-8602, Japan\label{aff195}
\and
Institute for Particle Physics and Astrophysics, Dept. of Physics, ETH Zurich, Wolfgang-Pauli-Strasse 27, 8093 Zurich, Switzerland\label{aff196}
\and
Department of Astrophysical Sciences, Peyton Hall, Princeton University, Princeton, NJ 08544, USA\label{aff197}
\and
Fakult\"at f\"ur Physik, Universit\"at Bielefeld, Postfach 100131, 33501 Bielefeld, Germany\label{aff198}
\and
Space physics and astronomy research unit, University of Oulu, Pentti Kaiteran katu 1, FI-90014 Oulu, Finland\label{aff199}
\and
International Centre for Theoretical Physics (ICTP), Strada Costiera 11, 34151 Trieste, Italy\label{aff200}
\and
Center for Computational Astrophysics, Flatiron Institute, 162 5th Avenue, 10010, New York, NY, USA\label{aff201}
\and
David A. Dunlap Department of Astronomy \& Astrophysics, University of Toronto, 50 St George Street, Toronto, Ontario M5S 3H4, Canada\label{aff202}
\and
SRON Netherlands Institute for Space Research, Landleven 12, 9747 AD, Groningen, The Netherlands\label{aff203}}    
   
\abstract{
We present the method used to identify the \rt{$15\,000$} high-grade galaxy--galaxy strong lens candidates in Euclid Data Release 1 (DR1). Given that lenses are intrinsically rare systems and the DR1 data set is vast, this lens search was conducted in a number of stages: a catalogue-level cut; a machine-learning and ensemble search; a two-stage citizen-science search, and finally grading of lens candidates by strong lens researchers. In this paper, we describe each search stage and their respective results. 
This search was enabled by the automated classification of over 
\rt{$68\,$million}  
objects by neural networks, more than 
\rt{$4.5\,$million} 
classifications of 
\rt{$300\,000$} 
objects by citizen-scientist volunteers, and the inspection of over
\rt{$30\,000$}  
objects by strong lens researchers. 
We find this multi-stage approach helps maximise the number of lenses identified, while minimising inspection requirements of both citizen scientists and strong lens researchers. 
We measure a completeness in our comprehensive search of \rt{$0.73_{-0.15}^{+0.08}$} to grade A (`definite') systems and \rt{$0.34_{-0.06}^{+0.05}$} to grade A+B (`definite' + `probable') systems. Our search therefore identifies the majority of the definitive strong lens systems in the DR1 area with the remainder \rt{likely to be fainter} or more ambiguous cases.
We find that researcher grading remains subjective; in this work we implement calibration to account for biases of individual inspectors and highlight the benefit of lens modelling (ECLIPSE-C) in determining the nature of a lens candidate. 
This paper forms part of the ECLIPSE series, here focussing on the discovery pipeline used to identify lens candidates in Euclid DR1; an overview of the lenses discovered is presented in ECLIPSE-A.
}
%
%
%
%
   \titlerunning{ECLIPSE B -- Discovery pipeline overview}
   \authorrunning{Euclid Collaboration: Holloway et al.}
   
   \maketitle
   \nolinenumbers
%
%
%
%
\section{Introduction}\label{Sect: Intro}
Gravitational lensing, the deflection of light from a background source by a foreground mass, can be used as a probe of both astrophysics and cosmology. Strong gravitational lenses are systems in which this deflection is sufficiently strong that multiple images of the same background source are produced. Such systems can be used to constrain 
the Hubble constant \citep{Refsdal1964,Treu2022,TDCOSMO2025}, 
evolution of dark energy \citep{Collett2014,Li2024,Shajib2025},
initial mass function (IMF, \citealp{Leier2016,Newman2017,Sonnenfeld2025}), galaxy mass profiles \citep{Koopmans2006,Sonnenfeld2013}, and probe the large-scale distribution of matter via line-of-sight shear \citep{Fleury2026}, 
amongst many other science applications. However, such systems are very rare with extragalactic strong lensing requiring the close angular alignment of the background source (e.g. a galaxy, quasar, or supernova), and a foreground massive galaxy (e.g. a galaxy, galaxy group, or cluster). 

The \Euclid satellite will detect 1.5 billion galaxies over the Euclid Wide Survey (EWS, \citealp{Scaramella-EP1,EuclidSkyOverview}). With an area of $14\,000\deg^2$ \citep{EuclidSkyOverview}, an \IE PSF with a full-width at half-maximum of \rt{\ang{;;0.16}} \citep{EuclidSkyVIS,DR1-TP002}, as well as three near-infrared bands providing crucial colour information \citep{EuclidSkyNISP}, the survey will revolutionise the field of strong lensing and is expected to identify more than $100\,000$ galaxy--galaxy lenses \citep{Collett2015}. However, given their complexity and wide range of configurations, identifying these systems is a significant challenge. Historically, strong lenses were identified using arc-finding algorithms \citep{Seidel2007,Gavazzi2014,Sonnenfeld2018}. However, successively larger data volumes meant that citizen-science \citep{Marshall2016,More2016}, and more recently machine-learning methods (ML, e.g., \citealp{Schuldt2025,González2025,Q1-SP053}), have been relied upon. 

The Euclid Quick Data Release (Q1, \citealp{Q1cite,Q1-TP001}), comprising  $63.1\deg^2$ of imaging, provided an opportunity to test state-of-the-art strong lens classifiers and demonstrate the revolutionary potential of lens finding in \Euclid. The systematic search was presented in the Strong Lens Discovery Engine series (\citealp{Q1-SP048}, \citealp{Q1-SP052}, \citealp{Q1-SP053}, \citealp{Q1-SP054}, \citealp{Q1-SP059}, \citealp{Ecker2026}, SLDE A--F respectively) and identified 
approximately $600$ 
high-grade lens systems. 

The Euclid Data Release 1 (DR1; \citealp{DR1cite,DR1-TP001}), which consists of $\rt{2000}\deg^2$ of area, presents a unique opportunity to identify over $10\,000$ galaxy-scale lens systems across a homogeneous data set. This represents a significant advancement in the field of strong lensing, but identifying these lenses presents a number of challenges. Firstly, the data volume is vast, requiring scalable search techniques to analyse such large data sets. Secondly, the total number of detectable lenses in DR1 is comparable to the inspection budget of strong lens researchers, $\sim10^4$, i.e., the total number of systems that inspectors could plausibly grade.\footnote{Following the completion of a lens search, it is common for researchers to grade lens candidates according to their likelihood that they are truly lenses. Given this is a time-consuming process this can become a limiting factor for a large-scale lens search.} The sample flagged for expert grading must therefore have very high purity to ensure that the majority of lens candidates are identified. This paper presents the discovery pipeline used to tackle these challenges, from ML to citizen inspection to grading of high-scoring candidates by strong lens researchers. This discovery pipeline was tuned to maximise the lens purity of systems inspected by experts, without compromising on the completeness of high-grade lenses. 

The DR1 lens search differs from the Q1 search in a number of ways. While both searches include an ML classification, citizen-science classification and expert grading stage, the DR1 search includes two citizen-science inspection stages to further refine the lens sample. Secondly, the DR1 ML classification utilised networks retrained on a much larger sample of real lenses, including lenses within the DR1 area. Thirdly the DR1 classification pipeline measured and equitably accounted for the variation in performance of the ML classifiers in order to maximise the number of lenses identified. Finally, the DR1 galaxy selection was loosened with respect to Q1 (Sect.\,\ref{Sect: Catalogue Cuts}) to ensure low-redshift lenses were not excluded \citep{Ecker2026}.

This paper is part of the ECLIPSE series that describes the DR1 lens search and catalogue. \citet{\Lines} presents the galaxy--galaxy strong lens catalogue, focussing on the nature of the lenses identified, and their properties. \citet{\Nightingale} presents modelling results for the high-confidence lenses, and \citet{\Nersesian} describes the photometric redshifts and physical properties of the strong lens candidates. We point to further papers for detailed description of aspects of the search and lens catalogue. \citet{\Li} and \Ecker{} present the \Euclid double-source plane lenses and edge-on lenses, respectively and Euclid Collaboration: Wielders et al. (in prep.) presents automated modelling of the multi-galaxy lens sample, i.e. strong lenses with more than one lens galaxy.
\Vincken{} compares the behaviour of the different ML lens classifiers and \citet{\Fogliardi} investigates the impact of training choices on ML lens classification performance. Alongside the modelling results, \citet{\Vinckensegmentation} presents segmentation maps, which mask the lens, source, and background objects in the surroundings of each lens candidate.

In comparison, this paper focusses on the lens search strategy, answering why and how the DR1 lens search was conducted as it was. Given the scale of the lens search, this differs from typical lens searches, and was informed by the inspection budget of both citizen-scientists and researchers. We also investigate the lens recovery rate with respect to previously-identified lens systems, and the inspection requirements to identify even greater numbers of lenses. 

This paper is structured as follows. In Sect.\,\ref{Sect: Method} we describe the different aspects of the classification pipeline including the individual inspection stages and classification methods used.
In Sect.\,\ref{Sect: Results} we describe the results of each stage of the inspection pipeline, and in Sect.\,\ref{Sect: Discussion} we discuss the lens selection function, completeness, and future challenges for lens searches. We conclude in Sect.\,\ref{Sect: Conclusions}.

\section{Method}\label{Sect: Method}
\subsection{Method overview}
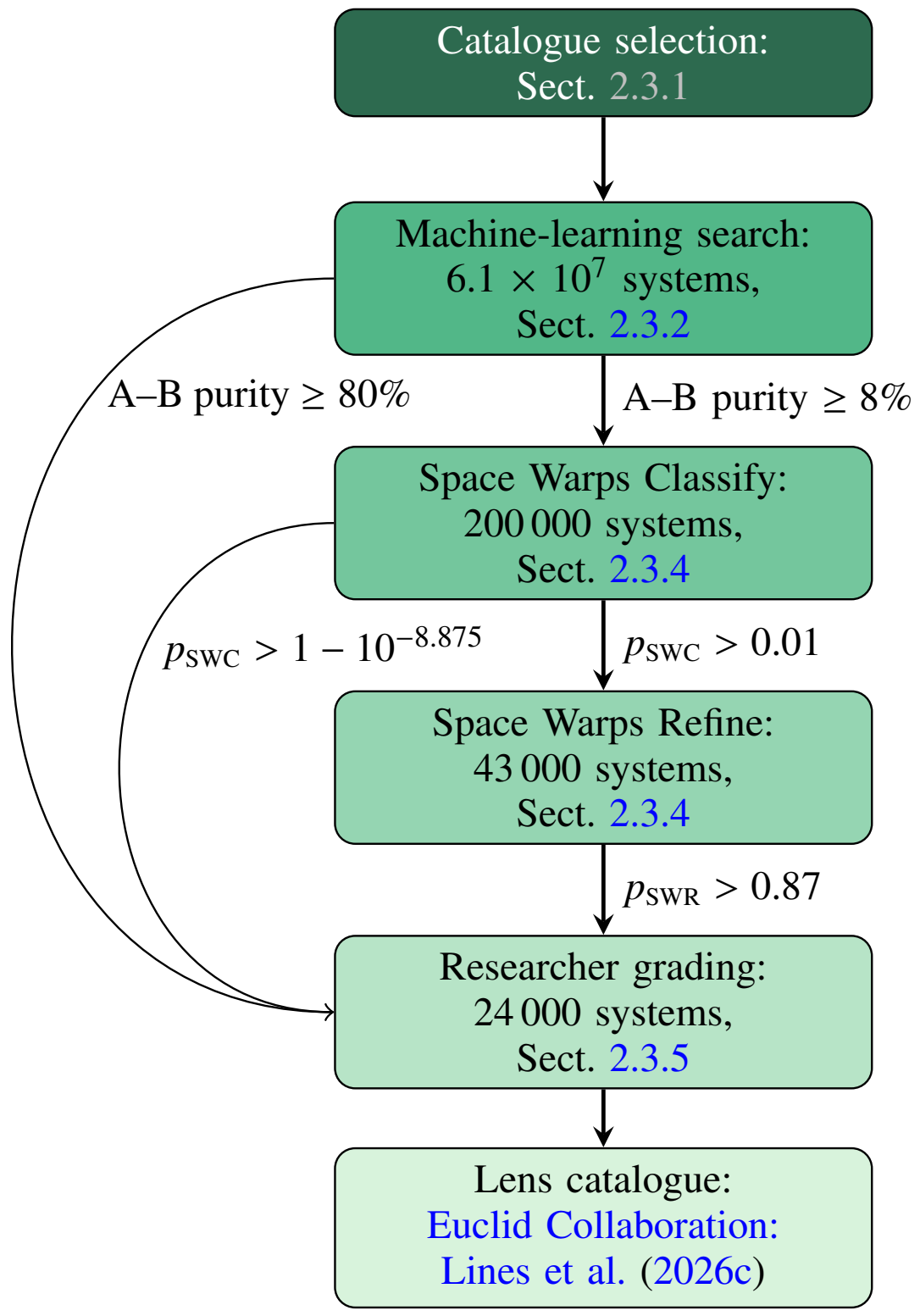
\begin {figure}[t]
\centering 
\begin{adjustbox}{width=0.8\textwidth}
\begin{tikzpicture}[node distance=0.5cm] 
\node(Galaxy_Selection)[flowbox={color1}{white}] {Catalogue selection:\\
            Sect.
            {\hypersetup{linkcolor=lightgray}\ref{Sect: Catalogue Cuts}}
            };
\node(ML_Search)[flowbox={color2}{black}, 
                below of=Galaxy_Selection,  
                yshift=-1cm] {Machine-learning search:\\
                \rt{$6.1\times10^7$ systems},\\
                Sect. \ref{Sect: ML Intro}}; 
\node(SWC_Search)[flowbox={color3}{black},
                 below of=ML_Search,
                 yshift=-1.2cm] {
                 Space Warps Classify:\\
                 \rt{200\,000 systems},\\
                 Sect. \ref{Sect: SW Intro}}; 
\node(SWR_Search)[flowbox={color4}{black},
                below of=SWC_Search,
                yshift=-1.2cm]{
                Space Warps Refine:\\
                \rt{43\,000 systems},\\
                Sect. \ref{Sect: SW Intro}}; 
\node(GJ_Inspection)[flowbox={color5}{black},, 
                     below of=SWR_Search,yshift=-1.2cm]{
                 Researcher grading:\\
                 \rt{24\,000 systems},\\
                 Sect. \ref{Sect: GJ Intro}
                     }; 
\node(Lens_Catalogue)[flowbox={color6}{black},
                      below of=GJ_Inspection,
                      yshift=-1cm] {Lens catalogue:\\\citet{\Lines}}; 
\draw [arrow] (Galaxy_Selection) -- (ML_Search); 

\draw [arrow] (ML_Search) -- node[midway, right, text width = 4cm, align=left]{A--B purity $\geq\rt{8\%}$} (SWC_Search); 

\draw[->] (SWC_Search.west)
  to[out=180, in=180, looseness=1.5]
  node[pos=0.33,right,align=left] {\tiny $p_{\rm{\sfont{SWC}}}%
  >\rt{1-10^{-8.875}}$} (GJ_Inspection.west);

\draw[->] (ML_Search.west) 
    to[out=180, in=180, looseness=1.5] 
    node[pos=0.255,right,align=left] {\tiny A--B purity $\geq\rt{80\%}$} (GJ_Inspection.west);

\draw [arrow] (SWC_Search) -- node[midway, right, text width = 8cm, align=left]{$p_{\rm{\sfont{SWC}}}
>\rt{0.01}$} (SWR_Search);
\draw [arrow] (SWR_Search) -- node[midway, right, text width = 8cm, align=left]{$p_{\rm{\sfont{SWR}}}>\rt{0.87}$} (GJ_Inspection);
\draw [arrow] (GJ_Inspection) -- (Lens_Catalogue); 
\end{tikzpicture} 
\end{adjustbox}
\caption{Flow chart of the strong lens detection pipeline, along with the number of systems classified at each stage. Deep tiles and those overlapping with Q1 are excluded from the quoted figures along with a $250\deg^2$ region of DR1 classified separately. We define $p_{\rm{\sfont{SWC}}}\equiv P({\rm{Lens}}|\{C_{{\rm{SWC}},i}\})$ and $p_{\rm{\sfont{SWR}}}\equiv P(G\in\{\rm{A,B}\}|\{C_{{\rm{SWR}},\textit{i}}\})$. The left-hand arrows indicate shortcut inspection pathways from early classification stages (ML and \SWCt{}) to GJ grading. These shortcuts helped to ensure that the highest-scoring systems were inspected by experts and were not missed by intermediate inspection stages. To be passed along these shortcuts, systems had to pass a higher score or purity threshold than required for the complete inspection pathway.}
\label{Inspection_Flowchart}
\end{figure}
Here we give a high-level overview of the discovery pipeline, and refer to individual sections for further details. Given the vast DR1 data set, the strong lens search was conducted using a range of lens-finding methods in a classification pipeline, in which likely lens candidates were passed to the next classification phase if they passed certain lens likelihood thresholds. The first of these stages was a catalogue-level cut (Sect.\,\ref{Sect: Catalogue Cuts}), designed to identify bright galaxies within the \Euclid MER \citep{DR1-TP009} catalogue. Objects that passed these cuts were then applied to multiple ML models (Sect.\,\ref{Sect: ML Intro}), trained to identify lenses. Systems which were scored highly by one or more of these networks were then inspected by citizen-scientists as part of the Space Warps lens finding project\footnote{\url{www.spacewarps.org}} (SW, Sect.\,\ref{Sect: SW Intro}). The citizen-science inspection was formed of two parts, a binary classification stage, hereafter \SWCt{}, and a second refinement stage, hereafter \SWRt{}. Finally, systems that received high scores from the SW volunteers (at either SW stage) were passed to strong lens researchers for grading in a phase termed `Galaxy Judges' (GJ, Sect.\,\ref{Sect: GJ Intro}). We show the inspection workflow in Fig.\,\ref{Inspection_Flowchart}. This staged approach ensured that the most likely lens candidates were inspected and graded by researchers, while less likely systems were excluded by increasingly tight constraints.

The search was conducted in three stages: a $200\deg^2$ search (Sect.\,\ref{Sect: 200deg Search Intro}), a $50\deg^2$ search (Sect.\,\ref{Sect: 50deg Search Intro}), and a more comprehensive search of the remaining data set (Sect.\,\ref{Sect: Systematic Search Intro}). The first of these was used to identify lens candidates with which to further improve the ML classifiers. The second was used to measure the performance of these re-trained networks, and the third searched the remaining area, guided by the findings of the $50\deg^2$ search.

\subsection{Image processing}
The strong lens search detailed in this work was applied to DR1 that comprises \rt{$2000\deg^2$} of the sky \citep{DR1-TP001}. Both the ML and human inspection stages used $10\arcsec\times10\arcsec$ cutouts (suitable for galaxy-scale strong lenses). These cutouts were generated using three of the \Euclid bands, \IE (from the VIS instrument) and \YE/\JE (both from the NISP instrument). For the visual inspection, these cutouts used either arcsinh or mid-tone transfer function (MTF) scaling (see SLDE-A). These cutouts were generated from the stacked MER images \citep{DR1-TP009} accessed via the ESA Science Archive Service and ESA Datalabs \citep{Datalabscite} platform. The ML classifiers primarily used \IE images, while the human classifiers primarily inspected \IE, \IE/\YE, \IE/\JE, and \IE/\YE/\JE cutouts with arcsinh or MTF stretch applied on a per-band basis prior to forming the RGB image. Given the differing instrumental pixel scales of the VIS and NISP instruments 
(\ang{;;0.1} and \ang{;;0.3} respectively)
as well as the larger NISP PSF, the \IE band was used for the image luminosity, while the NISP bands determined the hue and saturation, thereby retaining the finer resolution in the colour cutouts.

\subsection{DR1 search inspection pipeline}\label{Sect: Inspection Pipeline}

\subsubsection{Object selection}\label{Sect: Catalogue Cuts}
Galaxy--galaxy strong lens systems are typically composed of a massive early-type lens galaxy and a late-type source galaxy. Both objects are thus extended sources, with the lens usually a bright object, given that it must be sufficiently massive to produce detectable strong lensing. In order to reduce the number of systems with which to search, we applied the following cuts to the MER \citep{DR1-TP009} catalogue.
\begin{itemize}
    \item \texttt{VIS\_DET} = 1. This ensures an \IE detection.
    \item \texttt{MU\_MAX} > 15.0. \texttt{MU\_MAX} refers to the peak surface brightness above the background, and was used to remove saturated stars.
    \item \texttt{MUMAX\_MINUS\_MAG} > $-2.6$. This effectively compares the peak surface brightness of a source and its total magnitude, forming a compactness measure and was used to remove point-like objects.
    \item \IE < 23 or \HE < 22.5. This was used to select bright objects.
\end{itemize}
Excluding deep tiles and those overlapping with the Q1 area, this selection produced \rt{68 million} objects.
We note that this selection differs from the galaxy selection used in the Q1 SLDE-A paper in that it removes two previous selection criteria,
\begin{itemize}
    \item \texttt{SPURIOUS\_PROB}<0.05,
    \item \texttt{GAIA\_CROSSMATCH},
\end{itemize}
and adjusts the magnitude criterion from $\IE\leq22.5$ to include objects bright in the \HE band. In particular, the effect of the \texttt{GAIA\_CROSSMATCH} criterion was investigated by SLDE-F \citep{Ecker2026}, which found that this selection removed low-redshift lens candidates. Meanwhile, the more lenient magnitude cut was adjusted to help identify higher-redshift lens systems.
The simplicity of these catalogue selection cuts, in particular the absence of any colour cuts, ensures the selection retains rarer but scientifically interesting lens systems such as spiral lens galaxies or red lensed arcs.

\subsubsection{Machine-learning models}\label{Sect: ML Intro}
As part of the strong lens search, a number of ML models were developed to identify lens candidates. ML models have the significant benefit of being able to be applied to millions of images, far surpassing the inspection budgets of human classifiers. In previous works many such models have proven to have extremely high performance, capable of producing large samples of lenses at high purity (e.g., \citealp{Canameras2024,Schuldt2025}; SLDE-C). 
We describe the networks used as part of the comprehensive DR1 lens search below but direct readers to \Vincken{} for discussion on the composition, training data and behaviour of the different networks used. The training sets for most networks included lens candidates and non-lenses found in the Q1 SLDE lens search, along with lenses identified as part of a rapid $200\deg^2$ search of DR1 data (Sect.\,\ref{Sect: 200deg Search Intro}) and some simulated lenses. We show a sample of these simulated lenses in Appendix \ref{Ap: Simulation_Comparison}. The training sets differed between networks to enable them to be optimised to the particular architecture and foundation model (where relevant) of each model, and to ensure a diversity of lenses were found.
The following networks were used.
\begin{itemize}
    \item NN1: This network, named \texttt{AstroVink} and presented in \citet{Vincken2026}, uses a Vision Transformer architecture. In particular it uses the \texttt{DINOv3} encoder \citep{Simeoni2025} which was pre-trained on large, general (non-astronomical) image data sets, and was subsequently fine-tuned to identify strong lenses.
    \item NN2: This network, adapted from \citet{Wilde2022}, uses a OU-100 convolutional neural network architecture applied to \IE cutouts with arcsinh processing. 
    \item NN3: This network, adapted from \citet{DominquezSanchez2018} and \citet{ManjonGarcia2021} uses a four-layer convolutional neural network architecture. This was applied to \IE cutouts using arcsinh stretch.
    \item NN4: This network is built on the \texttt{Zoobot} foundation model \citep{Walmsley2023} using a ConvNeXT architecture, and is fine-tuned to identify strong lenses \citep{Q1-SP053,Q1-SP085}. The foundation model was trained using classifications from the Galaxy Zoo citizen-science project \citep{Lintott2008, Walmsley2025}, enabling it to identify galaxies of different morphologies from a range of astronomical surveys. The network uses \IE cutouts with arcsinh processing. 
    \item NN5: This network also utilises a retrained \texttt{Zoobot} network, but this time with a deeper ConvNeXt-Base architecture. NN5 is in fact composed of an ensemble of two individual networks, an \IE-only network and an \IE{}+\JE{}+\HE{} network, combined via a logit-mean. These individual networks both use FITS input data rather than the compressed JPEG files used by networks NN1--4. The development, performance, and characteristics of this network are discussed in detail in \citet{\Fogliardi}.
\end{itemize}

\subsubsection{Ensemble classification}
As well as applying individual networks to the \Euclid data, we also used an ensemble classifier (made up of multiple of the ML models listed above) as part of the lens search. Previous studies (e.g., \citealp{Andika2023,Holloway2024,Schuldt2025}) have shown that such ensembles can improve the purity of the resultant lens sample. Based on the performance of the individual networks on a test data set (described in Sect.\,\ref{Sect: 50deg Search Intro}), we generated an ensemble composed of \rt{networks 1 and 5}, via the independent classifier method described in \citet{Holloway2024} and SLDE-E. This method calibrates each classifier against a known ground truth, in this case, $50\%$ of the grade A+B systems identified as part of the $50\deg^2$ calibration search (Sect.\,\ref{Sect: 50deg Search Intro}) to produce a mapping from network score to posterior probability that a given system is a grade A+B lens. The probabilities from each network are then combined via Bayes theorem to produce a final posterior, which is used as the ensemble score. We discuss the performance of this classifier in Sect.\,\ref{Sect: 50deg Search Results}.

\subsubsection{Citizen-science}\label{Sect: SW Intro}
While each of the ML models individually performs very well (see Sect.\,\ref{Sect: 50deg Search Results} for a complete description of their performance), they produce a large number of high-scoring systems. Inspecting all of these systems would be beyond the inspection capability of individual researchers but remains well within that of a citizen-science project. Space Warps (\citealp{Marshall2016,More2016}) is a long-running citizen-science project to identify strong gravitational lenses. To date it has conducted systematic lens searches over multiple surveys including HSC \citep{Sonnenfeld2020}, CFHTLS \citep{Marshall2016,More2016}, and \Euclid (SLDE-A, SLDE-E) receiving millions of classifications. In the majority of these searches, volunteers have been asked to undertake a binary classification task, answering whether a given system shows a likely lens candidate or not. Volunteers are provided with tutorial and educational pages, along with feedback for training images to help them correctly identify lens systems (Appendix \ref{Ap: SW Training}). The training images consisted of simulated lenses and expert-labelled non-lenses (e.g., spiral galaxies and LRGs; \citealp{Q1-SP052}). The volunteers' classifications were processed via the Space Warps Analysis Pipeline (SWAP; \citealp{Marshall2016}).

This pipeline can be summarised as follows. Based on their responses to training images, each volunteer is assigned a `skill'. Each galaxy cutout (known as a `subject') is assigned a corresponding score\footnote{While the formalism used here may suggest these correspond directly to lens probabilities, in practice these are treated as scores since we use simulated lens systems for training rather than real lens systems and so a score of e.g. $P(\Lens|\{C_i\})=0.9$ does not necessarily correspond to a purity of $90\%$.} according to the likelihood that it is a gravitationally lensed system. This score is updated after each successive classification by a volunteer by an amount dependent on their skill. This skill weighting is similar to that used in polling research, for example to weight the respondents to reflect the prevalence of different groups in the wider population \citep{Kalton2003,Kolenikov2016}. In this work however, the weighting is used to enable volunteers with a greater skill to have a larger impact on the scores of each system.

For binary classification, the user's skill is represented by a $2\times2$ matrix, $\mathbf{S}$, denoting the probabilities that they correctly identify a lens as a lens, a non-lens as a non-lens, and the probabilities of the corresponding mis-classifications.
The elements of the skill matrix, $S_{ij}$, for each volunteer are defined by
\begin{equation}
    S_{ij} \equiv \frac{1+\NCGij}{2+\NGj}\,,
\end{equation}
where $\NCGij$ is the number of times the volunteer has assigned a classification $C_i\in\{\Lens,\Nonlens\}$ to a training system of type $G_j\in\{\Lens,\Nonlens\}$, and $\NGj$ is the total number of training systems of type $G_j$ that the volunteer has seen. Accordingly, a new user who has seen no training subjects is assigned a skill $S_{ij}=0.5$ which produces no change in the subject scores. 
From Bayes' theorem the posterior probability of a system being type $G_j$ following a classification $C_i$ is given by
\begin{equation}
    P(G_j|C_i) = \frac{P(G_j)\, P(C_i|G_j)}{P(C_i)}\,.
\end{equation}
Using the skill matrix, we approximate $P(C_i|G_j)\approx S_{ij}$. Since the probabilities must sum to unity, this posterior is equivalent to
\begin{equation}\label{Eq:Grade Posterior}
    P(G_j|C_i) = \frac{P(G_j)\, P(C_i|G_j)}{\sum_k P(G_k)\, P(C_i|G_k)}\,.
\end{equation}
For the case of binary classification (i.e. \SWCt{}), we define $p_{\rm{\sfont{SWC}}}\equiv P(\Lens|\{C_{{\rm{SWC}},i}\})$ as the \SWCt{} score. Each system starts with a prior lens probability $P_0(G_j=\Lens)=5\times10^{-4}$, and the posterior is updated following each classification with the posterior for the $n$th classification becoming the prior for the $(n+1)$th classification. 
As such, the subject score evolves as
\begin{equation}
    P_{n+1}(G_j|\{C_1,...,C_{n+1}\}) = \frac{P_n(G_j|\{C_1,...,C_n\})\, P(C_{n+1}|G_j)}{\sum_k P(G_k|\{C_1,...,C_n\})\, P(C_{n+1}|G_k)}\,,
\end{equation}
so that volunteers with a greater skill cause a greater shift in the subject score.
Systems were removed from the classifications stream if their score fell below $10^{-5}$ and they had received at least 6 classifications, or following 25 classifications regardless of score.

The binary classification project, \SWCt{} provides significant information on the likelihood that a system is a lens. However, as shown by the results of \citet{Marshall2016} and \Refine{}, conducting a second, refinement stage can help further purify the lens sample. This was also demonstrated by SLDE-E, which found that higher-grade lens candidates received successively higher scores from the volunteers. \Refine{} outlined an extension to SWAP in which volunteers provided grades to each lens candidate system rather than simple lens/non-lens classifications, dividing systems into definite (A), probable (B), possible (C), and non-lens (X) systems. In this case, the SWAP algorithm remains the same but the classifications $\{C_i\}$ and ground-truth $\{G_j\}$ become $\in\{\rm{A,B,C,X}\}$. We refer to such classification as \SWRt{}. For the ground-truth we used the consensus researcher grades from \Euclid systems graded previously (e.g. from Q1). In order to produce a high purity sample, and to reduce the requirements on researcher grading, systems receiving a high score from \SWCt{} were then shown to \SWRt{}, from which again the highest scoring systems were subsequently selected for researcher grading. Systems were removed from the \SWRt{} classification stream following 10 classifications by volunteers.

\subsubsection{Grading by researchers}\label{Sect: GJ Intro}
High-scoring systems were assigned a grade by members of the \Euclid Strong Lensing Science Working Group. As in the grading for \SWRt{}, this separated systems into definite, probable, possible, and non-lenses.
This grading used the following criteria.
\begin{itemize}
\item A; a sure lens. It shows clear lensing features, and no additional information is needed.
\item B; a probable lens. It shows lensing features, but additional information is required to verify it as a definite lens.
\item C; a possible lens. It shows lensing features, but they can be explained without resorting to gravitational lensing.
\item X; definitely not a lens.
\end{itemize}
For the $50\deg^2$ and full DR1 searches (i.e., excluding the $200\deg^2$ search), the grades from individual experts were calibrated to account for optimism/pessimism on an individual grader level. This was done in three stages, firstly calculating a preliminary grade for each system, then determined the `skill' of each grader relative to these grades, and finally updating the grades for each system to reflect each expert's skill. 

In practice, we calculated the preliminary grades and expert skill simultaneously. For a given grader $k$, their skill was calculated as follows. Scores were assigned to all classifications according to the following scheme: $\rm{A}=3,\rm{B}=2,\rm{C}=1,\rm{X}=0$. The mean score $s$ was then calculated for each subject classified by grader $k$, using the classifications of all graders except $k$. The score $s$ was then converted to a preliminary grade via the following thresholds\footnote{The grades published in the ECLIPSE catalogue correspond to the grades boundaries stated here. However, for the $200\deg^2$ and $50\deg^2$ searches, grade boundaries of 
A: $2 \leq s$, 
B: $1.5 \leq s < 2.0$,
C: $1 \leq s < 1.5$,
X: $s < 1.0$
were used initially. Therefore, plots of model performance (Fig.\,\ref{Lens_fraction_vs_rank} and Fig.\,\ref{Insp_requirements}) use these original boundaries, since these were used to determine the ML score thresholds to inspection. The change was made to increase the purity of the grade A sample and place the thresholds halfway between the (integer) values assigned to the classifications. We highlight that the B-C boundary remains unchanged and so this change doesn't significantly affect the combined total of grade A+B systems. The published catalogue uses the half-integer thresholds described above, including revised grades for the $200\deg^2$ and $50\deg^2$ searches for consistency.}
\\[-2ex]
\begin{equation}
\left\{
\begin{array}{l}
  \rm{A}, 2.5 \leq s; \\
  \rm{B}, 1.5 \leq s < 2.5; \\
  \rm{C}, 0.5 \leq s < 1.5; \\
  \rm{X}, s < 0.5\,.
\end{array}
\right.
\end{equation}
The skill matrix of expert $k$ was then calculated according to 
\begin{equation}\label{Eq: GJ_SWR_skill}
    P(C_k|G_j) \approx S_{{\rm{SWR}},kj} \equiv \frac{1+N_{\rm{C_{\it{k}}|G_{\it{j}}}}}{4+\NGj}\,.
\end{equation}
This was repeated for all graders. Finally, calibrated grades were recalculated for all systems, now accounting for each expert's skill, via Eq.\,(\ref{Eq:Grade Posterior}) using a flat prior $P_0(G_j)=0.25$. We discuss the differences between calibrated and uncalibrated grades for identical systems further in Appendix \ref{Ap:Effect_of_Calibration_on_Grades}.

We highlight here that lens grading is very subjective. As will be shown in Sect.\,\ref{Sect: Challenges_and_opportunities}, researchers often disagree as to the nature of strong lens systems, and as such some graders are much more optimistic than others. While in the future, lens grades could be replaced by lens probabilities using large spectroscopic surveys (e.g., the 4MOST Strong Lensing Spectroscopic Legacy Survey, 4SLSLS, \citealp{Collett2023}). These could be calibrated using a test set of real systems with a known ground truth (i.e. spectroscopic confirmation or refutation) thus enabling a mapping from uncalibrated score to lens probability. However, researcher-assigned grades currently remain widely used in the literature and as such are adopted in this work.

\subsection{DR1 search phases}\label{Sect: Inspection Phases}
Due to the large volume of DR1 data, and to optimise the lens search strategy, the DR1 search was split into three phases: a rapid $200\deg^2$ search to identify lenses with which to further improve the ML models; a $50\deg^2$ search to evaluate the performance of the retrained models; and a final search of the remaining DR1 area. We detail these individual searches below.

\subsubsection{$200\deg^2$ rapid search}\label{Sect: 200deg Search Intro}
As shown in \citet{Q1-SP085}, lens-finding networks can benefit from larger data sets of real lenses. Preliminary versions of the networks were run over cutouts spanning $200\deg^2$ for which the high-scoring systems underwent expert grading. \rt{701} grade A systems were identified along with \rt{544} grade B systems. We publish calibrated grades for these systems in the ECLIPSE catalogue. During the grading process the experts were shown systems from Q1 alongside the systems in the $200\deg^2$ area. For this search the experts were calibrated using their responses to the Q1 systems (Eq.\,\ref{Eq: GJ_SWR_skill}), instead of using the leave-one-out method described in Sect.\,\ref{Sect: GJ Intro} which was used in the subsequent search stages.
The lens candidates identified in this search were then used to supplement the training sets of the lens-finding networks in subsequent searches.

Retraining the networks using lenses identified by previous iterations of those networks created the possibility of a circular training loop, biasing the networks towards particular types of lenses. This was mitigated by volunteer inspection of systems randomly chosen from the catalogue selection (see below) rather than high-scoring ML systems, enabling biases to be investigated. Furthermore, the inclusion of simulated lenses in the ML training sets ensured they contained a wide variety of lens configurations beyond the real \Euclid systems already found. 

\subsubsection{$50\deg^2$ calibration search}\label{Sect: 50deg Search Intro}
Due to the change in performance caused by retraining the networks, we performed a test of the networks over a random non-contiguous $50\deg^2$ area within DR1. The purpose of this test was to measure and calibrate the networks' performance. While higher scores for each network indicated a greater degree of confidence that a system is a lens, an ML score threshold of 0.9 from each network could produce lens candidate samples of significantly different purity. Therefore this test aimed to produce an accurate mapping from network score to sample purity, thereby accounting for differing over/under-confidence between networks based on their raw outputs.

For each network, we sampled \num{6000} systems from those receiving the highest scores from that network. This was composed of $100\%$ of the top scored \num{2000}, $50\%$ of the next \num{4000} and $25\%$ of the next \num{8000} for each of the \rt{5} networks. We also selected around $30\,000$ systems randomly from the original galaxy selection. These systems (totalling approximately $60\,000$) were shown to volunteers as part of a \SWCt{}. The $50\deg^2$ \SWCt{} search was launched on the 16th of February, 2026, and ran for one month. During this time, \rt{796} volunteers made a total of \rt{490\,000} classifications.  A total of \rt{\num{3100}} high-scoring systems (in this case, with $p_{\rm{\sfont{SWC}}}\equiv P(\Lens|\{C_{{\rm{SWC}},i}\})>10^{-5}$), were then shown to experts for grading. Given the small area of the search, we did not perform an intermediate \SWRt{} search here.

\subsubsection{\rt{$2000\deg^2$} DR1 comprehensive search}\label{Sect: Systematic Search Intro}
Following the network retraining and calibration described above, we proceeded to apply the networks to the DR1 area and inspected all highly scored systems not previously inspected. We restricted our search to cutouts in the EWS, i.e. excluding those which were part of the Euclid Deep Survey.
The systems were ranked according to their ML score, in decreasing order (best candidates first). We define the rank $R_{8\%}$ as the rank at which the grade A+B lens purity is 8\%, based on the results of the $50\deg^2$ data set (the purity is naturally higher for $R\leq R_{8\%}$). In practice, the value of $R_{8\%}$ was estimated by extrapolating the results of the $50\deg^2$ calibration search to the full DR1 search (Fig.\,\ref{Lens_fraction_vs_rank}). This was done for each network and the ensemble. The resulting \rt{\num{160000}} systems for which $R\leq R_{8\%}$ across any network or ensemble were then allocated for inspection by SW volunteers, alongside \num{50000} randomly selected objects which passed the catalogue selection. These `randoms' were included to help measure the ML selection function, i.e. what lens candidates the ML classifiers may miss. A further \num{30000} systems were inspected based on a preliminary version of the ensemble which was updated subsequently to derive from NN1 and NN5 when data processing permitted. The $R_{8\%}$ threshold was chosen based on the trade-off of inspection-cost and lens-reward (Sect.\,\ref{Sect: 50deg Search Results}, Fig. \ref{Insp_requirements}) and the inspection capacity of SW-Classify, $\sim10^5$. Systems passing an even higher ML score threshold, $R_{80\%}$ were inspected directly by experts, but were also passed to \SWCt{} for completeness.

Of the systems inspected by \SWCt{}, those receiving a score $p_{\rm{\sfont{SWC}}}>1-10^{-8.875}$ were shown directly to experts. This threshold was based on the lens purity measured from previous SW searches (Q1 and the $50\deg^2$ search), where systems above this threshold had a grade A+B purity of $50$--$85\%$. Systems receiving a score $p_{\rm{\sfont{SWC}}}>0.01$ were instead shown to volunteers as part of \SWRt{} as part of a refinement stage. This threshold was chosen based on the expected inspection capacity of \SWRt{}, $\sim10^4$, given this classification stage was more time consuming than \SWCt{}.

From \SWRt{}, only systems receiving a posterior grade \rt{$P(G_j\in\{\rm{A,B}\}|\{C_{{\rm{SWR}},\textit{i}}\})>0.87$} were passed to experts for grading. Systems in the $200\deg^2$ or $50\deg^2$ areas which were not inspected by GJ or SW in the previous searches but passed the score thresholds for inspection used here were also inspected at this point to ensure consistency across the complete DR1 area. In total, \rt{52\,000} systems were passed to \SWRt{} and \rt{25\,000} were graded by experts. 

This inspection pipeline was designed to maximise the number of lenses identified, while ensuring that the inspection requirements for volunteers and researchers remained viable. 
The number of systems inspected in the final stage (researcher grading) was only \rt{$0.04\%$} of the \rt{$6.8\times10^7$} systems classified in total. This was enabled by the high performance of the ML classifiers and the volunteer classification stages, which ensured only high-purity lens samples required grading by researchers.

\section{Results}\label{Sect: Results}
In this section we present the results of the lens search in DR1 data. The ECLIPSE lens catalogue resulting from these searches is presented in \citet{\Lines}, while here we focus on the performance of the discovery pipeline and the resulting visual inspections.

\subsection{$50\deg^2$ calibration search}\label{Sect: 50deg Search Results}
\begin{figure}
    \centering
    \includegraphics[width=\columnwidth]{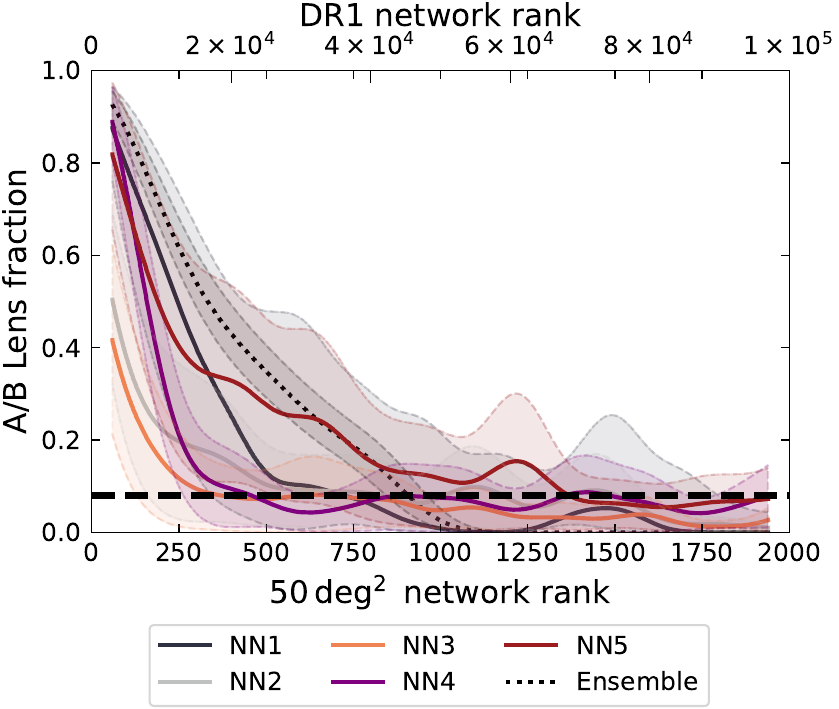}
    \caption{Lens purity versus rank of systems in the $50\deg^2$ calibration search (Sect.\,\ref{Sect: 50deg Search Intro}) for all networks and the ensemble. The shaded regions show $\pm 1\sigma$ uncertainties which were calculated via bootstrapping. The horizontal line indicates the $8\%$ grade A+B purity cutoff used for each network to determine which systems were directed to \SWCt{} for inspection. 
    }
    \label{Lens_fraction_vs_rank}
\end{figure}
\begin{figure}
    \centering
     \includegraphics[width=\columnwidth]{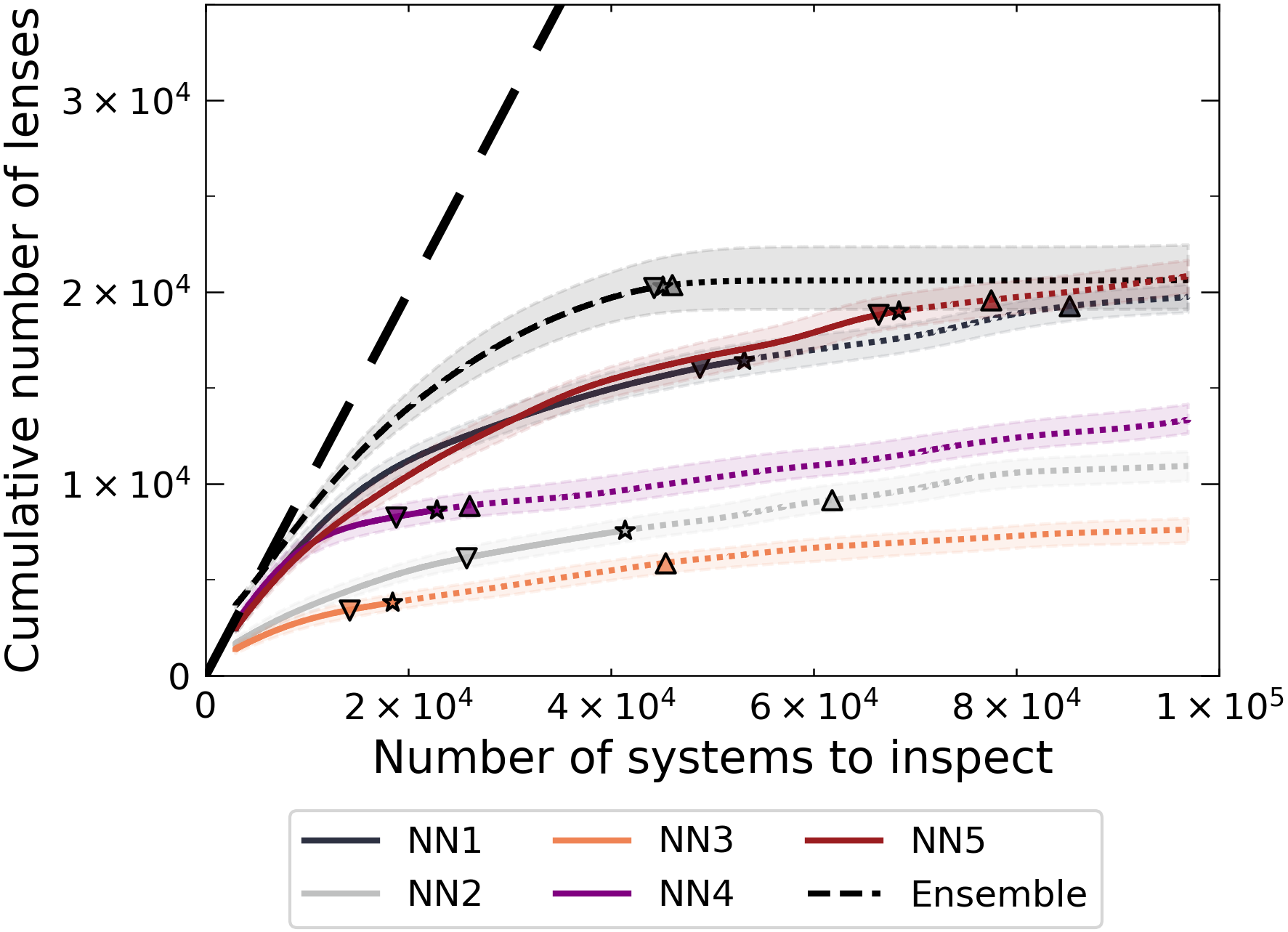}
    \caption{Expected number of lenses discoverable by each network, as a function of the number of objects required to be inspected to retrieve these, based on the network's performance on the $50\deg^2$ area. The dotted lines indicate regions after which the network purity dropped below the \SWCt{} threshold set for the comprehensive DR1 search. Given the overlap in high-scoring systems, the total number of high-scoring systems is less than the sum of the number flagged by each network. The 6\%, 8\%, and 10$\%$ running purity thresholds for each classifier are marked by $\blacktriangledown$,$\star$, and $\blacktriangle$, respectively and the long-dashed line indicates the one-to-one relation, whereby the network purity is $100\%$.}
    \label{Insp_requirements}
\end{figure}
The $50\deg^2$ lens search was used to determine the prioritisation strategy for inspection for the larger DR1 search. Figure \ref{Lens_fraction_vs_rank} shows the grade A+B lens purity of the networks and ensemble classifier as a function of rank within the $50\deg^2$ data set. For comparison, we also provide an alternative axis scaled to the total size of the DR1 area. We find that the ensemble provides a higher lens fraction than the individual networks, but that many networks have very high performance individually. Systems with a network score corresponding to a lens purity above \rt{$8\%$} were selected for inspection by \SWCt{}. Figure \ref{Insp_requirements} shows the expected number of lenses identifiable from each network, as a function of the number of systems to inspect for each ML model along with the ensemble. As illustrated, a significant proportion of the lenses are identified below a rank of $20\,000$, while there are diminishing returns to inspecting progressively further down the ranks. The purity-based approach to system prioritisation provides a performance-based metric to determine which systems should be passed to \SWCt{}; a similar approach has been adopted by previous works (e.g. \citealp{Schuldt2025}) using a distinct test set on which to first measure network performance prior to inspection. 

The ensemble classifier of NN1 and \rt{NN5} has the highest performance out of all classifiers used in the search. However, different networks favoured different sorts of lens systems. Therefore, to ensure a diverse range of lenses were identified, the ensemble was treated like the other networks when prioritising which systems to inspect, i.e., systems with an ML or ensemble score corresponding to a purity $>8\%$ were inspected by \SWCt{}.

\subsection{\rt{$2000\deg^2$} DR1 comprehensive search}\label{Sect: DR1 Search Results}
The \SWCt{} project began on the 21st of April, 2026, and accrued a total of 
\rt{3.4 million}  
classifications over 
\rt{5}  
months. The follow-up \SWRt{} project began on the 21st of May, 2026, and received a total of \rt{\num{570000}} classifications over \rt{5} months. Concurrently with these searches, high-scoring systems were inspected by researchers as part of GJ. A summary of the number of systems classified by each stage of the discovery pipeline is given in Table \ref{T: Comprehensive search stats}.
\begin{table}
\caption{Number of systems classified by each stage of the comprehensive search.}
\label{T: Comprehensive search stats}
\begin{tabular}{ccccccc}
\hline
\hline & \\[-1.5ex]
&\multicolumn{3}{c}{Inspection Stage}&\\
& ML stage & \SWCt{}  & \SWRt{} \\
\midrule
$\rightarrow$GJ & $8.5\times10^3$ & $2.5\times10^3$ & $4.9\times10^3$\\
$\rightarrow$\SWCt{} & $1.4\times10^5$ & / & /\\
$\rightarrow$\SWRt{}&/& $2.8\times10^4$ & /\\
 Reject & $6.0\times10^7$ & $1.1\times10^5$ & $2.3\times10^4$\\
 \hline
\end{tabular}
\tablefoot{The quoted values exclude the $200\deg^2$ and $50\deg^2$ areas, and excludes systems already passed to GJ or rejected by previous classification stages. 
Similarly, the quoted figures include only systems that met the score thresholds at the preceding classification stage. 
}
\end{table}

\begin{figure}
    \centering
    \includegraphics[width=0.9\columnwidth]{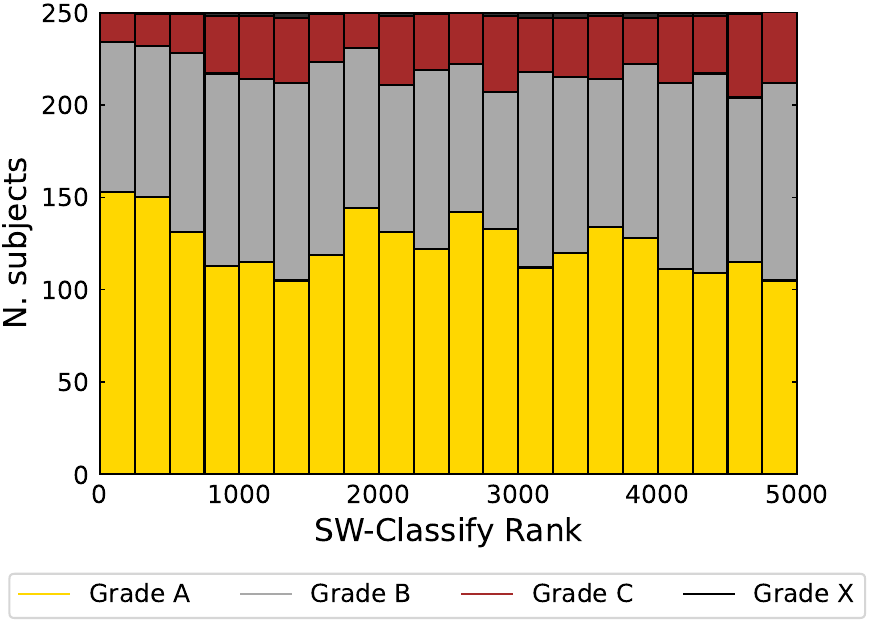}
    \caption{GJ grade as a function of \SWCt{} rank from the comprehensive search, excluding the $200\deg^2$ and $50\deg^2$ areas. Systems receiving higher scores from \SWCt{} are progressively more likely to be possible lens candidates (grade A--C), and progressively more likely to be of a higher grade.
    }
    \label{SWC_rank_plot}
\end{figure}
\begin{figure}
    \centering
    \includegraphics[width=0.9\columnwidth]{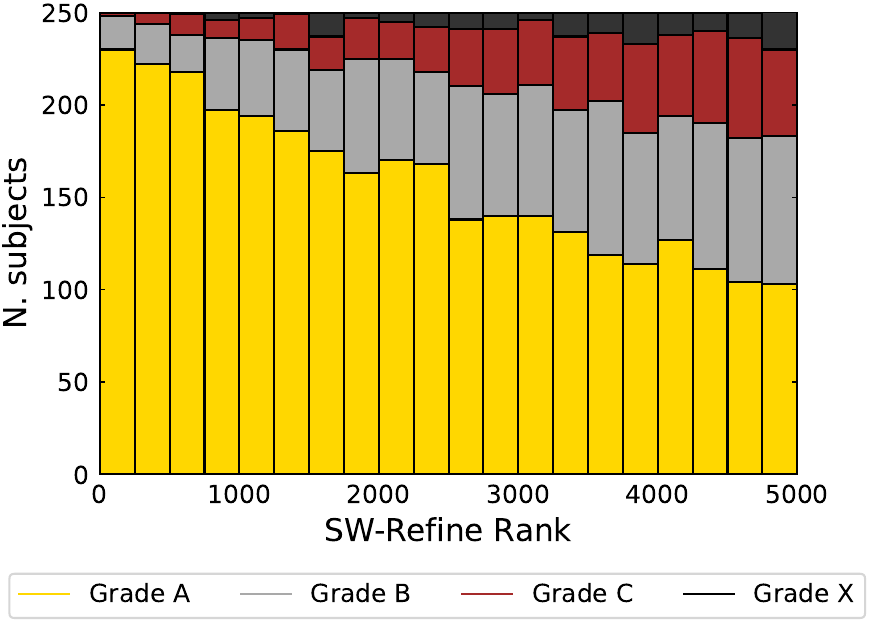}
    \caption{GJ grade as a function of \SWRt{} rank from the comprehensive search, excluding the $200\deg^2$ and $50\deg^2$ areas. The purity of these systems is higher than that achieved by \SWCt{} alone (Fig. \ref{SWC_rank_plot}), demonstrating the benefit of a second refinement stage.}
    \label{SWR_rank_plot}
\end{figure}
Figure \ref{SWC_rank_plot} shows the proportion of systems assigned each grade as a function of score rank in \SWCt{}; 
an equivalent plot for the highest ranked systems shown to \SWRt{} is given in Fig. \ref{SWR_rank_plot}. Highly ranked systems have a greater fraction of grade A+B candidates in \SWRt{} than in \SWCt{}, demonstrating that such a refinement stage was worthwhile.

\section{Discussion}\label{Sect: Discussion}

\subsection{Inspection selection function}
\subsubsection{Recovery of simulated lenses}
\begin{figure*}
    \centering
    \includegraphics[width=\textwidth]{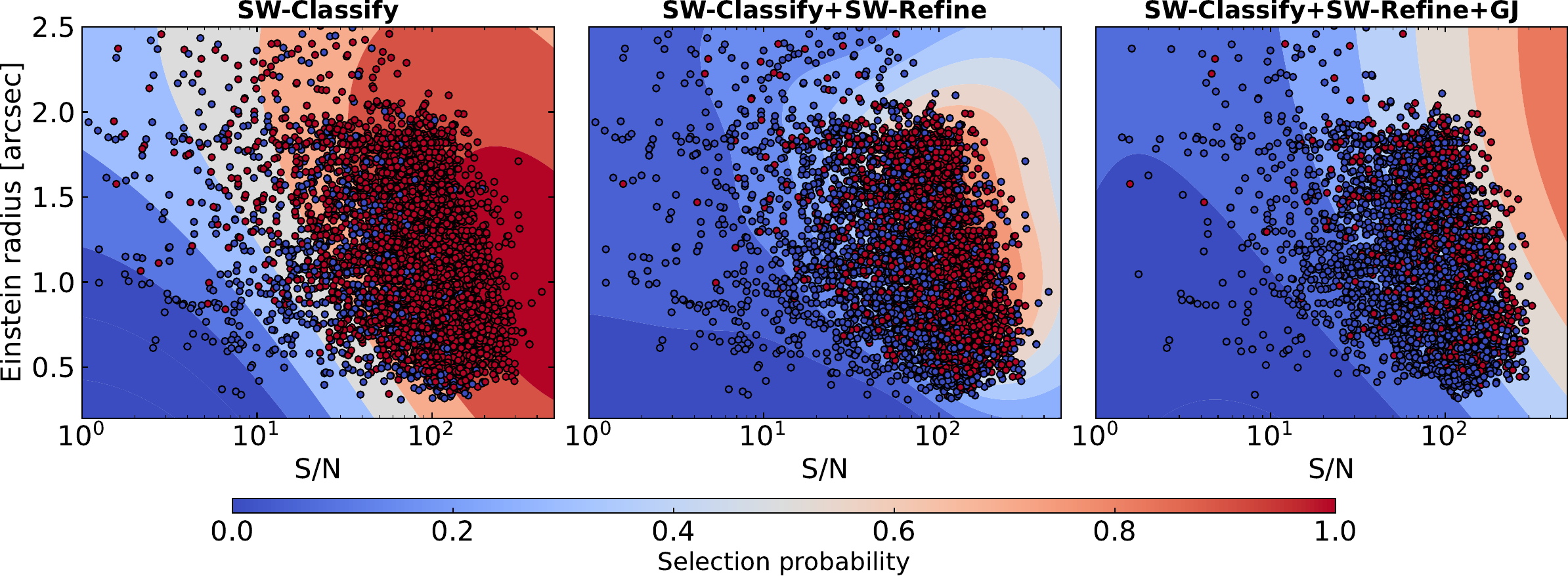}
    \caption{Selection function based on simulated lenses inserted into the inspection pipeline. The selection function plotted is cumulative, i.e. systems are only shown as identified (red) if they were also identified by all previous stages. Systems are plotted as having been identified by GJ if they were classified as grade A or B. The points show individual simulated lenses, coloured red if they passed the selection threshold to be inspected by the subsequent stage, or blue otherwise. The contours show a smoothed selection function, generated by applying a Gaussian-process regressor from the \texttt{scikit-learn} library \citep{Pedregosa2011} to the individual data allowing the broad trends to be seen. We show a collage of simulations spanning the range of Einstein radius-S/N parameter space in Appendix \ref{Ap: Simulation_Comparison}. Due to inspection constraints, only a random subset of simulated lenses that passed a given stage were passed to the next stage.}
    \label{Inspection_Selection_Function}
\end{figure*}

Throughout the inspection process we included simulated lenses to help determine the detection selection function -- which lenses are identified and which are missed. Figure \ref{Inspection_Selection_Function} shows this selection as a function of Einstein radius and signal-to-noise ratio (S/N), split into each stage of the inspection. The S/N was determined by taking the maximum value of the ratio of the cumulative lensed source flux to the cumulative sum of the root mean square background noise. We find that the selections are broadly similar across each stage -- systems with a larger S/N are much more likely to be flagged than fainter systems. There is an additional correlation with Einstein radius, whereby larger systems are more likely to be identified, but for a sufficiently low S/N such systems are still likely to be missed. 

Introducing multiple inspection stages, i.e., \SWCt{}, \SWRt{}, and GJ, compounds the selection effect, since most lens candidates must pass each stage to be included in the final catalogue. To mitigate this, we introduced small direct-to-GJ pipelines (as displayed in Fig. \ref{Inspection_Flowchart}), whereby systems with the highest ML and \SWCt{} scores were classified by GJ directly without relying on \SWCt{} or \SWRt{} scores respectively. This helped to ensure that the clearest lenses were not missed since, as shown by Fig.\,\ref{Inspection_Selection_Function}, the selection probability at each inspection stage was not $100\%$ even for high S/N systems.

Finally, we note that the true selection function of the lenses presented in \citet{\Lines} will depend on parameters beyond simply Einstein radius and S/N, for example, on the colour difference between the lens and source galaxies. We will present further discussion of this along with the complete selection function (i.e., including the influence of the ML classifiers) in future work.

\subsubsection{Search completeness}
The inspection pipeline was designed to ensure that high-scoring objects (from either the ML or SW) were directed to GJ for classification. The objects inspected by SW were drawn from the high-scoring ML sample, and those inspected by GJ were drawn from the high-scoring ML and SW samples. However, both of these required setting score thresholds for which systems would pass to the next stage. This necessarily introduced some loss of lens candidates which fell below these thresholds. We show a selection of systems (both lenses and non-lenses) flagged and missed by each classification stage in Appendix \ref{Ap: Systems_flagged_vs_missed}.
In ECLIPSE-A, we estimate the total completeness to grade A systems to be \rt{$78^{+8}_{-16}\%$}, and to grade A+B systems to be \rt{$42^{+7}_{-8}$\%}. The complete derivation of this is given in Appendix \ref{Ap: Search Completeness}, but in summary this was measured using three random data sets; a sample of \num{50000} systems shown to \SWCt{} regardless of their ML scores and two samples of \num{1000} and \num{2000} objects randomly drawn from the \SWCt{} and \SWRt{} data sets respectively which were shown to GJ irrespective of their SW scores. Out of the \num{50000} randoms, \rt{five} grade A and \rt{two} grade B systems passed the ML$\rightarrow$GJ threshold directly. A \rt{single} further grade B system failed the ML$\rightarrow$\SWCt{} threshold but, since the randoms were shown to SW regardless, passed the SW$\rightarrow$GJ thresholds. 

\begin{figure}
    \centering
    \includegraphics[width=\columnwidth]{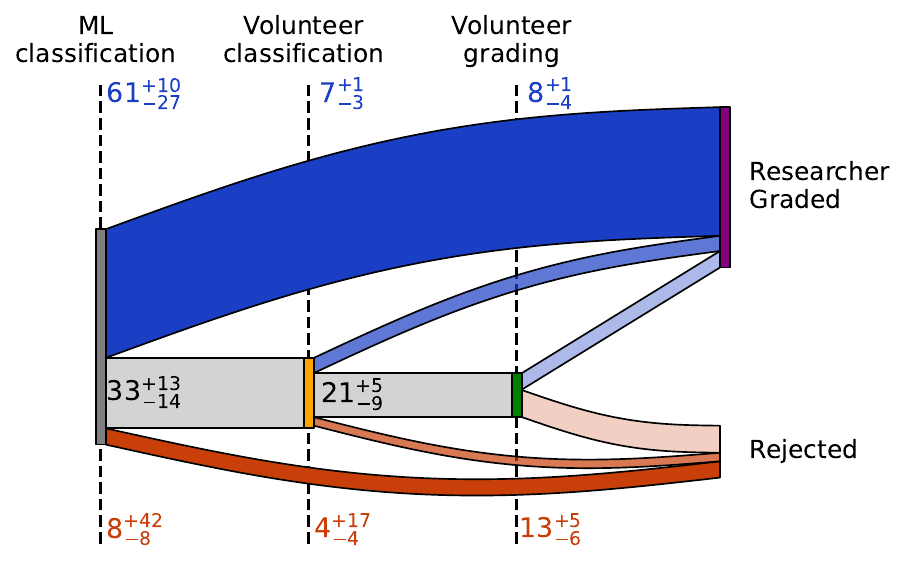}
    \caption{Illustration of the \rt{grade A} search completeness at each stage. The blue regions indicate lens candidates which were passed to GJ for grading, while the red regions indicate systems which failed to pass the score thresholds for GJ grading. The grey regions indicate systems in interim inspection stages. The `Volunteer classification/grading' stages refer to \SWCt{} and \SWRt{} respectively. The values above and below the diagram indicate how many out of 100 lenses would be accepted (above) or rejected (below) at each stage, along with $\pm2\sigma$ uncertainties.
    }
    \label{Search_Completeness_true}
\end{figure}

In Fig.\,\ref{Search_Completeness_true} we show the pathway of \rt{grade A} lenses in the search pipeline. We find that $60\%$ of lenses were identified by the ML$\rightarrow$GJ pathway. This is indicative of the high performance of the ML classifiers and the benefits of retraining with large samples of real data. The remainder were only identified by the SW volunteers. From Fig.\,\ref{Search_Completeness_true}, the largest percentage of lens candidates were rejected due to failing to pass the \SWRt{}$\rightarrow$GJ threshold $p_{\rm{\sfont{SWC}}}>0.87$. This threshold was driven by the inspection budget of the experts. 

\begin{figure}
    \centering
    \includegraphics[width=\columnwidth]{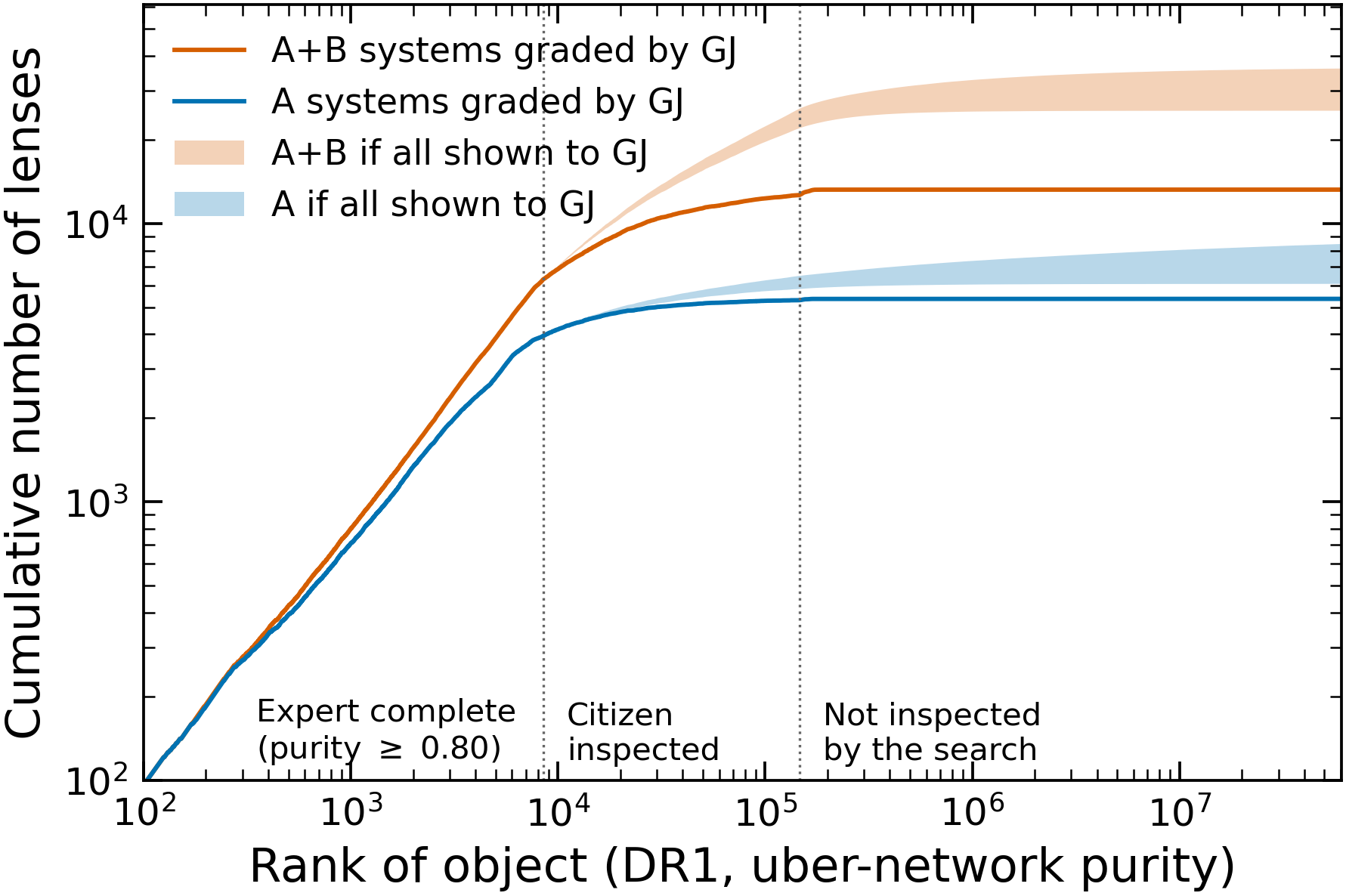}
    \caption{Numbers of lenses discovered as a function of images inspected. The numbers beyond the citizen complete threshold and their distribution in uber-network score are extremely poorly known, since \rt{0} grade A and only \rt{1} grade B lenses in the \rt{50\,000} randoms fell below our ML$\rightarrow$SW score threshold and also passed our SW score thresholds. We limit the totals to the comprehensive search area, i.e. excluding the $200\deg^2$ and $50\deg^2$ regions.
    }
    \label{Comp_vs_insp}
\end{figure}

Via the same method, we also estimate the number of systems which would require inspection to identify a given number of lenses. This was again estimated using the above three data sets of randomly selected objects.
Figure \ref{Comp_vs_insp} shows the number of lens candidates which would be identified as a function of the ML rank. This ML rank was derived using the measured performance of each network in the $50\deg^2$ calibration search, after which the ML scores could be mapped to a lens probability (the rank of which is given by the $x$-axis of Fig. \ref{Comp_vs_insp}) with extrapolation to lower ranks based on the completeness formalism of Appendix \ref{Ap: Search Completeness}.
We find the limiting factor to identifying a significantly larger number of lenses was expert inspection as thousands of grade A+B lenses were lost at the \SWRt{} stage.
Further improvements in model performance based on retraining post-DR1 will likely reduce this inspection requirement.

\subsubsection{Space Warps recovery of known lens candidates}
As detailed above, while the staged inspection process significantly reduced the inspection requirements of the volunteers and researchers, it also meant that low levels of incompleteness at each stage were compounded together in the final result. In this section we detail lens recovery rates based on the score thresholds used for each inspection stage, using a range of \Euclid lens samples as a proxy for measuring completeness. We measure the recovery rates across three lens samples: lens candidates from the Q1 search; the $200\deg^2$ DR1 inspection; and the $50\deg^2$ DR1 inspection. We focus on the SW recovery rate, since the networks used varied across these three searches and the completeness of the overall pipeline is discussed above.
We note that these lens samples themselves will not be complete, and thus these recovery rates should only be taken as approximate. 

While inspecting the $50\deg^2$ DR1 data set, \SWCt{} also examined a selection of Q1 lens candidates (Sect.\,\ref{Sect: 50deg Search Intro}).\footnote{While the calibration search over $50\deg^2$ focussed on DR1 data, to avoid confusion we refer to the Q1 area lens candidates shown as part of this search as Q1 candidates.} These same Q1 systems were shown as part of \SWRt{} during 2025 \citep{Holloway_Refine}. Using the same thresholds as for the full DR1 search, we find that 
\rt{$37\%$} 
of grade A+B lenses pass the \SWCt{} cut to be passed to GJ directly, while a further 
\rt{$55\%$}  
pass the corresponding cut to be inspected by \SWRt{} but failed the direct-to-GJ threshold. Of the latter \rt{$50\%$} 
passed the \SWRt{} threshold to be inspected by GJ. Overall, this gives a theoretical SW recovery rate of 
\rt{$64\%$}.

Considering the lens candidates identified during the $50\deg^2$ lens search, we find that 
\rt{26\%} 
pass the \SWCt{} threshold to be passed to GJ directly while a further 
\rt{$72\%$} 
pass the corresponding threshold to be passed to \SWRt{} but failed the direct-to-GJ threshold. Of the latter, 
\rt{25\%} 
passed the \SWRt{} threshold to be inspected by GJ. Overall, this gives a recovery rate of 
\rt{44\%}. 
We note that in previous searches the \SWCt{} threshold has been lower, for example $p_{\rm{\sfont{SWC}}}>10^{-5}$ in Q1, compared to \rt{$10^{-2}$} here. For context, the prior (the starting score prior to any classifications) is $p_{\rm{\sfont{SWC}}}=5\times10^{-4}$ and so the DR1 threshold moved above the prior. A higher threshold was required here so as to not overwhelm the inspection requirements of subsequent stages, but this reduces the resultant completeness.

Considering the $200\deg^2$ search which was originally only inspected by strong lens researchers, we find the following. Out of all the grade A+B systems, 
\rt{$22\%$} 
were scored by \SWCt{} higher than the direct-to-GJ threshold. A further 
\rt{$65\%$} 
of systems passed the threshold to be inspected by \SWRt{}. Out of those, 
\rt{$39\%$} 
were scored higher than the threshold to be sent to GJ. Overall, 
\rt{$48\%$} 
of systems passed the SW thresholds for GJ inspection. 

The above values suggest that there remain thousands of lenses yet to be discovered. This undiscovered sample is likely to have a higher proportion of ambiguous or unusual systems, given their failure to pass one or more of the above cuts. While we are using expert grades as the ground truth to estimate this lens recovery rate, we stress that these too are subjective; we discuss this further in the following section.

\subsection{Challenges and opportunities; next steps for lens finding}\label{Sect: Challenges_and_opportunities}
\begin{figure}
    \centering
    \includegraphics[width=\columnwidth]{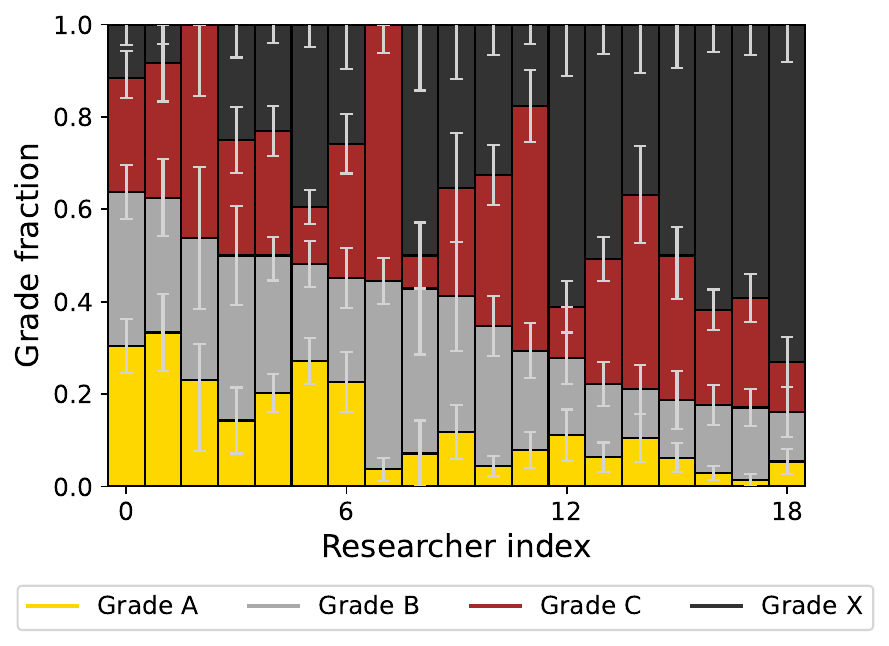}
    \caption{Comparison of the classification optimism between participants of the GJ expert $50\deg^2$ search grading. Each bar represents a different grader, with the size of each bar set according to their classification distribution, i.e., graders towards the left were much more optimistic in their grades than graders towards the right. The bars have been sorted according to the combined grade A+B fraction. To reduce bias due to graders inspecting different systems, this plot is restricted to the \rt{50} most classified systems and to the \rt{19} graders who classified more than \rt{10} of those. The uncertainties for each expert were calculated via bootstrapping their classifications.}
    \label{Grader_optimism_pessimism}
\end{figure}
Euclid DR1 heralds the start of a new era in lens finding, in which the detectable lens samples number in the tens-to-hundreds of thousands. This offers significant opportunity for strong lens science, but in this regime researcher inspection of even the highest purity lens candidates nears unfeasibility. 

The forthcoming \RomanST \citep{Roman2015} and Legacy Survey of Space and Time \citep{Ivezic2019} will face similar inspection challenges to that of \Euclid. 
The overlap region of \Euclid and \RomanST{}\footnote{\url{https://roman-docs.stsci.edu/roman-community-defined-surveys/high-latitude-wide-area-survey}} wide-area surveys will be $470\deg^2$, while those for \Euclid-LSST will be \num{7000}--\num{11000}$\,\deg^2$ \citep{Rhodes2017} dependent on final survey strategies. Lens searches in all of these surveys will benefit from lens discoveries in the others, since they will allow automated classifiers to be trained on known lenses prior to the commencement of systematic lens searches. As shown in \citet{Q1-SP085}, increasing the training set size can significantly improve the performance of a lens classifier. The DR1 lens candidates will enable further improvements beyond their current high performance as well as improvements in ensemble classification (Fig. \ref{Insp_requirements}). 
However, even with such improvements, visual inspection requirements will be significant. For the comprehensive DR1 search stage, systems received a median of 10 classifications from GJ, but some received fewer if they received consensus grade X or A classifications (with a minimum of 3 classifications for grade X systems). Reducing the number of inspectors required to grade each image, while an attractive proposition, risks mis-grading intermediate-grade lens candidates for which there is often the most disagreement (e.g. see \citealp{Rojas2023}).  Furthermore, unless early-stage lens classification stages achieve close to $100\%$ purity, samples flagged for inspection are likely to extend beyond $200\,000$ for the full EWS and other large-scale surveys. 

Beyond improved lens classification, such challenges may also be mitigated by rapid lens modelling, either by ML \citep{Q1-SP063} or forward-modelling codes \citep{\Nightingale}. This will help to provide additional verification as to the nature of a possible lens system beyond simple visual inspection. As part of ECLIPSE-C, model validation was performed to determine which lens candidates had successful models, and to flag those which following modelling were unlikely to be lenses. However, given the reliance on human inspection, it is possible that some false positives will be missed. As shown by Fig. \ref{Grader_optimism_pessimism}, there is significant variation in the optimism of researcher graders; some graders are much more likely to assign a system as grade A than others. Possible reasons for this include variation in experience of lens finding, classification fatigue (given the large number of images requiring inspection), and the subjective nature of the grading, for example distinguishing between a `probable' and `possible' lens. Calibration of graders against spectroscopically confirmed systems from 4SLSLS to produce lens probabilities rather than grades will help to mitigate this and reduce the reliance on subjective grades.

\section{Conclusions}\label{Sect: Conclusions}
In this paper we have presented the classification pipeline used to identify \rt{15\,072} grade A+B lens candidates in Euclid DR1. This was achieved through the classification of \rt{68\,million} objects by \rt{five} ML classifiers, the classification of \rt{300\,000} of these systems by citizen-scientist volunteers, and, finally, the grading of \rt{30\,000} of these systems by strong lens researchers. Our conclusions are as follows.
\begin{itemize}
\item ML lens classifiers can identify thousands of lenses at high purity, reaching \rt{74\%} grade A+B lenses at an estimated completeness of \rt{21\%} (based on the direct-to-GJ pipeline, without citizen-science inspection). For a survey area of $\num{7500}\deg^2$ and $\num{14000}\deg^2$ (corresponding to Euclid DR2 and the full EWS respectively), this will translate into 
\rt{\num{35000} (\num{65000})} 
high-scoring systems of which 
\rt{\num{25000} (\num{48000})} 
will be high-grade lenses though it is likely that the performance of the ML lens classifiers will improve further in the future.
\item A staged inspection process, through multiple citizen-science inspection stages prior to researcher grading enables a greater completeness than ML-only classification would allow. This staged approach will take on increasing importance for future large-scale data releases such as Euclid DR2, the \RomanST{} and LSST. Such a process reduces otherwise unviable inspection requirements on researchers and enables rigorous human verification of hundreds of thousands of systems.
\item Based on calibration samples inserted into the inspection pipeline, we estimate a completeness of $0.73^{+0.08}_{-0.15}$ to grade A systems in the comprehensive search area, and $0.81^{+0.09}_{-0.17}$ across the full DR1 area. The lenses which were missed are likely to have lower S/N or have a smaller Einstein radius, or both.
\item Researcher grading remains a subjective process. In this work we calibrated the researcher grades based on the consensus responses of the researcher crowd. Calibration of researchers using spectroscopically confirmed lenses and transitioning to lens probabilities rather than grades (once permitted by large-scale spectroscopic surveys via DESI, \citealp{DESI2026_DR1}, and follow-up via 4SLSLS, \citealp{Collett2023}) will help mitigate this further.
\end{itemize}

The ECLIPSE catalogue will usher in a new phase of strong lens science through the study of large and homogeneous populations of strong lenses with space-based data. The discovery pipeline presented here will be scalable to the larger data releases of \Euclid (DR2 and the full EWS) following the anticipated improvements in ML performance now possible following DR1. With tens of thousands of lenses still to discover, it is a truly exciting time for strong lensing.

\begin{acknowledgements}
We gratefully acknowledge the Zooniverse team for their invaluable work on the host platform for Space Warps. 
This project would not be possible without the immense contributions of citizen-scientists. 
Those contributing classifications for this project are listed in Appendix \ref{Ap: SW Volunteers} and on \href{https://www.zooniverse.org/projects/aprajita/space-warps-esa-euclid/about/team}{`The Team'} webpage.
We greatly appreciate their commitment, enthusiasm, and time.
PH is funded by the Royal Society (RF\textbackslash{}ERE\textbackslash{}210033) and acknowledges travel support provided by Euclid:UK through STFC grant APP91126.
NL is supported through a graduate studentship from the UKRI STFC and the University of Portsmouth, and acknowledges travel support provided by Euclid:UK through STFC grant APP91126 and by STFC for UK participation in the Rubin LSST through grant ST/X001334/1.
This project has received funding from the European Research Council (ERC) under the European Union’s Horizon 2020 research and innovation programme (LensEra: grant agreement No 945536). TEC is funded by the Royal Society through a University Research Fellowship. Numerical computations were done on the Sciama High Performance Compute (HPC) cluster which is supported by the ICG, SEPNet and the University of Portsmouth.
JWN is supported by an STFC/UKRI Ernest Rutherford Fellowship, Project Reference: ST/X003086/1 and acknowledges travel supported by Euclid:UK through STFC grant APP91126.
AS acknowledges support from the National Key R\&D Program of China (No. 2023YFA1607800, 2023YFA1607802), and from the Yangyang Development Fund.
AN gratefully acknowledges the support of the Belgian Federal Science Policy Office (BELSPO) for the provision of financial support in the framework of the PRODEX Programme of the European Space Agency (ESA) under contract number 4000143347.
AG acknowledges support by the SNSF through PostDoc Mobility Return grant P5R5-2\_235360.
MG acknowledges support from the Polish National Agency for Academic Exchange grant No. BPN/BEK/2024/1/00325/DEC/1.
R.C. acknowledges support from the French government under the France 2030 investment plan, as part of the Initiative d'Excellence d'Aix-Marseille Universit\'e -- A*MIDEX (AMX-23-CEI-088).
JP is supported by the ACME, ELSA, and OSCARS projects. ‘ACME: Astrophysics Centre for Multimessenger studies in Europe’, ‘ELSA: Euclid Legacy Science Advanced analysis tools’, and ‘OSCARS: Open Science Clusters’ Action for Research and Society’ are funded by the European Union under grant agreement nos 101131928, 101135203, and 101129751, respectively; ELSA is also funded by Innovate UK grant 10093177. Views and opinions expressed are however those of the authors only and do not necessarily reflect those of the European Union.
WJRE has received funding from the European Research Council (ERC) under the European Union’s Horizon 2020 research and innovation programme
(LensEra: grant agreement No 945536). WJRE acknowledges the support by the Italian Ministry of University and Research (grant FIS2023-01611, CUP C53C25000300001).
QH was supported by the Dutch Research Council (NWO) through the grant Open Competitie ENW M21-1.
Part of this research was carried out at the Jet Propulsion Laboratory, California Institute of Technology, under a contract with the National Aeronautics and Space Administration (80NM0018D0004).
DJ acknowledges support by the First Rand Foundation, South Africa, and the Centre National de la Recherche Scientifique of France.
Support was provided by Schmidt Sciences, LLC, for J. H-H. C.
This work received travel support from Euclid:UK through STFC grant APP91126.
JAAB’s research was supported by an appointment to the NASA Postdoctoral Program at the NASA Jet Propulsion Laboratory administered by Oak Ridge Associated Universities under contract with NASA.
"ELSA: Euclid Legacy Science Advanced analysis tools" (Grant Agreement no. 101135203) is funded by the European Union. Views and opinions expressed are however those of the author(s) only and do not necessarily reflect those of the European Union or Innovate UK. Neither the European Union nor the granting authority can be held responsible for them. UK participation is funded through the UK Horizon guarantee scheme under Innovate UK grant 10093177
CJS acknowledges funding from the NASA Euclid General Investigator Program Grant No. 80NSSC25K7735.
%
MB acknowledges support from the ERC Grant FIRSTLIGHT \#101053208, Slovenian national research agency ARIS through grant P1-0188, and ESA PRODEX Experiment Arrangements No. 4000146646 and 4000149972. 
The work of PJM was supported by DOE Contract DE-AC02-76SF00515.
\AckDRone
\AckEC  
\end{acknowledgements}

%
%

\bibliography{Euclid, DR1, References} 

@ARTICLE{DR1-SL06,
        author ={{Euclid Collaboration: Fogliardi}, M.  and others},
        title = "{Euclid DR1: Enhancing Deep Learning models: the impact of image processing and training data selection for Strong Lens detection in Euclid}",
        journal = {A\&A, hopefully submitted},
        year = 2026
        }

@ARTICLE{DR1-SL07,
        author ={{Euclid Collaboration: Vincken}, S.H. and others},
        title = "{Euclid DR1: AstroVink-T-REX: Strong lens segmentation as input for mass modeling pipelines}",
        journal = {A\&A, hopefully submitted},
        year = 2026
        }

@ARTICLE{DR1-TP001,
       author = {{Euclid Collaboration: Aussel}, H. and others},
        title = "{Euclid Data Release 1 (DR1). Data release overview}",
      journal = {arXiv e-prints},
         year = 2026,
        month = nov,
          eid = {arXiv:2611.16510},
        pages = {arXiv:2611.16510},
archivePrefix = {arXiv},
       eprint = {2611.16510},
 primaryClass = {astro-ph.CO},
       adsurl = {https://ui.adsabs.harvard.edu/abs/2026arXiv261116510E}
}

@ARTICLE{DR1-TP002,
       author = {{Euclid Collaboration: McCracken}, H.~J. and {Benson}, K. and {Cropper}, M. and others},
        title = "{Euclid Data Release 1 (DR1): VIS processing and data products}",
      journal = {A\&A, submitted},
         year = 2026,
        month = nov,
          eid = {arXiv:2611.15890},
        pages = {arXiv:2611.15890},
archivePrefix = {arXiv},
       eprint = {2611.15890},
 primaryClass = {astro-ph.CO},
       adsurl = {https://ui.adsabs.harvard.edu/abs/2026arXiv261115890E}
}

@ARTICLE{DR1-TP009,
       author = {{Euclid Collaboration: K{\"u}mmel}, M. and {Romelli}, E. and {Gracia-Carpio}, J. and others},
       title = "{Euclid Data Release 1 (DR1): The MERge processing}",
      journal = {A\&A, submitted},
         year = 2026,
        month = nov,
          eid = {arXiv:2611.17832},
        pages = {arXiv:2611.17832},
archivePrefix = {arXiv},
       eprint = {2611.17832},
 primaryClass = {astro-ph.IM},
       adsurl = {https://ui.adsabs.harvard.edu/abs/2026arXiv261117832E}
}

@ARTICLE{DR1-SL01,
       author = {{Euclid Collaboration: Lines}, N.~E.~P. and {\ignoreme{CCC}Holloway}, P. and {Li}, T. and others},
        title = "{Euclid Data Release 1 (DR1). Euclid Lens Identification Pipeline and Scientific Exploration (ECLIPSE) A: Strong lens catalogue overview}",
      journal = {A\&A, submitted},
         year = 2026,
        month = nov,
          eid = {arXiv:2611.15883},
        pages = {arXiv:2611.15883},
archivePrefix = {arXiv},
       eprint = {2611.15883},
 primaryClass = {astro-ph.GA},
       adsurl = {https://ui.adsabs.harvard.edu/abs/2026arXiv261115883E}
}

@ARTICLE{DR1-SL03,
	       author = {{Euclid Collaboration: Nightingale}, J.~W. and {Vincken}, S.~H. and {Ecker}, L.~R. and others},
        title = "{Euclid Data Release 1 (DR1). Euclid Lens Identification Pipeline and Scientific Exploration (ECLIPSE) C: Lens modelling pipeline overview}",
      journal = {A\&A, submitted},
         year = 2026,
        month = nov,
          eid = {arXiv:2611.15888},
        pages = {arXiv:2611.15888},
archivePrefix = {arXiv},
       eprint = {2611.15888},
 primaryClass = {astro-ph.GA},
       adsurl = {https://ui.adsabs.harvard.edu/abs/2026arXiv261115888E}
}

@ARTICLE{DR1-SL04,
       author = {{Euclid Collaboration: Nersesian}, A. and {Nightingale}, J.~W. and {Sluse}, D. and others},
        title = "{Euclid Data Release 1 (DR1). Euclid Lens Identification Pipeline and Scientific Exploration (ECLIPSE) D: Photometric redshifts and the first census of strong lenses in the galaxy evolution framework}",
      journal = {A\&A, submitted},
         year = 2026,
        month = nov,
          eid = {arXiv:2611.15877},
        pages = {arXiv:2611.15877},
archivePrefix = {arXiv},
       eprint = {2611.15877},
 primaryClass = {astro-ph.GA},
       adsurl = {https://ui.adsabs.harvard.edu/abs/2026arXiv261115877E}
}

@ARTICLE{Q1-TP001,
       author = {{Euclid Collaboration: Aussel}, H. and {Tereno}, I. and {Schirmer}, M. and others},
       title = "{Euclid Quick Data Release (Q1). I. Data release overview}",
      journal = {\aap},
         year = 2026,
        month = jun,
       volume = {711},
          eid = {A1},
        pages = {A1},
          doi = {10.1051/0004-6361/202554610},
       adsurl = {https://ui.adsabs.harvard.edu/abs/2026A&A...711A...1E}
}

@ARTICLE{Q1-SP048,
       author = {{Euclid Collaboration: Walmsley}, M. and {Holloway}, P. and {Lines}, N.~E.~P. and others},
        title = "{Euclid Quick Data Release (Q1). XXVI. The Strong Lensing Discovery Engine A -- System overview and first lens sample}",
      journal = {\aap},
         year = 2026,
        month = jun,
       volume = {711},
          eid = {A26},
        pages = {A26},
          doi = {10.1051/0004-6361/202555141},
       adsurl = {https://ui.adsabs.harvard.edu/abs/2026A&A...711A..26E}
}

@ARTICLE{Q1-SP052,
       author = {{Euclid Collaboration: Rojas}, K. and {Collett}, T.~E. and {Acevedo Barroso}, J.~A. and others},
        title = "{Euclid Quick Data Release (Q1). XXVII. The Strong Lensing Discovery Engine B -- Early strong lens candidates from visual inspection of high-velocity dispersion galaxies}",
      journal = {\aap},
         year = 2026,
        month = jun,
       volume = {711},
          eid = {A27},
        pages = {A27},
          doi = {10.1051/0004-6361/202554605},
       adsurl = {https://ui.adsabs.harvard.edu/abs/2026A&A...711A..27E}
}

@ARTICLE{Q1-SP053,
       author = {{Euclid Collaboration: Lines}, N.~E.~P. and {\ignoreme{AAA}Collett}, T.~E. and {Walmsley}, M. and others},
        title = "{Euclid Quick Data Release (Q1). XXVIII. The Strong Lensing Discovery Engine C -- Finding lenses with machine learning}",
      journal = {\aap},
         year = 2026,
        month = jun,
       volume = {711},
          eid = {A28},
        pages = {A28},
          doi = {10.1051/0004-6361/202554542},
       adsurl = {https://ui.adsabs.harvard.edu/abs/2026A&A...711A..28E}
}

@ARTICLE{Q1-SP054,
       author = {{Euclid Collaboration: Li}, T. and {Collett}, T.~E. and {Walmsley}, M. and others},
        title = "{Euclid Quick Data Release (Q1). XXIX. The strong-lensing discovery engine D -- Double-source-plane lens candidates}",
      journal = {\aap},
         year = 2026,
        month = jun,
       volume = {711},
          eid = {A29},
        pages = {A29},
          doi = {10.1051/0004-6361/202554543},
       adsurl = {https://ui.adsabs.harvard.edu/abs/2026A&A...711A..29E}
}

@ARTICLE{Q1-SP059,
       author = {{Euclid Collaboration: Holloway}, P. and {\ignoreme{AAA}Verma}, A. and {Walmsley}, M. and others},
        title = "{Euclid Quick Data Release (Q1). XXX. The Strong Lensing Discovery Engine E -- Ensemble classification of strong gravitational lenses: lessons for Data Release 1}",
      journal = {\aap},
         year = 2026,
        month = jun,
       volume = {711},
          eid = {A30},
        pages = {A30},
          doi = {10.1051/0004-6361/202554617},
       adsurl = {https://ui.adsabs.harvard.edu/abs/2026A&A...711A..30E}
}

@ARTICLE{Q1-SP063,
       author = {{Euclid Collaboration: Busillo}, V. and {Tortora}, C. and {Metcalf}, R.~B. and others},
        title = "{Euclid Quick Data Release (Q1). XXXI. LEMON -- LEns MOdelling with Neural networks. Automated and fast modelling of Euclid gravitational lenses with singular isothermal ellipsoid mass profile}",
      journal = {\aap},
         year = 2026,
        month = jun,
       volume = {711},
          eid = {A31},
        pages = {A31},
          doi = {10.1051/0004-6361/202554538},
       adsurl = {https://ui.adsabs.harvard.edu/abs/2026A&A...711A..31E}
}

@ARTICLE{Q1-SP085,
       author = {{Euclid Collaboration: Lines}, N.~E.~P. and {\ignoreme{BBB}Collett}, T.~E. and {Holloway}, P. and others},
        title = "{Euclid Quick Data Release (Q1). XXXII. From simulations to sky: Advancing machine-learning lens detection with real Euclid data}",
      journal = {\aap},
         year = 2026,
        month = jun,
       volume = {711},
          eid = {A32},
        pages = {A32},
          doi = {10.1051/0004-6361/202558456},
       adsurl = {https://ui.adsabs.harvard.edu/abs/2026A&A...711A..32E}
}

@ARTICLE{Holloway_Refine,
       author = {{Euclid Collaboration: Holloway}, P. and {\ignoreme{BBB}Verma}, A. and {Collett}, T.~E. and others},
       title = "{Euclid Quick Data Release (Q1). Refining the strong lens sample with Space Warps crowd-sourced science}",
      journal = {A\&A, submitted},
         year = 2026,
        month = nov,
          eid = {arXiv:2611.06667},
        pages = {arXiv:2611.06667},
          doi = {10.48550/arXiv.2611.06667},
archivePrefix = {arXiv},
       eprint = {2611.06667},
 primaryClass = {astro-ph.GA},
       adsurl = {https://ui.adsabs.harvard.edu/abs/2026arXiv261106667E}
}

@ARTICLE{EuclidSkyOverview,
author = {{Euclid Collaboration: Mellier}, Y. and {Abdurro'uf} and {Acevedo~Barroso}, J.A. and others},
	title = {Euclid - I. Overview of the Euclid mission},
	DOI= "10.1051/0004-6361/202450810",
	url= "https://doi.org/10.1051/0004-6361/202450810",
	journal = {A\&A},
	year = 2025,
	volume = 697,
	pages = "A1",
}

@ARTICLE{EuclidSkyVIS,
author = {{Euclid Collaboration: Cropper}, M. and {Al-Bahlawan}, A. and {Amiaux}, J. and others},
	title = {Euclid - II. The VIS instrument},
	DOI= "10.1051/0004-6361/202450996",
	url= "https://doi.org/10.1051/0004-6361/202450996",
	journal = {A\&A},
	year = 2025,
	volume = 697,
	pages = "A2",
}

@ARTICLE{EuclidSkyNISP,
author = {{Euclid Collaboration: Jahnke}, K. and {Gillard}, W. and {Schirmer}, M. and others},
	title = {Euclid - III. The NISP Instrument},
	DOI= "10.1051/0004-6361/202450786",
	url= "https://doi.org/10.1051/0004-6361/202450786",
	journal = {A\&A},
	year = 2025,
	volume = 697,
	pages = "A3",
}

@ARTICLE{EuclidSkyFlagship,
author = {{Euclid Collaboration: Castander}, F. and {Fosalba}, P. and {Stadel}, J. and others},
	title = {Euclid - V. The Flagship galaxy mock catalogue: A comprehensive simulation for the Euclid},
	DOI= "10.1051/0004-6361/202450853",
	url= "https://doi.org/10.1051/0004-6361/202450853",
	journal = {A\&A},
	year = 2025,
	volume = 697,
	pages = "A5",
}

@misc{Q1cite,
author = "{Euclid Quick Release Q1}",
howpublished = "\url{https://doi.org/10.57780/esa-2853f3b}",
year = 2025
}

@misc{DR1cite,
author = "{Euclid Data Release 1}",
howpublished = "\url{https://doi.org/10.57780/esa-d079c51}",
year = 2026
}

@INCOLLECTION{Datalabscite,
       author = {{Navarro}, Vicente and {del Rio}, Sara and {Angel Diego}, Miguel and {Lopez-Caniego}, Marcos and {Marinic}, Filip and {Kruk}, Sandor and {Reerink}, Jan and {Arviset}, Christophe},
        title = "{ESA Datalabs: Digital Innovation in Space Science}",
    booktitle = {Space Data Management. Studies in Big Data},
         year = 2024,
       volume = {141},
        pages = {1-13},
          doi = {10.1007/978-981-97-0041-7_1},
       adsurl = {https://ui.adsabs.harvard.edu/abs/2024sdm..book....1N}
}

@ARTICLE{Scaramella-EP1,
       author = {{Euclid Collaboration: Scaramella}, R. and {Amiaux}, J. and {Mellier}, Y. and others},
        title = "{Euclid preparation. I. The Euclid Wide Survey}",
      journal = {\aap},
         year = 2022,
        month = jun,
       volume = {662},
          eid = {A112},
        pages = {A112},
          doi = {10.1051/0004-6361/202141938},
archivePrefix = {arXiv},
       eprint = {2108.01201},
 primaryClass = {astro-ph.CO},
       adsurl = {https://ui.adsabs.harvard.edu/abs/2022A&A...662A.112E}
}

@ARTICLE{Collett2015,
       author = {{Collett}, Thomas E.},
        title = "{The Population of Galaxy-Galaxy Strong Lenses in Forthcoming Optical Imaging Surveys}",
      journal = {\apj},
         year = 2015,
        month = sep,
       volume = {811},
       number = {1},
          eid = {20},
        pages = {20},
          doi = {10.1088/0004-637X/811/1/20},
archivePrefix = {arXiv},
       eprint = {1507.02657},
 primaryClass = {astro-ph.CO},
       adsurl = {https://ui.adsabs.harvard.edu/abs/2015ApJ...811...20C}
}

@ARTICLE{TianDR1dspl,
       author = {{Euclid Collaboration: Li}, T. and {Lines}, N.~E.~P. and {Holloway}, P. and {Collett}, T.~E. and {Nightingale}, J.~W. and {Vincken}, S.~H. and {Bowden}, D.~J. and {Schuldt}, S. and {Fogliardi}, M. and et al.},
        title = "{Euclid Data Release 1 (DR1). Double-source-plane lens candidates}",
      journal = {A\&A, submitted},
         year = 2026,
        month = nov,
          eid = {arXiv:2611.20166},
        pages = {arXiv:2611.20166},
archivePrefix = {arXiv},
       eprint = {2611.20166},
 primaryClass = {astro-ph.GA},
       adsurl = {https://scixplorer.org/abs/2026arXiv261120166E}
}

@ARTICLE{Vincken2026,
       author = {{Euclid Collaboration: Vincken}, S.~H. and {\ignoreme{AAA}Rojas}, K. and {Melchior}, M. and {Lines}, N.~E.~P. and {Collett}, T.~E. and {Verma}, A. and {Holloway}, P. and {Despali}, G. and {Schuldt}, S. and {Metcalf}, R.~B. and {Gavazzi}, R. and {Courbin}, F. and {Acevedo Barroso}, J.~A. and {Cl{\'e}ment}, B. and {Li}, T. and {Sluse}, D. and {Wilde}, J. and {Melo}, A. and {Sonnenfeld}, A. and {Tortora}, C. and {Thai}, T.~T. and {Millon}, M. and {Spiniello}, C. and {Manj{\'o}n-Garc{\'\i}a}, A. and {Meneghetti}, M. and {Nagam}, B.~C. and {Altieri}, B. and {Andreon}, S. and {Auricchio}, N. and {Baccigalupi}, C. and {Baldi}, M. and {Balestra}, A. and {Bardelli}, S. and {Battaglia}, P. and {Biviano}, A. and {Branchini}, E. and {Brescia}, M. and {Camera}, S. and {Capobianco}, V. and {Carbone}, C. and {Carretero}, J. and {Casas}, S. and {Castellano}, M. and {Castignani}, G. and {Cavuoti}, S. and {Cimatti}, A. and {Colodro-Conde}, C. and {Congedo}, G. and {Conselice}, C.~J. and {Conversi}, L. and {Copin}, Y. and {Costille}, A. and {Courtois}, H.~M. and {Cropper}, M. and {Da Silva}, A. and {Degaudenzi}, H. and {De Lucia}, G. and {Dolding}, C. and {Dole}, H. and {Douspis}, M. and {Dubath}, F. and {Dupac}, X. and {Dusini}, S. and {Escoffier}, S. and {Farina}, M. and {Faustini}, F. and {Ferriol}, S. and {Finelli}, F. and {Fosalba}, P. and {Fotopoulou}, S. and {Frailis}, M. and {Franceschi}, E. and {Fumana}, M. and {Galeotta}, S. and {George}, K. and {Gillis}, B. and {Giocoli}, C. and {Gracia-Carpio}, J. and {Grazian}, A. and {Grupp}, F. and {Guzzo}, L. and {Haugan}, S.~V.~H. and {Hoar}, J. and {Holmes}, W. and {Hook}, I.~M. and {Hormuth}, F. and {Hornstrup}, A. and {Jahnke}, K. and {Jhabvala}, M. and {Joachimi}, B. and {Kermiche}, S. and {Kiessling}, A. and {Kubik}, B. and {K{\"u}mmel}, M. and {Kunz}, M. and {Kurki-Suonio}, H. and {Le Brun}, A.~M.~C. and {Ligori}, S. and {Lilje}, P.~B. and {Lindholm}, V. and {Lloro}, I. and {Mainetti}, G. and {Maiorano}, E. and {Mansutti}, O. and {Marcin}, S. and {Marggraf}, O. and {Martinelli}, M. and {Martinet}, N. and {Marulli}, F. and {Massey}, R.~J. and {Medinaceli}, E. and {Mei}, S. and {Merlin}, E. and {Meylan}, G. and {Mora}, A. and {Moresco}, M. and {Moscardini}, L. and {Munari}, E. and {Nakajima}, R. and {Neissner}, C. and {Nichol}, R.~C. and {Niemi}, S.-M. and {Nightingale}, J.~W. and {Padilla}, C. and {Paltani}, S. and {Pasian}, F. and {Pedersen}, K. and {Percival}, W.~J. and {Pettorino}, V. and {Pires}, S. and {Polenta}, G. and {Poncet}, M. and {Popa}, L.~A. and {Raison}, F. and {Renzi}, A. and {Rhodes}, J. and {Riccio}, G. and {Romelli}, E. and {Roncarelli}, M. and {Saglia}, R. and {Sakr}, Z. and {Sapone}, D. and {Sartoris}, B. and {Schirmer}, M. and {Schneider}, P. and {Secroun}, A. and {Seidel}, G. and {Sihvola}, E. and {Simon}, P. and {Sirignano}, C. and {Sirri}, G. and {Stanco}, L. and {Tallada-Cresp{\'\i}}, P. and {Taylor}, A.~N. and {Tereno}, I. and {Tessore}, N. and {Toft}, S. and {Torradeflot}, F. and {Tutusaus}, I. and {Valenziano}, L. and {Valiviita}, J. and {Vassallo}, T. and {Verdoes Kleijn}, G. and {Veropalumbo}, A. and {Wang}, Y. and {Weller}, J. and {Zacchei}, A. and {Zamorani}, G. and {Zucca}, E. and {Ballardini}, M. and {Bozzo}, E. and {Burigana}, C. and {Cabanac}, R. and {Calabrese}, M. and {Cappi}, A. and {Castro}, T. and {Escartin Vigo}, J.~A. and {Gabarra}, L. and {Hemmati}, S. and {Macias-Perez}, J. and {Maoli}, R. and {Mart{\'\i}n-Fleitas}, J. and {Mauri}, N. and {Monaco}, P. and {Nucita}, A.~A. and {Pezzotta}, A. and {P{\"o}ntinen}, M. and {Risso}, I. and {Scottez}, V. and {Sereno}, M. and {Tenti}, M. and {Tucci}, M. and {Viel}, M. and {Wiesmann}, M. and {Akrami}, Y. and {Andika}, I.~T. and {Angora}, G. and {Anselmi}, S. and {Archidiacono}, M.},
        title = "{Euclid Quick Data Release (Q1). AstroVink: A vision transformer approach to find strong gravitational lens systems}",
      journal = {A\&A, submitted},
         year = 2026,
        month = apr,
          eid = {arXiv:2604.21977},
        pages = {arXiv:2604.21977},
archivePrefix = {arXiv},
       eprint = {2604.21977},
 primaryClass = {astro-ph.IM},
       adsurl = {https://ui.adsabs.harvard.edu/abs/2026arXiv260421977E}
}

@misc{ManjonGarcia2021,
    school   = {University of Cantabria (Spain)},
    author   = {Manjón-García,A},
    year     = {2021},
    howpublished="\url{https://hdl.handle.net/10902/24245}",
    title="Automated techniques for the detection and analysis of the gravitational lensing effect in new generation galaxy surveys"
}

@ARTICLE{DominquezSanchez2018,
       author = {{Dom{\'\i}nguez S{\'a}nchez}, H. and {Huertas-Company}, M. and {Bernardi}, M. and {Tuccillo}, D. and {Fischer}, J.~L.},
        title = "{Improving galaxy morphologies for SDSS with Deep Learning}",
      journal = {\mnras},
         year = 2018,
        month = feb,
       volume = {476},
       number = {3},
        pages = {3661-3676},
          doi = {10.1093/mnras/sty338},
archivePrefix = {arXiv},
       eprint = {1711.05744},
 primaryClass = {astro-ph.GA},
       adsurl = {https://ui.adsabs.harvard.edu/abs/2018MNRAS.476.3661D}
}

@ARTICLE{Wilde2022,
       author = {{Wilde}, Joshua and {Serjeant}, Stephen and {Bromley}, Jane M. and {Dickinson}, Hugh and {Koopmans}, L{\'e}on V.~E. and {Metcalf}, R. Benton},
        title = "{Detecting gravitational lenses using machine learning: exploring interpretability and sensitivity to rare lensing configurations}",
      journal = {\mnras},
         year = 2022,
        month = may,
       volume = {512},
       number = {3},
        pages = {3464-3479},
          doi = {10.1093/mnras/stac562},
archivePrefix = {arXiv},
       eprint = {2202.12776},
 primaryClass = {astro-ph.GA},
       adsurl = {https://ui.adsabs.harvard.edu/abs/2022MNRAS.512.3464W}
}

@ARTICLE{Ecker2026,
       author = {{Euclid Collaboration: Ecker}, L.~R. and {Fabricius}, M. and {Seitz}, S. and {Saglia}, R. and {Lines}, N.~E.~P. and {Holloway}, P. and {Li}, T. and {Verma}, A. and {Balzer}, F. and {Jin}, Q. and {Manj{\'o}n-Garc{\'\i}a}, A. and {Vincken}, S.~H. and {Wilde}, J. and {Acevedo Barroso}, J.~A. and {Nightingale}, J.~W. and {Rojas}, K. and {Schuldt}, S. and {Walmsley}, M. and {Collett}, T.~E. and {Despali}, G. and {Sonnenfeld}, A. and {Tortora}, C. and {Metcalf}, R.~B. and {Bender}, R. and {Saulder}, C. and {Baeten}, E. and {Cornen}, C. and {Delley}, D. and {Finner}, K. and {Galan}, A. and {Gavazzi}, R. and {Johnson}, L.~C. and {Leuzzi}, L. and {Macmillan}, C. and {Marshall}, P.~J. and {Millon}, M. and {More}, A. and {Moustakas}, L.~A. and {Pearson}, J. and {Pippert}, J.-N. and {Scarlata}, C. and {Sluse}, D. and {Spiniello}, C. and {Thai}, T.~T. and {Ulivi}, L. and {Wang}, Han. and {Xu}, X. and {Courbin}, F. and {Meneghetti}, M. and {Aghanim}, N. and {Altieri}, B. and {Andreon}, S. and {Auricchio}, N. and {Baccigalupi}, C. and {Baldi}, M. and {Balestra}, A. and {Bardelli}, S. and {Battaglia}, P. and {Biviano}, A. and {Branchini}, E. and {Brescia}, M. and {Camera}, S. and {Ca{\~n}as-Herrera}, G. and {Capobianco}, V. and {Carbone}, C. and {Carretero}, J. and {Casas}, S. and {Castellano}, M. and {Castignani}, G. and {Cavuoti}, S. and {Chambers}, K.~C. and {Cimatti}, A. and {Colodro-Conde}, C. and {Congedo}, G. and {Conselice}, C.~J. and {Conversi}, L. and {Copin}, Y. and {Costille}, A. and {Courtois}, H.~M. and {Cropper}, M. and {Da Silva}, A. and {Degaudenzi}, H. and {De Lucia}, G. and {Dolding}, C. and {Dole}, H. and {Dubath}, F. and {Dupac}, X. and {Dusini}, S. and {Ealet}, A. and {Escoffier}, S. and {Farina}, M. and {Farinelli}, R. and {Faustini}, F. and {Ferriol}, S. and {Finelli}, F. and {Fosalba}, P. and {Fotopoulou}, S. and {Frailis}, M. and {Franceschi}, E. and {Fumana}, M. and {Galeotta}, S. and {George}, K. and {Gillard}, W. and {Gillis}, B. and {Giocoli}, C. and {G{\'o}mez-Alvarez}, P. and {Gracia-Carpio}, J. and {Grazian}, A. and {Grupp}, F. and {Guzzo}, L. and {Haugan}, S.~V.~H. and {Hoekstra}, H. and {Holmes}, W. and {Hormuth}, F. and {Hornstrup}, A. and {Jahnke}, K. and {Jhabvala}, M. and {Joachimi}, B. and {Keih{\"a}nen}, E. and {Kermiche}, S. and {Kiessling}, A. and {Kubik}, B. and {K{\"u}mmel}, M. and {Kunz}, M. and {Kurki-Suonio}, H. and {Le Brun}, A.~M.~C. and {Le Mignant}, D. and {Ligori}, S. and {Lilje}, P.~B. and {Lindholm}, V. and {Lloro}, I. and {Mainetti}, G. and {Maino}, D. and {Maiorano}, E. and {Mansutti}, O. and {Marcin}, S. and {Marggraf}, O. and {Martinelli}, M. and {Martinet}, N. and {Marulli}, F. and {Massey}, R.~J. and {Medinaceli}, E. and {Mei}, S. and {Mellier}, Y. and {Merlin}, E. and {Meylan}, G. and {Mora}, A. and {Moresco}, M. and {Moscardini}, L. and {Nakajima}, R. and {Neissner}, C. and {Nichol}, R.~C. and {Niemi}, S.-M. and {Padilla}, C. and {Paltani}, S. and {Pasian}, F. and {Pedersen}, K. and {Percival}, W.~J. and {Pettorino}, V. and {Pires}, S. and {Polenta}, G. and {Poncet}, M. and {Pozzetti}, L. and {Raison}, F. and {Renzi}, A. and {Rhodes}, J. and {Riccio}, G. and {Rix}, H.-W. and {Romelli}, E. and {Roncarelli}, M. and {Rossetti}, E. and {Sakr}, Z. and {S{\'a}nchez}, A.~G. and {Sapone}, D. and {Sartoris}, B. and {Schneider}, P. and {Schrabback}, T. and {Secroun}, A. and {Seidel}, G. and {Serrano}, S. and {Simon}, P. and {Sirignano}, C. and {Sirri}, G. and {Stanco}, L. and {Steinwagner}, J. and {Tallada-Cresp{\'\i}}, P. and {Taylor}, A.~N. and {Teplitz}, H.~I. and {Tereno}, I. and {Tessore}, N. and {Toft}, S. and {Toledo-Moreo}, R. and {Torradeflot}, F. and {Tutusaus}, I. and {Valenziano}, L. and {Valiviita}, J. and {Vassallo}, T. and {Wang}, Y. and {Weller}, J.},
        title = "{Euclid Quick Data Release (Q1). The Strong Lensing Discovery Engine F -- Bright and low-redshift strong lenses}",
      journal = {arXiv e-prints},
         year = 2026,
        month = mar,
          eid = {arXiv:2603.28580},
        pages = {arXiv:2603.28580},
          doi = {10.48550/arXiv.2603.28580},
archivePrefix = {arXiv},
       eprint = {2603.28580},
 primaryClass = {astro-ph.GA},
       adsurl = {https://ui.adsabs.harvard.edu/abs/2026arXiv260328580E}
}

@ARTICLE{Marshall2016,
       author = {{Marshall}, Philip J. and {Verma}, Aprajita and {More}, Anupreeta and {Davis}, Christopher P. and {More}, Surhud and {Kapadia}, Amit and {Parrish}, Michael and {Snyder}, Chris and {Wilcox}, Julianne and {Baeten}, Elisabeth and {Macmillan}, Christine and {Cornen}, Claude and {Baumer}, Michael and {Simpson}, Edwin and {Lintott}, Chris J. and {Miller}, David and {Paget}, Edward and {Simpson}, Robert and {Smith}, Arfon M. and {K{\"u}ng}, Rafael and {Saha}, Prasenjit and {Collett}, Thomas E.},
        title = "{SPACE WARPS - I. Crowdsourcing the discovery of gravitational lenses}",
      journal = {\mnras},
         year = 2016,
        month = jan,
       volume = {455},
       number = {2},
        pages = {1171-1190},
          doi = {10.1093/mnras/stv2009},
archivePrefix = {arXiv},
       eprint = {1504.06148},
 primaryClass = {astro-ph.IM},
       adsurl = {https://ui.adsabs.harvard.edu/abs/2016MNRAS.455.1171M}
}

@ARTICLE{More2016,
       author = {{More}, Anupreeta and {Verma}, Aprajita and {Marshall}, Philip J. and {More}, Surhud and {Baeten}, Elisabeth and {Wilcox}, Julianne and {Macmillan}, Christine and {Cornen}, Claude and {Kapadia}, Amit and {Parrish}, Michael and {Snyder}, Chris and {Davis}, Christopher P. and {Gavazzi}, Raphael and {Lintott}, Chris J. and {Simpson}, Robert and {Miller}, David and {Smith}, Arfon M. and {Paget}, Edward and {Saha}, Prasenjit and {K{\"u}ng}, Rafael and {Collett}, Thomas E.},
        title = "{SPACE WARPS- II. New gravitational lens candidates from the CFHTLS discovered through citizen science}",
      journal = {\mnras},
         year = 2016,
        month = jan,
       volume = {455},
       number = {2},
        pages = {1191-1210},
          doi = {10.1093/mnras/stv1965},
archivePrefix = {arXiv},
       eprint = {1504.05587},
 primaryClass = {astro-ph.CO},
       adsurl = {https://ui.adsabs.harvard.edu/abs/2016MNRAS.455.1191M}
}

@ARTICLE{Sonnenfeld2020,
       author = {{Sonnenfeld}, Alessandro and {Verma}, Aprajita and {More}, Anupreeta and {Baeten}, Elisabeth and {Macmillan}, Christine and {Wong}, Kenneth C. and {Chan}, James H.~H. and {Jaelani}, Anton T. and {Lee}, Chien-Hsiu and {Oguri}, Masamune and {Rusu}, Cristian E. and {Veldthuis}, Marten and {Trouille}, Laura and {Marshall}, Philip J. and {Hutchings}, Roger and {Allen}, Campbell and {O'Donnell}, James and {Cornen}, Claude and {Davis}, Christopher P. and {McMaster}, Adam and {Lintott}, Chris and {Miller}, Grant},
        title = "{Survey of Gravitationally-lensed Objects in HSC Imaging (SuGOHI). VI. Crowdsourced lens finding with Space Warps}",
      journal = {\aap},
         year = 2020,
        month = oct,
       volume = {642},
          eid = {A148},
        pages = {A148},
          doi = {10.1051/0004-6361/202038067},
archivePrefix = {arXiv},
       eprint = {2004.00634},
 primaryClass = {astro-ph.IM},
       adsurl = {https://ui.adsabs.harvard.edu/abs/2020A&A...642A.148S}
}

@ARTICLE{Holloway2024,
       author = {{Holloway}, Philip and {Marshall}, Philip J. and {Verma}, Aprajita and {More}, Anupreeta and {Ca{\~n}ameras}, Raoul and {Jaelani}, Anton T. and {Ishida}, Yuichiro and {Wong}, Kenneth C.},
        title = "{A Bayesian approach to strong lens finding in the era of wide-area surveys}",
      journal = {\mnras},
         year = 2024,
        month = may,
       volume = {530},
       number = {2},
        pages = {1297-1310},
          doi = {10.1093/mnras/stae875},
archivePrefix = {arXiv},
       eprint = {2311.07455},
 primaryClass = {astro-ph.GA},
       adsurl = {https://ui.adsabs.harvard.edu/abs/2024MNRAS.530.1297H}
}

@ARTICLE{Ivezic2019,
       author = {{Ivezi{\'c}}, {\v{Z}}eljko and {Kahn}, Steven M. and {Tyson}, J. Anthony and {Abel}, Bob and {Acosta}, Emily and {Allsman}, Robyn and {Alonso}, David and {AlSayyad}, Yusra and {Anderson}, Scott F. and {Andrew}, John and {Angel}, James Roger P. and {Angeli}, George Z. and {Ansari}, Reza and {Antilogus}, Pierre and {Araujo}, Constanza and {Armstrong}, Robert and {Arndt}, Kirk T. and {Astier}, Pierre and {Aubourg}, {\'E}ric and {Auza}, Nicole and {Axelrod}, Tim S. and {Bard}, Deborah J. and {Barr}, Jeff D. and {Barrau}, Aurelian and {Bartlett}, James G. and {Bauer}, Amanda E. and {Bauman}, Brian J. and {Baumont}, Sylvain and {Bechtol}, Ellen and {Bechtol}, Keith and {Becker}, Andrew C. and {Becla}, Jacek and {Beldica}, Cristina and {Bellavia}, Steve and {Bianco}, Federica B. and {Biswas}, Rahul and {Blanc}, Guillaume and {Blazek}, Jonathan and {Blandford}, Roger D. and {Bloom}, Josh S. and {Bogart}, Joanne and {Bond}, Tim W. and {Booth}, Michael T. and {Borgland}, Anders W. and {Borne}, Kirk and {Bosch}, James F. and {Boutigny}, Dominique and {Brackett}, Craig A. and {Bradshaw}, Andrew and {Brandt}, William Nielsen and {Brown}, Michael E. and {Bullock}, James S. and {Burchat}, Patricia and {Burke}, David L. and {Cagnoli}, Gianpietro and {Calabrese}, Daniel and {Callahan}, Shawn and {Callen}, Alice L. and {Carlin}, Jeffrey L. and {Carlson}, Erin L. and {Chandrasekharan}, Srinivasan and {Charles-Emerson}, Glenaver and {Chesley}, Steve and {Cheu}, Elliott C. and {Chiang}, Hsin-Fang and {Chiang}, James and {Chirino}, Carol and {Chow}, Derek and {Ciardi}, David R. and {Claver}, Charles F. and {Cohen-Tanugi}, Johann and {Cockrum}, Joseph J. and {Coles}, Rebecca and {Connolly}, Andrew J. and {Cook}, Kem H. and {Cooray}, Asantha and {Covey}, Kevin R. and {Cribbs}, Chris and {Cui}, Wei and {Cutri}, Roc and {Daly}, Philip N. and {Daniel}, Scott F. and {Daruich}, Felipe and {Daubard}, Guillaume and {Daues}, Greg and {Dawson}, William and {Delgado}, Francisco and {Dellapenna}, Alfred and {de Peyster}, Robert and {de Val-Borro}, Miguel and {Digel}, Seth W. and {Doherty}, Peter and {Dubois}, Richard and {Dubois-Felsmann}, Gregory P. and {Durech}, Josef and {Economou}, Frossie and {Eifler}, Tim and {Eracleous}, Michael and {Emmons}, Benjamin L. and {Fausti Neto}, Angelo and {Ferguson}, Henry and {Figueroa}, Enrique and {Fisher-Levine}, Merlin and {Focke}, Warren and {Foss}, Michael D. and {Frank}, James and {Freemon}, Michael D. and {Gangler}, Emmanuel and {Gawiser}, Eric and {Geary}, John C. and {Gee}, Perry and {Geha}, Marla and {Gessner}, Charles J.~B. and {Gibson}, Robert R. and {Gilmore}, D. Kirk and {Glanzman}, Thomas and {Glick}, William and {Goldina}, Tatiana and {Goldstein}, Daniel A. and {Goodenow}, Iain and {Graham}, Melissa L. and {Gressler}, William J. and {Gris}, Philippe and {Guy}, Leanne P. and {Guyonnet}, Augustin and {Haller}, Gunther and {Harris}, Ron and {Hascall}, Patrick A. and {Haupt}, Justine and {Hernandez}, Fabio and {Herrmann}, Sven and {Hileman}, Edward and {Hoblitt}, Joshua and {Hodgson}, John A. and {Hogan}, Craig and {Howard}, James D. and {Huang}, Dajun and {Huffer}, Michael E. and {Ingraham}, Patrick and {Innes}, Walter R. and {Jacoby}, Suzanne H. and {Jain}, Bhuvnesh and {Jammes}, Fabrice and {Jee}, M. James and {Jenness}, Tim and {Jernigan}, Garrett and {Jevremovi{\'c}}, Darko and {Johns}, Kenneth and {Johnson}, Anthony S. and {Johnson}, Margaret W.~G. and {Jones}, R. Lynne and {Juramy-Gilles}, Claire and {Juri{\'c}}, Mario and {Kalirai}, Jason S. and {Kallivayalil}, Nitya J. and {Kalmbach}, Bryce and {Kantor}, Jeffrey P. and {Karst}, Pierre and {Kasliwal}, Mansi M. and {Kelly}, Heather and {Kessler}, Richard and {Kinnison}, Veronica and {Kirkby}, David and {Knox}, Lloyd and {Kotov}, Ivan V. and {Krabbendam}, Victor L. and {Krughoff}, K. Simon and {Kub{\'a}nek}, Petr and {Kuczewski}, John and {Kulkarni}, Shri and {Ku}, John and {Kurita}, Nadine R. and {Lage}, Craig S. and {Lambert}, Ron and {Lange}, Travis and {Langton}, J. Brian and {Le Guillou}, Laurent and {Levine}, Deborah and {Liang}, Ming and {Lim}, Kian-Tat and {Lintott}, Chris J. and {Long}, Kevin E. and {Lopez}, Margaux and {Lotz}, Paul J. and {Lupton}, Robert H. and {Lust}, Nate B. and {MacArthur}, Lauren A. and {Mahabal}, Ashish and {Mandelbaum}, Rachel and {Markiewicz}, Thomas W. and {Marsh}, Darren S. and {Marshall}, Philip J. and {Marshall}, Stuart and {May}, Morgan and {McKercher}, Robert and {McQueen}, Michelle and {Meyers}, Joshua and {Migliore}, Myriam and {Miller}, Michelle and {Mills}, David J.},
        title = "{LSST: From Science Drivers to Reference Design and Anticipated Data Products}",
      journal = {\apj},
         year = 2019,
        month = mar,
       volume = {873},
       number = {2},
          eid = {111},
        pages = {111},
          doi = {10.3847/1538-4357/ab042c},
archivePrefix = {arXiv},
       eprint = {0805.2366},
 primaryClass = {astro-ph},
       adsurl = {https://ui.adsabs.harvard.edu/abs/2019ApJ...873..111I}
}

@article{Walmsley2023, doi = {10.21105/joss.05312}, url = {https://doi.org/10.21105/joss.05312}, year = {2023}, publisher = {The Open Journal}, volume = {8}, number = {85}, pages = {5312}, author = {Mike Walmsley and Campbell Allen and Ben Aussel and Micah Bowles and Kasia Gregorowicz and Inigo Val Slijepcevic and Chris J. Lintott and Anna M. m. Scaife and Maja Jabłońska and Kosio Karchev and Denise Lanzieri and Devina Mohan and David O’Ryan and Bharath Saiguhan and Crisel Suárez and Nicolás Guerra-Varas and Renuka Velu}, title = {Zoobot: Adaptable Deep Learning Models for Galaxy Morphology}, journal = {Journal of Open Source Software} }

@ARTICLE{Lintott2008,
       author = {{Lintott}, Chris J. and {Schawinski}, Kevin and {Slosar}, An{\v{z}}e and {Land}, Kate and {Bamford}, Steven and {Thomas}, Daniel and {Raddick}, M. Jordan and {Nichol}, Robert C. and {Szalay}, Alex and {Andreescu}, Dan and {Murray}, Phil and {Vandenberg}, Jan},
        title = "{Galaxy Zoo: morphologies derived from visual inspection of galaxies from the Sloan Digital Sky Survey}",
      journal = {\mnras},
         year = 2008,
        month = sep,
       volume = {389},
       number = {3},
        pages = {1179-1189},
          doi = {10.1111/j.1365-2966.2008.13689.x},
archivePrefix = {arXiv},
       eprint = {0804.4483},
 primaryClass = {astro-ph},
       adsurl = {https://ui.adsabs.harvard.edu/abs/2008MNRAS.389.1179L}
}

@ARTICLE{TDCOSMO2025,
       author = {{TDCOSMO Collaboration} and {Birrer}, Simon and {Buckley-Geer}, Elizabeth J. and {Cappellari}, Michele and {Courbin}, Fr{\'e}d{\'e}ric and {Dux}, Fr{\'e}d{\'e}ric and {Fassnacht}, Christopher D. and {Frieman}, Joshua A. and {Galan}, Aymeric and {Gilman}, Daniel and {Huang}, Xiang-Yu and {Knabel}, Shawn and {Langeroodi}, Danial and {Lin}, Huan and {Millon}, Martin and {Morishita}, Takahiro and {Motta}, Veronica and {Mozumdar}, Pritom and {Paic}, Eric and {Shajib}, Anowar J. and {Sheu}, William and {Sluse}, Dominique and {Sonnenfeld}, Alessandro and {Spiniello}, Chiara and {Stiavelli}, Massimo and {Suyu}, Sherry H. and {Tan}, Chin Yi and {Treu}, Tommaso and {van de Vyvere}, Lyne and {Wang}, Han and {Wells}, Patrick and {Williams}, Devon M. and {Wong}, Kenneth C.},
        title = "{TDCOSMO 2025: Cosmological constraints from strong lensing time delays}",
      journal = {\aap},
         year = 2025,
        month = dec,
       volume = {704},
          eid = {A63},
        pages = {A63},
          doi = {10.1051/0004-6361/202555801},
archivePrefix = {arXiv},
       eprint = {2506.03023},
 primaryClass = {astro-ph.CO},
       adsurl = {https://ui.adsabs.harvard.edu/abs/2025A&A...704A..63T}
}

@ARTICLE{Treu2022,
       author = {{Treu}, Tommaso and {Suyu}, Sherry H. and {Marshall}, Philip J.},
        title = "{Strong lensing time-delay cosmography in the 2020s}",
      journal = {\aapr},
         year = 2022,
        month = dec,
       volume = {30},
       number = {1},
          eid = {8},
        pages = {8},
          doi = {10.1007/s00159-022-00145-y},
archivePrefix = {arXiv},
       eprint = {2210.15794},
 primaryClass = {astro-ph.CO},
       adsurl = {https://ui.adsabs.harvard.edu/abs/2022A&ARv..30....8T}
}

@ARTICLE{Li2024,
       author = {{Li}, Tian and {Collett}, Thomas E. and {Krawczyk}, Coleman M. and {Enzi}, Wolfgang},
        title = "{Cosmology from large populations of galaxy-galaxy strong gravitational lenses}",
      journal = {\mnras},
         year = 2024,
        month = jan,
       volume = {527},
       number = {3},
        pages = {5311-5323},
          doi = {10.1093/mnras/stad3514},
archivePrefix = {arXiv},
       eprint = {2307.09271},
 primaryClass = {astro-ph.CO},
       adsurl = {https://ui.adsabs.harvard.edu/abs/2024MNRAS.527.5311L}
}

@ARTICLE{Collett2014,
       author = {{Collett}, Thomas E. and {Auger}, Matthew W.},
        title = "{Cosmological constraints from the double source plane lens SDSSJ0946+1006}",
      journal = {\mnras},
         year = 2014,
        month = sep,
       volume = {443},
       number = {2},
        pages = {969-976},
          doi = {10.1093/mnras/stu1190},
archivePrefix = {arXiv},
       eprint = {1403.5278},
 primaryClass = {astro-ph.CO},
       adsurl = {https://ui.adsabs.harvard.edu/abs/2014MNRAS.443..969C}
}

@ARTICLE{Shajib2025,
       author = {{Shajib}, Anowar J. and {Smith}, Graham P. and {Birrer}, Simon and {Verma}, Aprajita and {Arendse}, Nikki and {Collett}, Thomas and {Daylan}, Tansu and {Serjeant}, Stephen and {LSST Strong Lensing Science Collaboration}},
        title = "{Strong gravitational lenses from the Vera C. Rubin Observatory}",
      journal = {Philosophical Transactions of the Royal Society of London Series A},
         year = 2025,
        month = may,
       volume = {383},
       number = {2295},
          eid = {20240117},
        pages = {20240117},
          doi = {10.1098/rsta.2024.0117},
archivePrefix = {arXiv},
       eprint = {2406.08919},
 primaryClass = {astro-ph.CO},
       adsurl = {https://ui.adsabs.harvard.edu/abs/2025RSPTA.38340117S}
}

@ARTICLE{Sonnenfeld2025,
       author = {{Sonnenfeld}, Alessandro},
        title = "{The SLACS strong lens sample, debiased: II. Lensing-only constraints on the stellar initial mass function and dark matter contraction in early-type galaxies}",
      journal = {\aap},
         year = 2025,
        month = may,
       volume = {697},
          eid = {A95},
        pages = {A95},
          doi = {10.1051/0004-6361/202452846},
archivePrefix = {arXiv},
       eprint = {2501.02054},
 primaryClass = {astro-ph.GA},
       adsurl = {https://ui.adsabs.harvard.edu/abs/2025A&A...697A..95S}
}

@ARTICLE{Leier2016,
       author = {{Leier}, Dominik and {Ferreras}, Ignacio and {Saha}, Prasenjit and {Charlot}, St{\'e}phane and {Bruzual}, Gustavo and {La Barbera}, Francesco},
        title = "{Strong gravitational lensing and the stellar IMF of early-type galaxies}",
      journal = {\mnras},
         year = 2016,
        month = jul,
       volume = {459},
       number = {4},
        pages = {3677-3692},
          doi = {10.1093/mnras/stw885},
archivePrefix = {arXiv},
       eprint = {1512.00462},
 primaryClass = {astro-ph.GA},
       adsurl = {https://ui.adsabs.harvard.edu/abs/2016MNRAS.459.3677L}
}

@ARTICLE{Newman2017,
       author = {{Newman}, Andrew B. and {Smith}, Russell J. and {Conroy}, Charlie and {Villaume}, Alexa and {van Dokkum}, Pieter},
        title = "{The Initial Mass Function in the Nearest Strong Lenses from SNELLS: Assessing the Consistency of Lensing, Dynamical, and Spectroscopic Constraints}",
      journal = {\apj},
         year = 2017,
        month = aug,
       volume = {845},
       number = {2},
          eid = {157},
        pages = {157},
          doi = {10.3847/1538-4357/aa816d},
archivePrefix = {arXiv},
       eprint = {1612.00065},
 primaryClass = {astro-ph.GA},
       adsurl = {https://ui.adsabs.harvard.edu/abs/2017ApJ...845..157N}
}

@ARTICLE{Sonnenfeld2013,
       author = {{Sonnenfeld}, Alessandro and {Treu}, Tommaso and {Gavazzi}, Rapha{\"e}l and {Suyu}, Sherry H. and {Marshall}, Philip J. and {Auger}, Matthew W. and {Nipoti}, Carlo},
        title = "{The SL2S Galaxy-scale Lens Sample. IV. The Dependence of the Total Mass Density Profile of Early-type Galaxies on Redshift, Stellar Mass, and Size}",
      journal = {\apj},
         year = 2013,
        month = nov,
       volume = {777},
       number = {2},
          eid = {98},
        pages = {98},
          doi = {10.1088/0004-637X/777/2/98},
archivePrefix = {arXiv},
       eprint = {1307.4759},
 primaryClass = {astro-ph.CO},
       adsurl = {https://ui.adsabs.harvard.edu/abs/2013ApJ...777...98S}
}

@ARTICLE{Koopmans2006,
       author = {{Koopmans}, L{\'e}on V.~E. and {Treu}, Tommaso and {Bolton}, Adam S. and {Burles}, Scott and {Moustakas}, Leonidas A.},
        title = "{The Sloan Lens ACS Survey. III. The Structure and Formation of Early-Type Galaxies and Their Evolution since z \raisebox{-0.5ex}\textasciitilde 1}",
      journal = {\apj},
         year = 2006,
        month = oct,
       volume = {649},
       number = {2},
        pages = {599-615},
          doi = {10.1086/505696},
archivePrefix = {arXiv},
       eprint = {astro-ph/0601628},
 primaryClass = {astro-ph},
       adsurl = {https://ui.adsabs.harvard.edu/abs/2006ApJ...649..599K}
}

@ARTICLE{Seidel2007,
       author = {{Seidel}, G. and {Bartelmann}, M.},
        title = "{Arcfinder: an algorithm for the automatic detection of gravitational arcs}",
      journal = {\aap},
         year = 2007,
        month = sep,
       volume = {472},
       number = {1},
        pages = {341-352},
          doi = {10.1051/0004-6361:20066097},
archivePrefix = {arXiv},
       eprint = {astro-ph/0607547},
 primaryClass = {astro-ph},
       adsurl = {https://ui.adsabs.harvard.edu/abs/2007A&A...472..341S}
}

@ARTICLE{Gavazzi2014,
       author = {{Gavazzi}, Rapha{\"e}l and {Marshall}, Philip J. and {Treu}, Tommaso and {Sonnenfeld}, Alessandro},
        title = "{RINGFINDER: Automated Detection of Galaxy-scale Gravitational Lenses in Ground-based Multi-filter Imaging Data}",
      journal = {\apj},
         year = 2014,
        month = apr,
       volume = {785},
       number = {2},
          eid = {144},
        pages = {144},
          doi = {10.1088/0004-637X/785/2/144},
archivePrefix = {arXiv},
       eprint = {1403.1041},
 primaryClass = {astro-ph.CO},
       adsurl = {https://ui.adsabs.harvard.edu/abs/2014ApJ...785..144G}
}

@ARTICLE{Sonnenfeld2018,
       author = {{Sonnenfeld}, Alessandro and {Chan}, James H.~H. and {Shu}, Yiping and {More}, Anupreeta and {Oguri}, Masamune and {Suyu}, Sherry H. and {Wong}, Kenneth C. and {Lee}, Chien-Hsiu and {Coupon}, Jean and {Yonehara}, Atsunori and {Bolton}, Adam S. and {Jaelani}, Anton T. and {Tanaka}, Masayuki and {Miyazaki}, Satoshi and {Komiyama}, Yutaka},
        title = "{Survey of Gravitationally-lensed Objects in HSC Imaging (SuGOHI). I. Automatic search for galaxy-scale strong lenses}",
      journal = {\pasj},
         year = 2018,
        month = jan,
       volume = {70},
          eid = {S29},
        pages = {S29},
          doi = {10.1093/pasj/psx062},
archivePrefix = {arXiv},
       eprint = {1704.01585},
 primaryClass = {astro-ph.GA},
       adsurl = {https://ui.adsabs.harvard.edu/abs/2018PASJ...70S..29S}
}

@ARTICLE{Schuldt2025,
       author = {{Schuldt}, S. and {Ca{\~n}ameras}, R. and {Shu}, Y. and {Andika}, I.~T. and {Bag}, S. and {Grillo}, C. and {Melo}, A. and {Suyu}, S.~H. and {Taubenberger}, S.},
        title = "{HOLISMOKES: XVI. Lens search in HSC-PDR3 with a neural network committee and post-processing for false-positive removal}",
      journal = {\aap},
         year = 2025,
        month = jul,
       volume = {699},
          eid = {A350},
        pages = {A350},
          doi = {10.1051/0004-6361/202554425},
archivePrefix = {arXiv},
       eprint = {2503.07733},
 primaryClass = {astro-ph.IM},
       adsurl = {https://ui.adsabs.harvard.edu/abs/2025A&A...699A.350S}
}

@ARTICLE{Andika2023,
       author = {{Andika}, Irham Taufik and {Suyu}, Sherry H. and {Ca{\~n}ameras}, Raoul and {Melo}, Alejandra and {Schuldt}, Stefan and {Shu}, Yiping and {Eilers}, Anna-Christina and {Jaelani}, Anton Timur and {Yue}, Minghao},
        title = "{Streamlined lensed quasar identification in multiband images via ensemble networks}",
      journal = {\aap},
         year = 2023,
        month = oct,
       volume = {678},
          eid = {A103},
        pages = {A103},
          doi = {10.1051/0004-6361/202347332},
archivePrefix = {arXiv},
       eprint = {2307.01090},
 primaryClass = {astro-ph.GA},
       adsurl = {https://ui.adsabs.harvard.edu/abs/2023A&A...678A.103A}
}

@ARTICLE{Collett2023,
       author = {{Collett}, T.~E. and {Sonnenfeld}, A. and {Frohmaier}, C. and {Glazebrook}, K. and {Sluse}, D. and {Motta}, V. and {Verma}, A. and {Anguita}, T. and {Koopmans}, L. and {Tortora}, C. and {Courbin}, F. and {Cabanac}, R. and {Frye}, B. and {Smith}, G.~P. and {Diego}, J.~M. and {Alteiri}, B. and {Lopez}, S. and {Fassnacht}, C. and {Cooray}, A. and {Goobar}, A. and {Ryczanowski}, D. and {Serjeant}, S. and {Richard}, J. and {Treu}, T. and {Moustakas}, L. and {Li}, R. and {Jacobs}, C. and {Lemon}, C. and {Marchetti}, L. and {Hartley}, P. and {Jullo}, E. and {Lee}, C. -H. and {Birrer}, S. and {Fritz}, A. and {Nightingale}, J. and {Napolitano}, N. and {Plazas}, A.~A. and {Kruk}, S. and {Spiniello}, C. and {Grillo}, C. and {Suyu}, S. and {Shajib}, A. and {Vernardos}, G. and {Dye}, S. and {Daylan}, T. and {Newman}, J. and {Schuldt}, S.},
        title = "{The 4MOST Strong Lensing Spectroscopic Legacy Survey (4SLSLS)}",
      journal = {The Messenger},
         year = 2023,
        month = mar,
       volume = {190},
        pages = {49-52},
          doi = {10.18727/0722-6691/5313},
       adsurl = {https://ui.adsabs.harvard.edu/abs/2023Msngr.190...49C}
}

@ARTICLE{Canameras2024,
       author = {{Ca{\~n}ameras}, R. and {Schuldt}, S. and {Shu}, Y. and {Suyu}, S.~H. and {Taubenberger}, S. and {Andika}, I.~T. and {Bag}, S. and {Inoue}, K.~T. and {Jaelani}, A.~T. and {Leal-Taix{\'e}}, L. and {Meinhardt}, T. and {Melo}, A. and {More}, A.},
        title = "{HOLISMOKES: XI. Evaluation of supervised neural networks for strong-lens searches in ground-based imaging surveys}",
      journal = {\aap},
         year = 2024,
        month = dec,
       volume = {692},
          eid = {A72},
        pages = {A72},
          doi = {10.1051/0004-6361/202347072},
archivePrefix = {arXiv},
       eprint = {2306.03136},
 primaryClass = {astro-ph.GA},
       adsurl = {https://ui.adsabs.harvard.edu/abs/2024A&A...692A..72C}
}

@ARTICLE{Roman2015,
       author = {{Spergel}, D. and {Gehrels}, N. and {Baltay}, C. and {Bennett}, D. and {Breckinridge}, J. and {Donahue}, M. and {Dressler}, A. and {Gaudi}, B.~S. and {Greene}, T. and {Guyon}, O. and {Hirata}, C. and {Kalirai}, J. and {Kasdin}, N.~J. and {Macintosh}, B. and {Moos}, W. and {Perlmutter}, S. and {Postman}, M. and {Rauscher}, B. and {Rhodes}, J. and {Wang}, Y. and {Weinberg}, D. and {Benford}, D. and {Hudson}, M. and {Jeong}, W.-S. and {Mellier}, Y. and {Traub}, W. and {Yamada}, T. and {Capak}, P. and {Colbert}, J. and {Masters}, D. and {Penny}, M. and {Savransky}, D. and {Stern}, D. and {Zimmerman}, N. and {Barry}, R. and {Bartusek}, L. and {Carpenter}, K. and {Cheng}, E. and {Content}, D. and {Dekens}, F. and {Demers}, R. and {Grady}, K. and {Jackson}, C. and {Kuan}, G. and {Kruk}, J. and {Melton}, M. and {Nemati}, B. and {Parvin}, B. and {Poberezhskiy}, I. and {Peddie}, C. and {Ruffa}, J. and {Wallace}, J.~K. and {Whipple}, A. and {Wollack}, E. and {Zhao}, F.},
        title = "{Wide-Field InfrarRed Survey Telescope-Astrophysics Focused Telescope Assets WFIRST-AFTA 2015 Report}",
      journal = {arXiv e-prints},
         year = 2015,
        month = mar,
          eid = {arXiv:1503.03757},
        pages = {arXiv:1503.03757},
          doi = {10.48550/arXiv.1503.03757},
archivePrefix = {arXiv},
       eprint = {1503.03757},
 primaryClass = {astro-ph.IM},
       adsurl = {https://ui.adsabs.harvard.edu/abs/2015arXiv150303757S}
}

@ARTICLE{Simeoni2025,
       author = {{Sim{\'e}oni}, Oriane and {Vo}, Huy V. and {Seitzer}, Maximilian and {Baldassarre}, Federico and {Oquab}, Maxime and {Jose}, Cijo and {Khalidov}, Vasil and {Szafraniec}, Marc and {Yi}, Seungeun and {Ramamonjisoa}, Micha{\"e}l and {Massa}, Francisco and {Haziza}, Daniel and {Wehrstedt}, Luca and {Wang}, Jianyuan and {Darcet}, Timoth{\'e}e and {Moutakanni}, Th{\'e}o and {Sentana}, Leonel and {Roberts}, Claire and {Vedaldi}, Andrea and {Tolan}, Jamie and {Brandt}, John and {Couprie}, Camille and {Mairal}, Julien and {J{\'e}gou}, Herv{\'e} and {Labatut}, Patrick and {Bojanowski}, Piotr},
        title = "{DINOv3}",
      journal = {arXiv e-prints},
         year = 2025,
        month = aug,
          eid = {arXiv:2508.10104},
        pages = {arXiv:2508.10104},
          doi = {10.48550/arXiv.2508.10104},
archivePrefix = {arXiv},
       eprint = {2508.10104},
 primaryClass = {cs.CV},
       adsurl = {https://ui.adsabs.harvard.edu/abs/2025arXiv250810104S}
}

@ARTICLE{Walmsley2025,
       author = {{Walmsley}, Mike and {Bamford}, Steven and {Dickinson}, Hugh and {G{\'e}ron}, Tobias and {Gordon}, Alexander J. and {Ferguson}, Annette M.~N. and {Fortson}, Lucy and {Kruk}, Sandor and {Lines}, Natalie and {Lintott}, Chris J. and {Masters}, Karen L. and {Mann}, Robert G. and {Pearson}, James and {Roberts}, Hayley and {Scaife}, Anna M.~M. and {Schuldt}, Stefan and {Simmons}, Brooke and {Smethurst}, Rebecca and {Speagle}, Josh and {Willett}, Kyle},
        title = "{Galaxy Zoo Evo: 1 million human-annotated images of galaxies}",
      journal = {arXiv e-prints},
         year = 2025,
        month = dec,
          eid = {arXiv:2512.23691},
        pages = {arXiv:2512.23691},
          doi = {10.48550/arXiv.2512.23691},
archivePrefix = {arXiv},
       eprint = {2512.23691},
 primaryClass = {astro-ph.IM},
       adsurl = {https://ui.adsabs.harvard.edu/abs/2025arXiv251223691W}
}

@ARTICLE{Rhodes2017,
       author = {{Rhodes}, Jason and {Nichol}, Robert C. and {Aubourg}, {\'E}ric and {Bean}, Rachel and {Boutigny}, Dominique and {Bremer}, Malcolm N. and {Capak}, Peter and {Cardone}, Vincenzo and {Carry}, Beno{\^\i}t and {Conselice}, Christopher J. and {Connolly}, Andrew J. and {Cuillandre}, Jean-Charles and {Hatch}, N.~A. and {Helou}, George and {Hemmati}, Shoubaneh and {Hildebrandt}, Hendrik and {Hlo{\v{z}}ek}, Ren{\'e}e and {Jones}, Lynne and {Kahn}, Steven and {Kiessling}, Alina and {Kitching}, Thomas and {Lupton}, Robert and {Mandelbaum}, Rachel and {Markovic}, Katarina and {Marshall}, Phil and {Massey}, Richard and {Maughan}, Ben J. and {Melchior}, Peter and {Mellier}, Yannick and {Newman}, Jeffrey A. and {Robertson}, Brant and {Sauvage}, Marc and {Schrabback}, Tim and {Smith}, Graham P. and {Strauss}, Michael A. and {Taylor}, Andy and {Von Der Linden}, Anja},
        title = "{Scientific Synergy between LSST and Euclid}",
      journal = {\apjs},
         year = 2017,
        month = dec,
       volume = {233},
       number = {2},
          eid = {21},
        pages = {21},
          doi = {10.3847/1538-4365/aa96b0},
archivePrefix = {arXiv},
       eprint = {1710.08489},
 primaryClass = {astro-ph.IM},
       adsurl = {https://ui.adsabs.harvard.edu/abs/2017ApJS..233...21R}
}

@article{Pedregosa2011,
  title={Scikit-learn: Machine Learning in {P}ython},
  author={Pedregosa, F. and Varoquaux, G. and Gramfort, A. and Michel, V.
          and Thirion, B. and Grisel, O. and Blondel, M. and Prettenhofer, P.
          and Weiss, R. and Dubourg, V. and Vanderplas, J. and Passos, A. and
          Cournapeau, D. and Brucher, M. and Perrot, M. and Duchesnay, E.},
  journal={Journal of Machine Learning Research},
  volume={12},
  pages={2825--2830},
  year={2011}
}

@ARTICLE{Fleury2026,
       author = {{Fleury}, Pierre and {Johnson}, Daniel and {Duboscq}, Th{\'e}o and {Hogg}, Natalie B. and {Larena}, Julien},
        title = "{Cosmology with the line-of-sight shear of strong gravitational lenses}",
      journal = {arXiv e-prints},
         year = 2026,
        month = mar,
          eid = {arXiv:2603.03441},
        pages = {arXiv:2603.03441},
          doi = {10.48550/arXiv.2603.03441},
archivePrefix = {arXiv},
       eprint = {2603.03441},
 primaryClass = {astro-ph.CO},
       adsurl = {https://ui.adsabs.harvard.edu/abs/2026arXiv260303441F}
}

@ARTICLE{Kalton2003,
       author = {{Kalton}, G. and {Flores-Cervantes}, I},
        title = "Weighting Methods",
      journal = {Journal of Official Statistics},
         year = 2003,
       volume = {19},
       number = {2},
        pages = {81-97},
}

@article{Kolenikov2016,
	author = {Kolenikov, Stas},
	journal = {Survey Practice},
	doi = {10.29115/SP-2016-0014},
	number = {3},
	year = {2016},
	month = {jul 31},
	title = {Post-stratification or non-response adjustment?},
	volume = {9},
}

@article{Metcalf2014,
    author = {Metcalf, R. Benton and Petkova, Margarita},
    title = {glamer – I. A code for gravitational lensing simulations with adaptive mesh refinement},
    journal = {MNRAS},
    volume = {445},
    number = {2},
    pages = {1942-1953},
    year = {2014},
    month = {12},
    issn = {0035-8711},
    doi = {10.1093/mnras/stu1859},
    url = {https://doi.org/10.1093/mnras/stu1859},
    eprint = {https://academic.oup.com/mnras/article-pdf/445/2/1942/18198053/stu1859.pdf},
}

@article{González2025,
doi = {10.3847/1538-4357/ae450c},
url = {https://doi.org/10.3847/1538-4357/ae450c},
year = {2026},
month = {apr},
publisher = {The American Astronomical Society},
volume = {1002},
number = {2},
pages = {116},
author = {González, J. and Holloway, P. and Collett, T. and Verma, A. and Bechtol, K. and Marshall, P. and More, A. and Acevedo Barroso, J. and Cartwright, G. and Martinez, M. and Li, T. and Rojas, K. and Schuldt, S. and Birrer, S. and Diehl, H. T. and Morgan, R. and Drlica-Wagner, A. and O’Donnell, J. H. and Zaborowski, E. and Nord, B. and Baeten, E. M. and Johnson, L. C. and Macmillan, C. and Abbott, T. M. C. and Aguena, M. and Allam, S. S. and Brooks, D. and Buckley-Geer, E. and Burke, D. L. and Carnero Rosell, A. and Carretero, J. and Cawthon, R. and Davis, T. M. and De Vicente, J. and Desai, S. and Doel, P. and Everett, S. and Flaugher, B. and Frieman, J. and García-Bellido, J. and Gaztanaga, E. and Giannini, G. and Gruen, D. and Gruendl, R. A. and Gutierrez, G. and Hinton, S. R. and Hollowood, D. L. and Honscheid, K. and James, D. J. and Kuehn, K. and Lahav, O. and Lee, S. and Lima, M. and Marshall, J. L. and Mena-Fernández, J. and Miquel, R. and Myles, J. and Pereira, M. E. S. and Pieres, A. and Plazas Malagón, A. A. and Roodman, A. and Samuroff, S. and Sanchez, E. and Sanchez Cid, D. and Santiago, B. and Sevilla-Noarbe, I. and Smith, M. and Suchyta, E. and Tarle, G. and Tucker, D. L. and Vikram, V. and Walker, A. R. and Weaverdyck, N. and and (DES Collaboration)},
title = {Discovering Strong Gravitational Lenses in the Dark Energy Survey with Interactive Machine Learning and Crowd-sourced Inspection with Space Warps},
journal = {ApJ}
}

@article{DESI2026_DR1,
doi = {10.3847/1538-3881/ae4c43},
url = {https://doi.org/10.3847/1538-3881/ae4c43},
year = {2026},
month = {apr},
publisher = {The American Astronomical Society},
volume = {171},
number = {5},
pages = {285},
author = {{DESI Collaboration} and Abdul Karim, M. and Adame, A. G. and Aguado, D. and Aguilar, J. and Ahlen, S. and Alam, S. and Aldering, G. and Alexander, D. M. and Alfarsy, R. and Allen, L. and Allende Prieto, C. and Alves, O. and Anand, A. and Andrade, U. and Armengaud, E. and Avila, S. and Aviles, A. and Awan, H. and Bailey, S. and Baleato Lizancos, A. and Ballester, O. and Bault, A. and Bautista, J. and Bean, R. and Behera, J. and BenZvi, S. and Beraldo e Silva, L. and Bermejo-Climent, J. R. and Beutler, F. and Bianchi, D. and Blake, C. and Blum, R. and Bolton, A. S. and Bonici, M. and Brieden, S. and Brodzeller, A. and Brooks, D. and Buckley-Geer, E. and Burtin, E. and Byström, A. and Canning, R. and Carnero Rosell, A. and Carr, A. and Carrilho, P. and Casas, L. and Castander, F. J. and Cereskaite, R. and Cervantes-Cota, J. L. and Chaussidon, E. and Chaves-Montero, J. and Chen, S. and Chen, X. and Circosta, C. and Claybaugh, T. and Cole, S. and Cooper, A. P. and Cousinou, M.-C. and Cuceu, A. and Davis, T. M. and Dawson, K. S. and de Belsunce, R. and de la Cruz, R. and de la Macorra, A. and de Mattia, A. and Deiosso, N. and Della Costa, J. and Demina, R. and Demirbozan, U. and DeRose, J. and Dey, A. and Dey, B. and Ding, J. and Ding, Z. and Doel, P. and Douglass, K. and Dowicz, M. and Ebina, H. and Edelstein, J. and Eisenstein, D. J. and Elbers, W. and Emas, N. and Escoffier, S. and Fagrelius, P. and Fan, X. and Fanning, K. and Favole, G. and Fawcett, V. A. and Fernández-García, E. and Ferraro, S. and Findlay, N. and Font-Ribera, A. and Forero-Romero, J. E. and Forero-Sánchez, D. and Frenk, C. S. and Gänsicke, B. T. and Galbany, L. and García-Bellido, J. and Garcia-Quintero, C. and Garrison, L. H. and Gaztañaga, E. and Gil-Marín, H. and Gloudemans, A. and Gnedin, O. Y. and Gontcho A Gontcho, S. and Gonzalez, D. and Gonzalez-Morales, A. X. and Gonzalez-Perez, V. and Gordon, C. and Graur, O. and Green, D. and Gruen, D. and Gsponer, R. and Guandalin, C. and Gutierrez, G. and Guy, J. and Hahn, C. and Han, J. J. and Han, J. and He, S. and Herrera-Alcantar, H. K. and Heydenreich, S. and Honscheid, K. and Hou, J. and Howlett, C. and Huterer, D. and Iršič, V. and Ishak, M. and Jacques, A. and Jiang, L. and Jimenez, J. and Jing, Y. P. and Joachimi, B. and Joudaki, S. and Joyce, R. and Jullo, E. and Juneau, S. and Karaçaylı, N. G. and Karim, T. and Kehoe, R. and Kent, S. and Khederlarian, A. and Kirkby, D. and Kisner, T. and Kitaura, F.-S. and Kizhuprakkat, N. and Kong, H. and Koposov, S. E. and Kremin, A. and Krolewski, A. and Lahav, O. and Lai, Y. and Lamman, C. and Lan, T.-W. and Landriau, M. and Lang, D. and Lange, J. U. and Lasker, J. and Le Goff, J.M. and Le Guillou, L. and Leauthaud, A. and Levi, M. E. and Li, S. and Li, T. S. and Liu, W. and Lodha, K. and Lokken, M. and Luo, Y. and Luo, Y. and Magneville, C. and Manera, M. and Manser, C. J. and Margala, D. and Martini, P. and Maus, M. and McCullough, J. and McDonald, P. and Medina, G. E. and Medina-Varela, L. and Meisner, A. and Mena-Fernández, J. and Menegas, A. and Meneses-Rizo, J. and Mezcua, M. and Miquel, R. and Montero-Camacho, P. and Moon, J. and Moustakas, J. and Muñoz-Gutiérrez, A. and Mu noz-Santos, D. and Myers, A. D. and Myles, J. and Nadathur, S. and Najita, J. and Napolitano, L. and Newman, J. A. and Nikakhtar, F. and Nikutta, R. and Niz, G. and Noriega, H. E. and Nugent, P. and Padmanabhan, N. and Paillas, E. and Palanque-Delabrouille, N. and Palmese, A. and Pan, J. and Pan, Z. and Parkinson, D. and Peacock, J. A. and Ibanez, M. P. and Percival, W. J. and Pérez-Fernández, A. and Pérez-Ràfols, I. and Peterson, P. and Piat, J. and Pieri, M. M. and Pinon, M. and Poppett, C. and Porredon, A. and Prada, F. and Pucha, R. and Qin, F. and Rabinowitz, D. and Raichoor, A. and Ramírez-Pérez, C. and Ramirez-Solano, S. and Rashkovetskyi, M. and Ravoux, C. and Ried Guachalla, B. and Riley, A. H. and Rocher, A. and Rockosi, C. and Rohlf, J. and Rosado-Marín, A. J. and Ross, A. J. and Ross, C. and Rossi, G. and Ruggeri, R. and Ruhlmann-Kleider, V. and Sabiu, C. G. and Said, K. and Sailer, N. and Saintonge, A. and Salcedo Hernandez, Y. and Samushia, L. and Sanchez, E. and Sanders, N. and Sandford, N. and Satyavolu, S. and Saulder, C. and Saydjari, A. K. and Schlafly, E. F. and Schlegel, D. and Scholte, D. and Schubnell, M. and Semenaite, A. and Seo, H. and Shafieloo, A. and Sharples, R. and Silber, J. and Sinigaglia, F. and Siudek, M. and Slepian, Z. and Smith, A. and Soumagnac, M. and Sprayberry, D. and Suárez-Pérez, J. and Swanson, J. and Tan, T. and Tarlé, G. and Taylor, P. and Thomas, G. and Tojeiro, R. and Turner, R. J. and Turner, W. and Ureña-López, L. A. and Vaisakh, R. and Valluri, M. and Valogiannis, G. and Vargas-Magaña, M. and Verde, L. and Vielzeuf, P. and Walther, M. and Wang, B. and Wang, M. S. and Wang, W. and Weaver, B. A. and Weaverdyck, N. and Wechsler, R. H. and Weinberg, D. H. and White, M. and Whitford, A. and Wolfson, M. and Yang, J. and Yèche, C. and Youles, S. and Yu, J. and Yuan, S. and Zaborowski, E. A. and Zarrouk, P. and Zhang, H. and Zhao, C. and Zhao, R. and Zheng, Z. and Zhou, C. and Zhou, R. and Zhou, Y. and Zou, H. and Zou, S. and Zu, Y.},
title = {Data Release 1 of the Dark Energy Spectroscopic Instrument},
journal = {AJ}
}

@article{Rojas2023,
    author = {Rojas, Karina and Collett, Thomas E and Ballard, Daniel and Magee, Mark R and Birrer, Simon and Buckley-Geer, Elizabeth and Chan, James H H and Clément, Benjamin and Diego, José M and Gentile, Fabrizio and González, Jimena and Joseph, Rémy and Mastache, Jorge and Schuldt, Stefan and Tortora, Crescenzo and Verdugo, Tomás and Verma, Aprajita and Daylan, Tansu and Millon, Martin and Jackson, Neal and Dye, Simon and Melo, Alejandra and Mahler, Guillaume and Ogando, Ricardo L C and Courbin, Frédéric and Fritz, Alexander and Herle, Aniruddh and Acevedo Barroso, Javier A and Cañameras, Raoul and Cornen, Claude and Dhanasingham, Birendra and Glazebrook, Karl and Martinez, Michael N and Ryczanowski, Dan and Savary, Elodie and Góis-Silva, Filipe and Arturo Ureña-López, L and Wiesner, Matthew P and Wilde, Joshua and Valim Calçada, Gabriel and Cabanac, Rémi and Pan, Yue and Sierra, Isaac and Despali, Giulia and Cavalcante-Gomes, Micaele V and Macmillan, Christine and Maresca, Jacob and Grudskaia, Aleksandra and O’Donnell, Jackson H and Paic, Eric and Niemiec, Anna and de la Bella, Lucia F and Bromley, Jane and Williams, Devon M and More, Anupreeta and Levine., Benjamin C},
    title = {The impact of human expert visual inspection on the discovery of strong gravitational lenses},
    journal = {MNRAS},
    volume = {523},
    number = {3},
    pages = {4413-4430},
    year = {2023},
    month = {08},
    issn = {0035-8711},
    doi = {10.1093/mnras/stad1680},
    url = {https://doi.org/10.1093/mnras/stad1680},
    eprint = {https://academic.oup.com/mnras/article-pdf/523/3/4413/50660202/stad1680.pdf},
}

@ARTICLE{Refsdal1964,
       author = {{Refsdal}, S.},
        title = "{On the possibility of determining Hubble's parameter and the masses of galaxies from the gravitational lens effect}",
      journal = {\mnras},
         year = 1964,
        month = jan,
       volume = {128},
        pages = {307},
          doi = {10.1093/mnras/128.4.307},
       adsurl = {https://ui.adsabs.harvard.edu/abs/1964MNRAS.128..307R}
}
%
%

%

\begin{appendix}

\section{Effect of calibration on expert grades} \label{Ap:Effect_of_Calibration_on_Grades}
\begin{figure}
    \centering
    \includegraphics[width=\columnwidth]{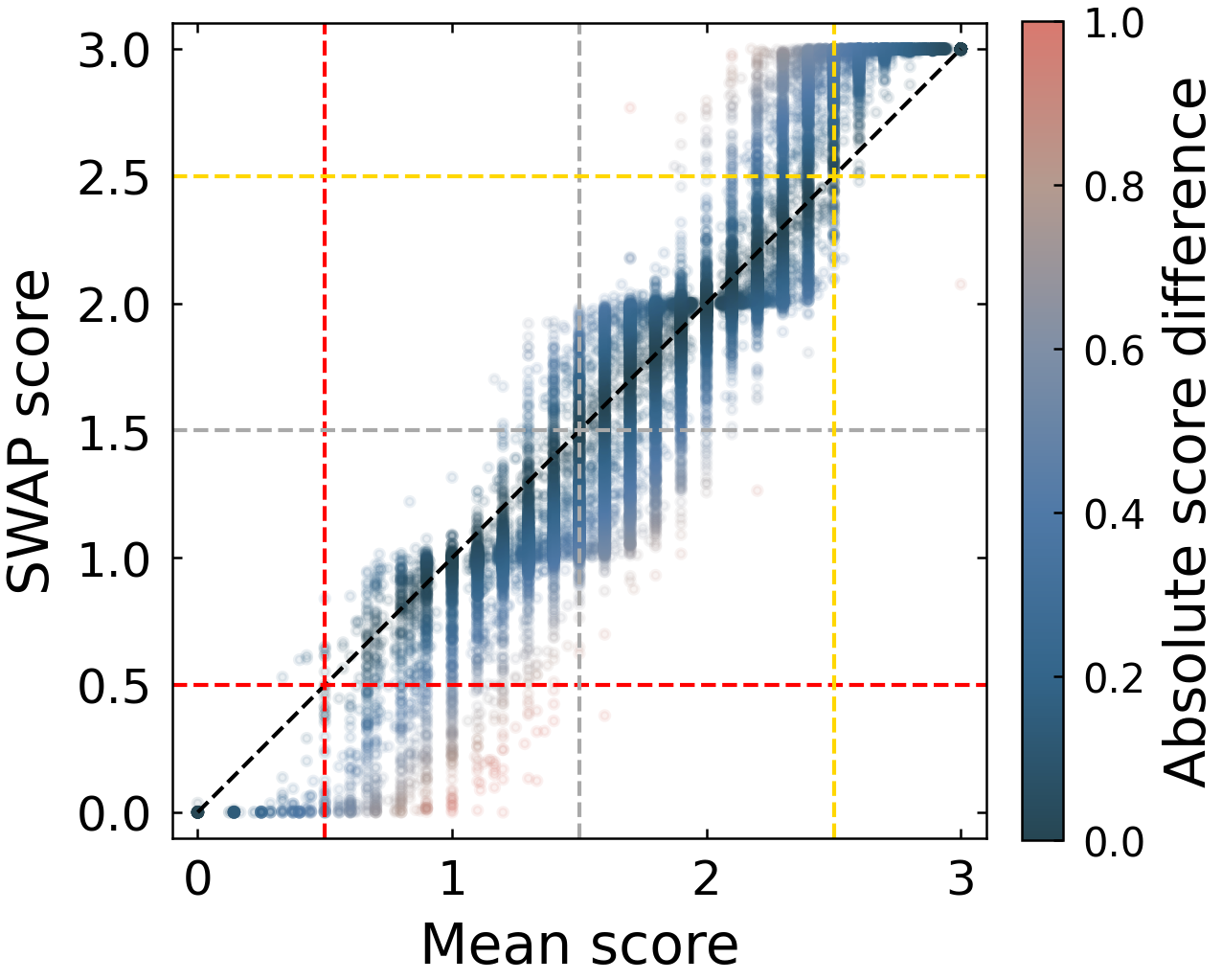}
    \caption{Comparison of mean score compared to calibrated score for systems in the comprehensive DR1 search. We plot the grade A/B, B/C, and C/X thresholds in each case (gold, grey, and red lines respectively).}
    \label{Mean_vs_calibrated_grade}
\end{figure}
Figure \ref{Mean_vs_calibrated_grade} shows a comparison of the mean and calibrated scores for systems in the comprehensive DR1 search. Systems that gain many grade A  classifications have a calibrated score very close to 3. This is because the corresponding grade posterior $P(G=\rm{A}|\{C_{\rm{GJ}}\})$ tends towards 1, and thus the calibrated grade tends towards 3. An equivalent effect is seen for grade X systems. Given the variation in responses across expert graders however, calibration is necessary to account for their varying degrees of optimism (Fig. \ref{Grader_optimism_pessimism}).

\section{Simulated lenses used for training}\label{Ap: Simulation_Comparison}
\begin{figure*}
    \centering
    \includegraphics[width=\textwidth]{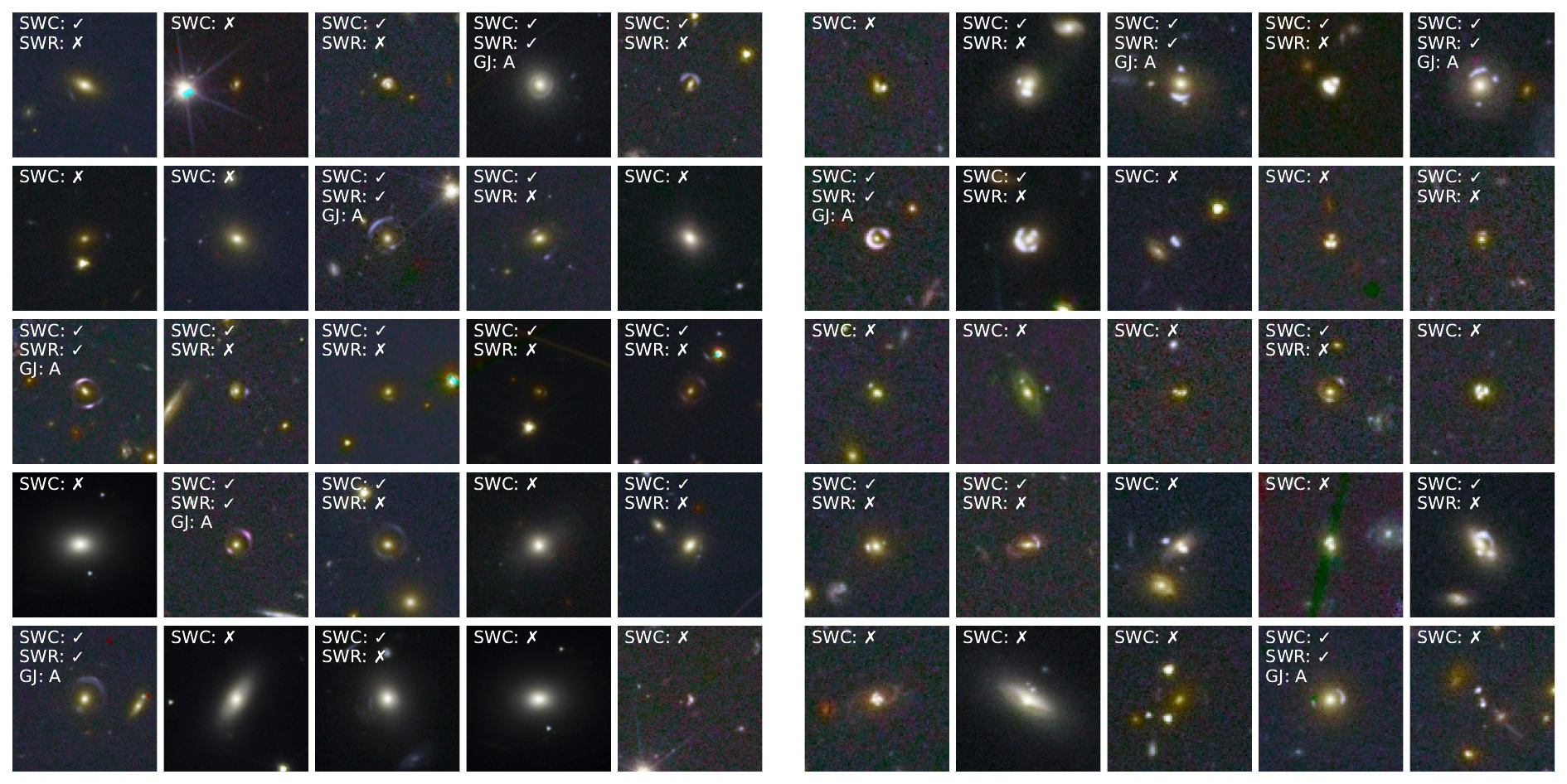}
    \caption{Selection of simulated lenses used as part of the training for both the ML models and SW volunteers. The S1 simulations were taken from \citet{Q1-SP052}, and the S2 simulations were generated using the \texttt{GLAMER} \citep{Metcalf2014} simulation code with simulated lens and source galaxies. We include flags for whether the simulations passed each inspection stage.}
    \label{Simulation_Comparison}
\end{figure*}
\begin{figure}
    \centering
    \includegraphics[width=\columnwidth]{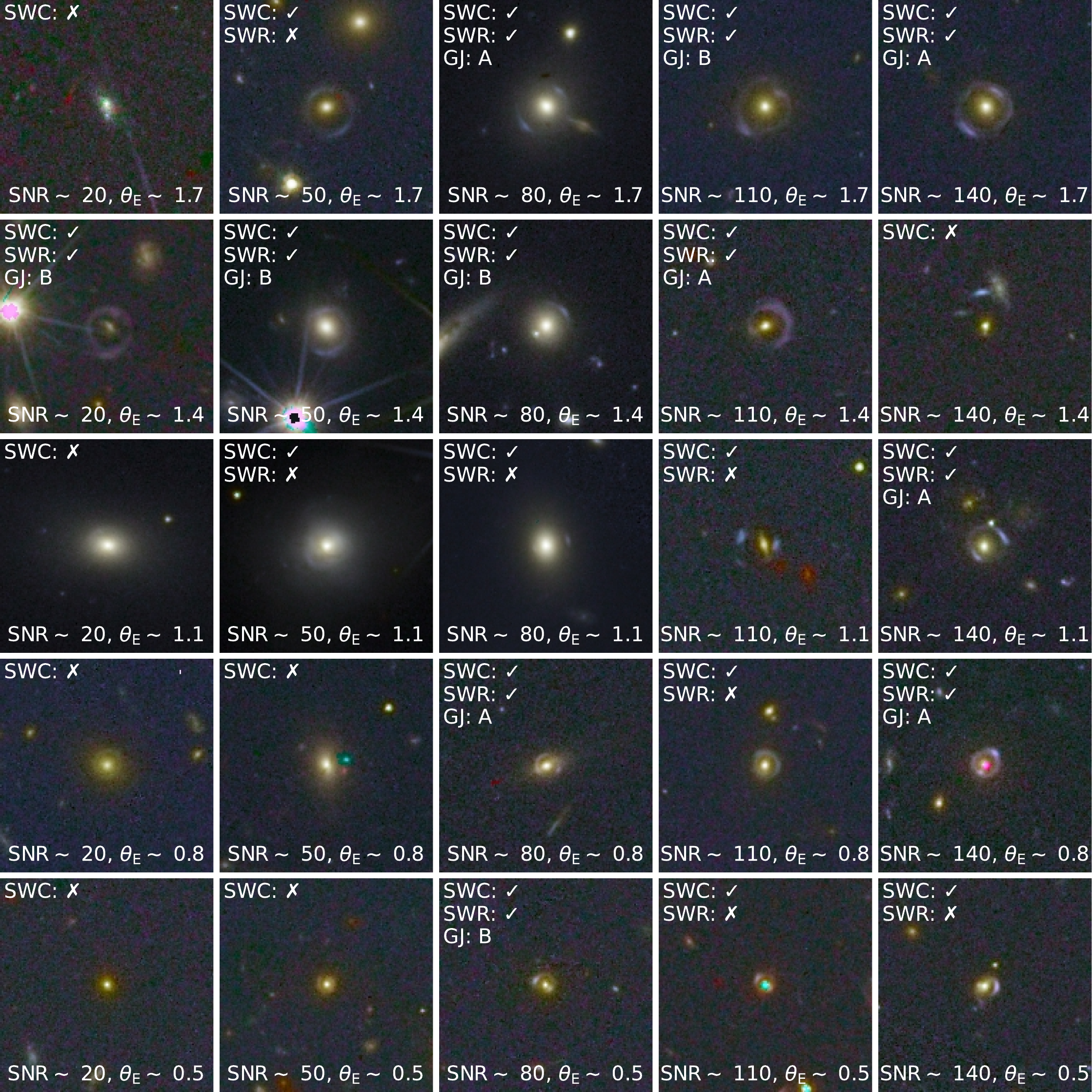}
    \caption{Selection of lens simulations plotted as a function of S/N ($x$-axis) and Einstein radius ($y$-axis), taken from the S1 data set \citealp{Q1-SP052}. We include flags for whether the simulations passed each inspection stage.}
    \label{Sim_Collage}
\end{figure}

Simulated lenses were used as part of the training data set for both the ML models and SW volunteers. Two sets of simulated lenses were used, S1 and S2. The first of these were taken from \citet{Q1-SP052}, which used real galaxies for both the lens and source. The second of these were generated using the \texttt{GLAMER} code \citep{Metcalf2014} with deflectors based on the Flagship simulation \citep{EuclidSkyFlagship} and simulated sources based on observed galaxies the HUDF. Both sets of simulations spanned a wide range of S/N and lens configurations. A sample of both of these is shown in Fig.\,\ref{Simulation_Comparison}, while a selection
of S1 systems as a function of Einstein radius and S/N is shown in Fig.\,\ref{Sim_Collage}. 

\section{Instructions provided to Space Warps volunteers}\label{Ap: SW Training}
The SW volunteers were provided with a number of tools to enable them to classify lensed objects. Newcomers to the website were shown tutorial pages showing a range of \Euclid lens candidates, with links to further subsets of lenses and non-lenses (e.g., lensed quasars, galaxy- and group-scale lenses, artefacts, and stars) which formed part of the field guide. The volunteers were asked to classify lens candidates regardless of where the candidate appeared in the image and, in \SWCt{}, were encouraged to be inclusive, i.e. to mark ambiguous cases as lenses. During \SWRt{}, the volunteers were asked to grade systems according to their likelihood that they were lenses on an A/B/C/X scale. The \SWRt{} tutorial contained examples of systems at each grade drawn from the expert-classified systems from the Q1 data release. In both \SWCt{} and \SWRt{} the volunteers received feedback for training images interspersed within the classification stream. In the case of \SWCt{} this included simulated lenses (Appendix \ref{Simulation_Comparison}) and known non-lenses (e.g. LRGs, spirals, or mergers, \citealp{Q1-SP052}). For \SWRt{} the volunteers received feedback for systems spanning the complete range of lens grades and customised to each training system, such as `This image shows an arc warped round the central galaxy. However, a counter-image (an image of the same colour as the arc, on the opposite side to the central galaxy) is not visible.  The researchers were mostly in agreement that this was a probable (grade B) lens. Researcher grades: A: 0\%, B: 90\%, C: 10\%, X: 0\%'. In this way, the volunteers could learn which particular features of an image were relevant when judging a systems likelihood to be a lens. The volunteers were not asked to only identify lensed arcs of a particular colour but were informed that blue arcs would appear more commonly than red arcs.

\section{Systems flagged and missed at each classification stage}\label{Ap: Systems_flagged_vs_missed}
\begin{figure*}
    \centering
        \includegraphics[width=\textwidth]{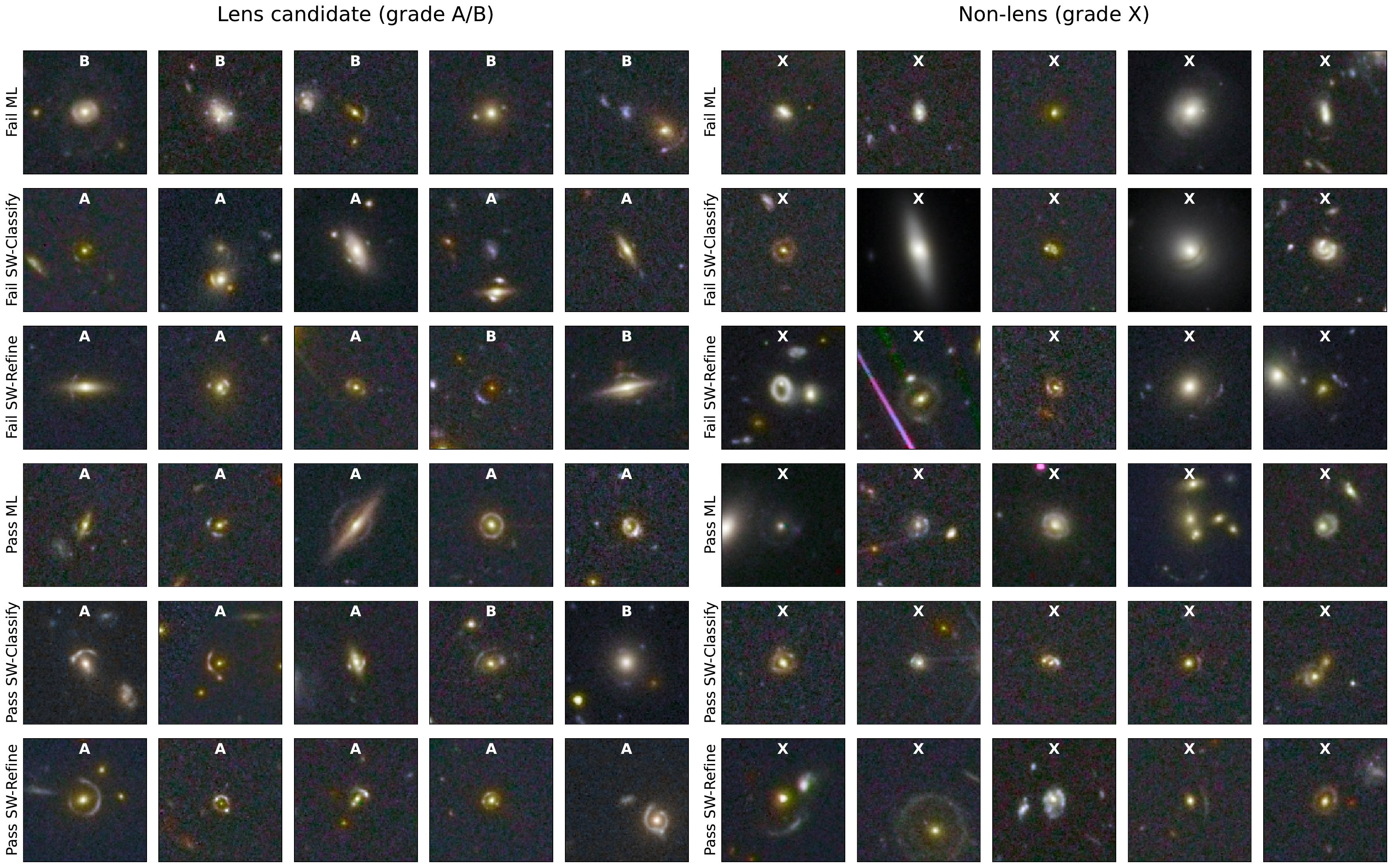}
    \caption{A selection of systems which pass/fail the score thresholds for each of the three classification methods. Objects graded as lens candidates by GJ are shown in the left-hand column (with grade A systems shown preferentially where available), and those labelled as non-lenses are shown in the right-hand column. The lens candidates are not necessarily centred within the cutout since recentering was performed after GJ - the cutouts shown here correspond directly to the locations of the cutouts classified by SW and the ML.}
    \label{Pipeline_Selection_Examples}
\end{figure*}

A selection of systems (both lens candidates and non-lenses) which were flagged for inspection or rejected by the ML and SW are shown in Fig. \ref{Pipeline_Selection_Examples}. Each classification stage (ML, \SWCt{}, and \SWRt{}) identifies many lens candidates, but also includes some false positives.  

\section{Derivation of the search completeness}\label{Ap: Search Completeness}
\newcommand*{\MLfoundlens}{\ensuremath{\textit{L}_{\rm{{ML \,found}}}}}
\newcommand*{\SWCfoundlens}{\ensuremath{\textit{L}_{\rm{{SWC \,found}}}}}
\newcommand*{\SWRfoundlens}{\ensuremath{\textit{L}_{\rm{{SWR \,found}}}}}
\newcommand*{\MLlostlens}{\ensuremath{\textit{L}_{\rm{{ML \,lost}}}}}
\newcommand*{\SWClostlens}{\ensuremath{\textit{L}_{\rm{{SWC \,lost}}}}}
\newcommand*{\SWRlostlens}{\ensuremath{\textit{L}_{\rm{{SWR \,lost}}}}}
\newcommand*{\eSW}{\ensuremath{\epsilon}}

\newcommand*{\MLrand}{\ensuremath{\textit{N}^{\rm{ML\,rand}}}}
\newcommand*{\SWCrand}{\ensuremath{\textit{N}^{\rm{SWC\,rand}}}}
\newcommand*{\SWRrand}{\ensuremath{{\textit{N}^{\rm{SWR\,rand}}}}}

\newcommand*{\MLrandlost}{\ensuremath{\textit{N}{\rm{^{ML\,rand}_{lost}}}}}
\newcommand*{\SWCrandlost}{\ensuremath{\textit{N}{\rm{^{SWC\,rand}_{lost}}}}}
\newcommand*{\SWRrandlost}{\ensuremath{{\textit{N}{\rm{^{SWR\,rand}_{lost}}}}}}

\newcommand*{\MLlost}{\ensuremath{\textit{N}_{\rm{ML\,lost}}}}
\newcommand*{\SWClost}{\ensuremath{\textit{N}_{\rm{SWC\,lost}}}}
\newcommand*{\SWRlost}{\ensuremath{{\textit{N}_{\rm{SWR\,lost}}}}}

\newcommand*{\MLrandlens}{\ensuremath{\textit{L}^{\rm{ML\,rand}}}}
\newcommand*{\SWCrandlens}{\ensuremath{\textit{L}^{\rm{SWC\,rand}}}}
\newcommand*{\SWRrandlens}{\ensuremath{{\textit{L}^{\rm{SWR\,rand}}}}}

\newcommand*{\MLrandlenslost}{\ensuremath{\textit{L}\rm{^{ML\,rand}_{lost}}}}
\newcommand*{\SWCrandlenslost}{\ensuremath{\textit{L}\rm{^{SWC\,rand}_{lost}}}}
\newcommand*{\SWRrandlenslost}{\ensuremath{\textit{L}\rm{^{SWR\,rand}_{lost}}}}

Here we estimate the completeness of the comprehensive DR1 search. We define the completeness as the fraction of lenses identified by the search compared to the total number of lenses which would have been identified had GJ inspected all images in the DR1 area (excluding the $200\deg^2$ and $50\deg^2$ areas). 
In other words, we assume that all lens candidates presented to GJ were successfully identified by the experts. In doing so, we attribute any incompleteness to systems that failed to exceed the score thresholds imposed on the ML models and/or SW.
We highlight that given we applied catalogue selections (requiring an \IE detection, etc.), this value does not represent the completeness to all detectable lenses in \Euclid, rather it is the completeness for the selection used here. 

Lenses could be identified via three pathways -- ML, \SWCt{}, and \SWRt{}. 
The number of lenses identified and lost by the search can be written 
\begin{equation}
L_{\rm{found}}=\MLfoundlens+\SWCfoundlens+\SWRfoundlens
\end{equation}
and 
\begin{equation}
L_{\rm{lost}}=\MLlostlens+\SWClostlens+\SWRlostlens\,,
\end{equation}
respectively. Therefore, the completeness $f$ can be written as
\begin{equation}
    f = \frac{L_{\rm{found}}}{L_{\rm{found}}+L_{\rm{lost}}}\,,
\end{equation}
where `found' denotes systems identified by a given search stage which had not been sent to GJ by a previous stage. Similarly, `lost' refers to systems rejected by a given search stage and excluding systems already rejected at previous stages to prevent double-counting. We define a separate completeness $\eSW$ as the completeness of the SW inspection, excluding the (likely obvious lens) systems which were sent to GJ directly based on their ML score
\begin{equation}
    \eSW = \frac{\SWCfoundlens+\SWRfoundlens}{\SWCfoundlens+\SWRfoundlens+\SWClostlens+\SWRlostlens}\,.
\end{equation}

To estimate the completeness of the search, we use three different sets of `random' objects. The first is a set of \num{50000} objects drawn randomly from the objects passing the catalogue selection, which were shown to \SWCt{} regardless of their ML scores. This can be used to estimate the number of systems which the ML classifiers miss. The second is a sample of around \num{1000} objects selected randomly from systems seen by \SWCt{}, and the third is a sample of around \num{2000} objects selected randomly from systems seen by \SWRt{}. These latter two data sets were shown in totality to GJ.

We determine the number of lens systems rejected at the ML stage, \MLlostlens{}, using the \num{50000} random sample. We model \MLlostlens{} by \rt{drawing samples from a} $\beta$ distribution in which there are 
\MLrandlenslost{} 
lenses amongst 
\MLrandlost{} 
systems in the random sample which were rejected by the ML. Accounting for a Jeffrey's prior (used throughout\footnote{The $1/2$ terms in the $\beta$ distributions are the result of the Jeffrey's prior.}) we have
\begin{equation}
\MLlostlens{}\sim\beta\left(\MLrandlenslost+\frac{1}{2},\MLrandlost-\MLrandlenslost+\frac{1}{2}\right)\,\frac{\MLlost{}}{\eSW}\,,
\end{equation}
where we define \MLlost{} as all systems in the area of the DR1 comprehensive search, which were rejected by the ML. Since this random sample was inspected by SW rather than directly by GJ, we must divide by the completeness of SW. We note this assumes that the SW volunteers classify objects which pass the ML$\rightarrow$\SWCt{} threshold in the same manner as those which fail the ML$\rightarrow$\SWCt{} cut.

We determine the number of systems rejected at the \SWCt{} stage, \SWClostlens{}, using the \SWCt{} random sample. \SWClostlens{} can also be described by \rt{samples from} a $\beta$ distribution. Strictly, the posteriors accounted for non-uniform sampling of the calibration randoms since higher scores were more densely sampled. In practice, the posteriors were generated individually for each sampling band, and then recombined using their known population sizes. For simplicity we quote the uniform sampling formalism here.
\begin{equation}
\SWClostlens\sim\beta\left(\SWCrandlenslost{}+\frac{1}{2},\SWCrandlost{}-\SWCrandlenslost{}+\frac{1}{2}\right)\,\SWClost{}\,,
\end{equation}
where we have defined \SWClost{} as the number of systems rejected by \SWCt{} which had previously failed the ML$\rightarrow$GJ cut and \SWCrandlost{} the number within the \SWCt{} random sample.

Finally, we determine the number of systems rejected at the \SWRt{} stage, 
\begin{equation}
\SWRlostlens{}\sim\beta\left(\SWRrandlenslost{}+\frac{1}{2},\SWRrandlost{}-\SWRrandlenslost{}+\frac{1}{2}\right)\,\SWRlost{}\,,
\end{equation}
using the \SWRt{} random sample, where \SWRlost{} denotes the number of systems rejected by \SWRt{} which had previously failed both the ML$\rightarrow$GJ and \SWCt{}$\rightarrow$GJ cuts and \SWRrandlost{} the equivalent number in the random sample.

The remaining terms, \MLfoundlens{}, \SWCfoundlens{}, and \SWRfoundlens{} could be taken directly from the catalogue. We provide a table of input and derived parameters in Table \ref{tab:completeness}. Since the search was conducted on cutouts, for which some overlapped, we apply a deduplication factor to determine the completeness to grade A (A+B) systems rather than to grade A (A+B) cutouts. This used a $7\arcsec$ maximum deduplication radius, and scaled the number of cutouts inversely to the average number of systems per cutout. In practice, this only made a small difference relative to the uncertainty in the completeness estimate.

Overall, we find a completeness of $0.73_{-0.15}^{+0.08}$ to grade A systems and $0.34_{-0.06}^{+0.05}$ to grade A+B systems in the comprehensive search area. 
Scaling this to the complete DR1 area ($2000\deg^2$, i.e. including the $50\deg^2$ and $200\deg^2$ search areas) would suggest a total of
7500 
grade A systems across the DR1 area. We report the discovery of \rt{6073} grade A systems across in ECLIPSE-A indicating a completeness of $0.81_{-0.17}^{+0.09}$ across the DR1 area, as reported in ECLIPSE-A.

\begin{table}
\centering
\caption{Input and inferred parameters used to measure the completeness of the DR1 comprehensive search.}
\begin{tabular}{ccc}
\hline
\hline & \\[-1.5ex]
Parameter & Grade A & Grade A+B\\
\midrule 
\MLfoundlens{} & 4000 & 6300\\[4pt]
\SWCfoundlens{}& 470 & 1600\\[4pt]
\SWRfoundlens{}& 510 & 2700\\[4pt]
\MLlostlens{}& $490^{+5300}_{-490}$& $5300^{+\num{16400}}_{-4900}$\\[4pt]
\SWClostlens{}& $270_{-240}^{+1500}$& $7100_{-3200}^{+4600}$\\[4pt]
\SWRlostlens{}& $860_{-230}^{+330}$& $7000_{-650}^{+780}$\\[4pt]
\eSW{} & $0.5_{-0.19}^{+0.11}$&$0.24_{-0.05}^{+0.05}$\\[4pt]
\MLrandlenslost{}&0&1\\[4pt]
\SWCrandlenslost{}$^{(+)}$&1&8\\[4pt]
\SWCrandlenslost{}$^{(-)}$&0&10\\[4pt]
\SWRrandlenslost{}$^{(+)}$&47&369\\[4pt]
\SWRrandlenslost{}$^{(-)}$&0&6\\[4pt]
\MLrandlost{}&\multicolumn{2}{c}{\num{49876}}\\[4pt]
\SWCrandlost{}$^{(+)}$&\multicolumn{2}{c}{52}\\[4pt]
\SWCrandlost{}$^{(-)}$&\multicolumn{2}{c}{140}\\[4pt]
\SWRrandlost{}$^{(+)}$&\multicolumn{2}{c}{781}\\[4pt]
\SWRrandlost{}$^{(-)}$&\multicolumn{2}{c}{61}\\[4pt]
\MLlost{}&\multicolumn{2}{c}{$6.0\times10^7$}\\[4pt]
\SWClost{}$^{(+)}$&\multicolumn{2}{c}{4346}\\[4pt]
\SWClost{}$^{(-)}$&\multicolumn{2}{c}{\num{100687}}\\[4pt]
\SWRlost{}$^{(+)}$&\multicolumn{2}{c}{\num{14315}}\\[4pt]
\SWRlost{}$^{(-)}$&\multicolumn{2}{c}{7464}\\[4pt]
\hline
\end{tabular}
\tablefoot{We highlight that the SWC, SWR, and $\eSW$ terms exclude systems either rejected by the previous ML or ML+SWC stages, or sent directly to GJ, i.e. they only include systems which remained indeterminate. These excluded systems would most likely be the most obvious lenses or non-lenses, which would drive down the reported recovery by SW. The uncertainties indicate $\pm2\sigma$ errors. We use $L$ to denote lenses, and $N$ to denote systems (lens or not). We denote $(+)$ and $(-)$ to denote sampling bands within the \SWCt{} and \SWRt{} randoms. We point readers to Table \ref{T: Comprehensive search stats} for the number of systems inspected at each stage. }
\label{tab:completeness}
\end{table}

\section{Space Warps volunteers} \label{Ap: SW Volunteers}
This project would not be possible without the immense contributions of citizen-scientists. Volunteers who classified more than 100 Euclid images on Space Warps whose names were available for publication are listed below. Zooniverse IDs for all those who classified more than one image are listed on \href{https://www.zooniverse.org/projects/aprajita/space-warps-esa-euclid/about/team}{`The Team'} webpage. We greatly appreciate their commitment, enthusiasm, and time.\\
\begin{spacing}{0.7}
\begin{CJK}{UTF8}{gbsn}
\tinier{\noindent A. Kirkpatrick, Aaira Sharjeel, Aaron Michael Cano, Abbey Jennings, Abby Clevenger, Abegail Chanyalew, Abiel Wong, Abigail Ginsberg, Abigail Zheng, Ada Rosenfield, Adam Setchfield, Adam Vitti, Addison Butz, Adeerah Nair, Adrian Astraeus Powers, Adrian Kai En Tee, Adriana Campos Dominik, Adrit Dongol, Agastya Srivastava, Agustina Villa Vidal, Aidan Conrad Gibbs, Aidan Duncan, Aiden Hummelgaard, Aidyn Cuenca, Akeela Valdes, Al Lamperti, Alaa Mahjoubi, Alan M. Gross, Alan Needham, Alan Pellé, Alan W Upchurch, Alberto Hernadez, Alejandro Morales, Aleks Stoyanov, Aleksandr Timofeev, Aleksandra Bijata, Aleksri Bykov, Alessandro Bejinaru, Alex Alves, Alex Anchain, Alex Arktos, Alex Ceballo, Alex Faber, Alex LaCoss, Alex Znyk, Alexander Benra, Alexander Folkers, Alexander Gagliano, Alexander Ivitsky, Alexander Ramer, Alexander Wehl Nielsen, Alexandra Attikova, Alexandre Hainaut, Alexandre Portefaix, Alexis Goupil, Alexis Louis C. Lebrun, Alexis Reed, Alexsey Gorbkov, Ali Alhawas, Ali Hasan Naser Hasan, Alicia Teo, Alicia Weissman, Alina Schulz, Alina Stefanescu, Alisa Karaneuskaya, Alison Watson, Alissa Strauf, Alkistis Kallinikou, Allie Bender, Allison Espinosa, Alma Quijano, Alondra Lara Cruz, Alyona Kobzeva, Alyssa Adam, Alyssa Ferrazzara, Amanda Demster, Amanda Peters, Amanda Peters, Amelia Booth, Amelia Evans, Amelina Anna, Amera Mostafa, Amit Kumar Tamrakar, Amoli Kakkar, Amy Nerkeliunas, Ana Fernandez Alvarez, Anaaya Chowdhary, Anakin Belarmino, Ananya Yogesh, Andelana O'Hara, Andrea Ceriana, Andrea Schimpke, Andreas Gruber, Andreas Lauth, Andreas Richnow, Andreas Thöny, Andres Eloy Martinez Rojas, Andrew Croager, Andrew Lackoff, Andrew M. Hodges, Andrew St.Clair, Andrew Stanley, Andrey CRB Gorokhov, András Kovács, André Holland, Andrés Abel Cosio Tello, Andy Middleton, Andy Spafford, Angela Jursevics, Angela Karunakaran, Angelin Vincent, Angelo Ricciardo, Angus Geidesz, Aniruddh Herle, Anja Verhöfen, Anmol Gandhi, Anna Fei, Anna Holub, Anna Ivanova, Anna Jungell, Anna Kozak, Anna Nikodym-Bilska, Anna Todorova, Annie Su, Anthony Charest, Anthony Michel, Anthony Yoder, Antonio David Garcia Gil, Antonio Pasqua, Antonín Unucka, Anupreeta More, Anvi Digra, Arabella Brosnan, Arief Al Hadid, Arkhipova Daria, Armando Campos, Armin L. Steyn, Arnaud Geollot, Arnaud Trybou, Arnav Singh, Arnav Srivastava, Arndt Stelter, Arne Peterson, Artem Barto, Artemiy Seleznev, Arthur Rodrigues Oliveira, Arto Inkala, Arto Koukkula, Arturo M. Gtz., Aruzhan Sungatkyzy, Aryan Kate, Aryan Pandey, Aryash Singh, Ashish Rai, Ashley McCann, Ashley Pullen, Ashton Taylor, Ashwini Axl Malayan, Asiah Ath-Thohiroh, Astro Yona VII, Athanasia Vlachou, Athene Bunch, Aubrey Tyson, Audrey Schollier, Augusto Gabanzo, Aurélien Genin, Autumn Faulk, Autumn Pearce, Ava Biasotti, Ava Johnston, Ava K Wozniak, Ava Vanhemert, Avanteesh Motheram, Avisha Gupta, Aviva Christie, Aya Mellaoui, Ayana Sinton, Ayesha Kashif Khan, Aymeric Galan, Baba Karthik Kalapatapu, Bailey Kinner, Balaji Prasath S, Barbara Pavlov, Barbara Téglás, Baris Mouland, Baron Michael Mcdonald, BeBe Wren Kiel, Belbahi Oualid, Ben Adamowski, Ben Apgar, Benjamin Cassan-de Gorostarzu, Benjamin Gill, Benjamin Graetz, Benjamin Keplinger, Benjamin Langley, Benjamin See, Bernd Nikolaus, Bill Richardson, Birte Rosiñol Prieto, Blaise Pyź, Bob Fletcher, Brad Brown, Brad M. Collins, Braden Sharette, Bradford Riley, Bradley Kavanagh, Brett Sorby, Brian Altmeyer, Brian Eckfeldt, Brian Myers, Brian Olivero, Brice Young, Brigitte Wollenweber-Ratzer, Brodie Dines, Brody Hoffman, Brody Ramirez, Bruce Binnie, Bruno Chiaranti, Bryan Ranharter, Bryan Shook, Bryan Tom, Bryce Greco, Byunghee Kim, C Unsworth, C. Graham, C. S. Tolliver, Caden Freimark, Cadence Edkins, Cadence Mestel, Caeley N Carmack, Callum Baird, Callum Blake, Camden Micci, Camelia Mallek, Canteloup Samy, Caoimhe Murphy, Carl Larsson, Carl Noack, Carla Brown, Carlos Gainzarain Armentia, Carlos Gonzalez, Carlos Romero, Carlos Santiago, Carlotta Mevert, Carly Gillespie, Carly Siedman, Carol Sabo, Casey Lambert, Casper Lehmann Aalund Vestergaard, Cassandra J Kurtz, Catherine Sui, Cathleen Robinson, Cathryn Collins, Chaithanya K Iruvaram, Charity Mystique Gregory, Charles Kime, Charlie Granger, Charlie Parks, Charlie Rayfield, Charlotte Smith, Charlotte Williams, Charlotte Williams, Cheng Yılın, Cherie Spielman, Cheyenne Best, Chhavy Loch, Chiara Cisternino, Chimalume Eze, Chioma Anyasodo, Chiriac Eduard Cristian, Chloe Hyndes, Chloe Reyes, Chonglin Ran, Chris Calleja, Chris Clarkson, Chris Ensign, Chris Harshaw, Chris McConnell, Chris Molloy, Chrissy Holwerda, Christian Malbroux, Christian Stork, Christine Rödel, Christine Steller, Christine V. Macmillan, Christophe Sims, Christopher B. Edwards, Christopher Doran, Christopher Pemberton, Christopher Tanner, Christopher Vittore, Ciro Sirio Perrella, Claire Bonnett, Claire Duguin, Claire Watts, Clara Stanton, Clare J Flood, Claude Cornen, Claude Revilliod, Claudio Pedrazzi, Clément Imbert--Youssef, Colby Jones, Colette Boever, Colin Bilsland, Colin Chandler, Collien Stone, Colter Kashian, Connor Bruce Campbell, Connor Choi, Connor J Gibson, Connor Wallace, Cooper Bell, Corina Trifu, Cory Nathan Fessler, Cory O'Brien, Coy Morris, Craig Janda, Crescenzo Tortora, Damien Cherouvrier, Dan Prins, Dani Muller, Daniel Cunningham, Daniel Gascón Jiménez, Daniel Hearnden, Daniel Lopes, Daniel Molina Fernandez, Daniel Motherway, Daniel Pantoja, Daniel Piaget, Daniel Pishyar, Daniel Robinson, Daniel Skarednov, Daniela Mantula, Danielle Hoss, DanielR. Gigot, Daniil Filatov, Danika Belzile, Danilo Blagojević, Danny Muskens, Danny Westphal, Daria Dawson, Darla Dangcil, Darrell Bettany, Dave Hardy, Dave Santos, David Aldred, David Alexander, David Browne, David Carozza, David Hockaday, David Klemmer, David Lafuente, David M Williams, David Meierhenry, David Mumma, David N. Saunders, David Osten, David Plesl, David Serrano, David Stegemann, David Svinning, David Truckenmiller, David W. Rodgers, David Warrilow, David “Tony” Fantone, Davin Brown, Dawn Sunrise, Dayne Svendsen, Declan Ora, Deepti Sharma, Denis Korostelevs, Denis Terskov, Deniz Ege Emre, Dennis Kuenzel, Dennis R. Patten, Sr., Dennis Stanescu, Dennis Ward, Dennis Werkes, Derek Blauth, Derek Jones, Derek Sperry, Dereon Robertson, Devanandana Mohan, Dhananjay Singla, Diana Osipov, Diego A. Díaz, Diego Ivan Reynolds Fernandez, Diego Lorenzo de los Ríos, Diego Pirrami, Dieter Haile, Dillan Hayes, Dina Guttmann, Dinesh Patel, Djoulane Bentarcha, Dmitriy Emelyanov, Dmitry Babin, Dmitry Ivanov, Doha Bedda, Domenico Avino, Dominic Savard, Dominikus Koelbl, Don J Vavasseur, Don T. Beakunt, Donald K. Lorenzo, Donna Charles, Donna Frostick, Dr Richard Nitzsche, Dr. Gundolf Lange, Duncan Jones, Durando Nathan, Dylan Jennings, E Chillington, E.V. Valentine, Edith Freundl, Eduardo da Costa Valadão, Edward Silvia, Edward Zhou, Eka Thea Sharma, Elad Ben-Ezra, Eleanor Aalders, Eleanor Cox, Eleni Hope, Elina Litvinova, Elisa Casale, Elisa Cipriani, Elisa Goni, Elisabeth Baeten, Elizabeth Beckwith, Ella Cowan, Elmar Delvos, Elmar Pohl, Elspeth Edwards, Elwood Downey, Emese Kocsis, Emily A Caffarel, Emily Bollich, Emily Jayne Bean, Emily Lutterloh, Emily Sharo, Emma Bazemore, Emma Cucca, Emma Eloise Thompson, Emma Gerntke, Emma Short, Emma Taylor, Eoin Neylon, Eric Balestra, Eric Cooper, Eric Fabrigat, Eric Henrique Marques da Silva, Eric Le, Eric Petersen, Erick F. Del Cid Borja, Erik Bowman, Erik Gasteiger, Erik Martinez, Erjon Buzo, Ernest Courtois, Ernest Jude P. Tiu, Esraa samy Saad, Essi Kultanen, Ethan Hernandez, Ethan Jeffris, Ethelinda McCall, Etienne Sevrin, Eva Gimenez Jelic, Evan Thompson, Eve England, Evelyn Kelley, Evelyne Stanton, Eyser Choquet, Fabio Rollo, Fabrice Bongiardini, Fabrizio Trara Genoino, Faiz Qamar, Fan Baoyi, Farha Ahmed Morsy, Federico Desii, Felicia Vinciguerra, Felix S. Firmbach, Fermor Baird, Fernanda Freitas, Fernando A. Alfaro, Filip Vdović, Fintan Sargent, Fiona Barbolak, Floortje van Santen, Florent Perrin, Florian Arina, Florian Busch, Florian Mansion, Flávia Nathalia da Silva Santos, Fox James, Frances and Stuart Holloway, Francesca Cottone, Francesco Simeoni, Franck Pochet, Frank Falor, Franziska Bauer, Fred Linley, Freddie Jr. O. Cabrejas, Frederick Wilner, Freya Isaksson, Fruzsina Jelen, G. Greene, Gabriel Taylor, Gabriela Madera Pena, Gabriele Pavan, Gael Lubin, Gail Gibson-Smith, Gaivoronskiy Stanislav Vadimovich, Galini Michaela Chronaki, Gary Buss, Gary Rowland, Gaurav Kumar, Gavin Rider, Gayathri Manoj, Geoff Kirby, George Deng, Gerald W. Nash Jr., Gerhard Petz, Gina Grogg, Giro Neil C. Generao, Giulia Despali, Gordon Eldon, Grace Scura, Grady Fitzpatrick, Graham Darlington, Graham Dungworth, Greg Balnis, Greg Jenkins, Greg Roth, Gregory Avramenko, Gregory Drzymkowski, Gregory Kaptcov, Griffin Cook, Guido Bucchi, Guido Lauer, Guilherme Pereira Brandão, Guillaume Penaud, Guillaume Quenouille, Gunnar Russell, Guoyou Sun（孙国佑）, Gyongyi Plank, Génesis Pimentel, Gérault Nikita, Günter Steinbach, H. A. Güler, H. W. Riley, Haley Willits, Hams Gadalla, Hanna Lebedzevich, Hannah Kempken, Hannes Lundvall, Haotian Yang, Haoxuan Zhang, Harold Price, Harry Moreland, Harshitha Vijayakumar, Haruka Yanagisawa, Haseeba Zia, Hassan Namwar, Hau Rava, Hayoon Kang, Heather Bassett, Heiko Gericke, Helen Wetherbee, Helena Yeung, Helga Haase, Hendrik Teixeira, Henry Clark, Henry Williams, Henryk Krawczyk, Heraldo Henrique Felix de Moraes, Hiago Jorge Silva, Hibah Arshad, Hiroki Fukuyama, Hiya Singh, Hrvoje Krizic, Huang Rui Guo, Hudson Fu, Hudson Varkony Cook, Hugo Mazer, Hugo Mendes, Hugo Podsieldik, Huub Mooren, Ian Carley, Ian Russo, Iaroslav Mokroguz, Icaro Melo Ferreira, Igor Alvise Justo, Ilhan Abdul Rahim, Ilja Dotšar, Ina Thamke, Ingo Holetzek, Ingrid Nurminen, Inika Sridhar, Ioana-Alexandra Nedela, Irina Aerith, Irina Stocker, Iris Ham, Isabella Ana Cuan, Isabella Eva Bigatti, Isabella Roberts, Isabella Skaggs, Isabella Wang, Isabelle Brander, Isabelle Graves, Isaiah Thomas Varghese, Ishana Parmar, Ishita Mukherjea, Isla Gray, Isobel Frew, Ivan A. Terentev, Ivana Varsányiová, İrem İlhan, J. Rocco, J.J. Carrol, J.T. Magee, Jaana Kemppainen, Jace Spence, Jackie Branc, Jackson Daniel Griffith, Jackson Keener, Jackson VanHall, Jacob Amato, Jacob Andrusyszyn, Jacob Brown, Jacob Fielding, Jacob Reed, Jacqueline Elliston, Jacqueline Hirschfeld, Jacquelyn Higinbotham, Jafar Hasanjee, Jaiden Lopez, Jakub Jarý, Jakub Michalak, James Auld, James Betts, James C Mullikin, James Edmunds Mace, James Lambourne, James Pearson, James Perdue, Jan Sjoerd de Vries, Janae Anderson, Jane Sloan, Janya Singh, Jared Thompson, Jari Taavitsainen, Jarne Loos, Jason Linke, Jason MacDonald, Jason Spencer, Jasper Buan, Jaxon Sowder, Jay Alexander, Jay Maleev, Jay Rexroad, Jay Valdes, Jean Nimes, Jean Rheault, Jean-Francois Peyron, Jean-Noel Terry, Jeff Taggart, Jeffrey Sgrignoli, Jehu Campos, Jelly Penalosa, Jen Cotrone, Jennifer D Mesa, Jennifer Rosebrock, Jennifer Saar, Jens Mario Schmidt, Jeremiah Obinna Odaba, Jeremias Peltola, Jeremie Muller, Jeremy Maniago, Jerried Ramon, Jerry Nguyen, Jesse Prenatt, Jessica Hallberg, Jessica Hevia, Jessica Keane, Jessica L Esteves, Jessie Huang, Jessie R. Hayes, Jesús Pasten Arvizu, Jetro Ketola, Jett Amsden, Jhoanna Jimena Chanta Flores, Jiashuo Zhang, Jieun Bok, Jim Locke, Jimmy Simpson, Jo Hofmann, Joanna Maria Dulowska, Joanne Villani, Joanne-Denise Nickel, Joceline Castrejon, Jocelyn Huang, Joe Mertes, Joe Reape, Joe Schmidt, Joey Serber, Johan Bertrand, Johan Le Menn, John Barcelona, John Bragg, John Davis, John Fellows, John Kotas, John Michael, John Parks, John R DuPont III, John V Morris, John Webb, Johnny B. Cowart Jr, Johnny Reid, Jon-Paul  Merry, Jonah Turner, Jonathan Grossauer, Jonathan Tielve, Jonathan Verrier, Jordan A. Simpson, Jordi Solaz, Jordyn Hutchins, Jorge Sobradelo Silva, Joris Beelen, Jose Diego, Jose Luis Navarro, Jose Palafox, Jose Pinto, Joseph Downs, Joseph Henderson, Josh Gebelein, Josh Hill, Josh Wilde, Joshua Fitzpatrick, Joshua McArthur, Joshua Ridgill, José Manuel Leal Quesada, José Victor de Matos Lima, Juan Alberto Cano Díez, Jude Garbe, Jude Wechsler, Judelin Joasilus, Judith Waschelitz, Judy Garstone, Juli Fowler, Julia Brazolim dos Santos, Julia Cavanaugh, Julia de Montfort, Julia H. Stickney, Julia Katona, Julian Asialo Kioa, Julian Collis, Juliana Newman, Julie Paynter, Julie Potasiewicz, Julien Blary, Julien Frutozo, Julien Lühning, Julien Schaer, Jur Vogelzang, Justin Klingele, Justine Lefebvre, Justus Maximilian Gebauer, Jérémie Cuny, Jérôme F. Delaunay, Jérôme Rafani, K Siva Srinath Reddy, Kai (Connor) Young, Kai Kaufmann, Kai Schindler, Kai Thomas, Kaitlyn Randall, Kaja Sinur, Kal Marshall, Kalvin Dewayne Rush Sr, Kamil Dębowski, Kamila Vesela, Kareem Hany, Karen L Jones, Karolína Minaříková, Karthika Kannan, Karthikeyan Duraisamy, Katharina Stowe Martin, Katherine Collins, Katherine Erin Calton, Kathleen Del Prince, Kathrin Loch, Katie Stigter, Katrina Wong, Katya Shlyakhova, Kaustubhi Joshi, Kaylie Scorza, Kazi Rejwan Islam, Kelsey Ryan, Kelson Sisk, Kelvin Salter, Kemo Sidibeh, Ken Davis, Ken Lyman, Ken Nakada, Ken O'Neill, Kendall Hall, Kevin J. Daly B.E., Ph.D., Kevin Moreau, Kevin Peter, Kevin Quesada, Khushi Lalit, Kieran Smith, Kieron Nate S. Presbitero, Kiersten Hoffman, Kievon Ferguson, Kiran Kumar, Kiran Nandkumar Suryarao, Kirill Sakharov, KK Maclean, Klajdi Qinami, Klaudia Solska, Klaus Rohrich, Kohsheen Kachru Naik, Kommal Tariq, Korbin Spencer, Kostula Papandrianu, Kristen Ess, Kristie Yorkston, Kristiina Rimpeläinen, Ksenia Nikonova, Ksenia Smirnova, Kumar Utukuri, Kyara Haffane, Kyle Hartman, Kyle Nelson, L. Van Warren, La Wun Eain, Laetitia Martel, Laila Gregory, Lamberto Veraldi, Lara Ertaul, Lara Maia, Lara Nelson, Laura Dionysius, Laura J. Martin, Laura Phillips, Laura Turiano, Lauren Li, Lauren Murphy, Laurie B. Gaines, Laëtitia Renard, Le Garrec Jean-François, Leah Sun, Leah Yang, Leen Rawad Hilal, Leighton Yanity, Leila Billingham, Leishangthem Debajit Singh, Lena Blomeyer, Leon Roman Ecker, Leona Kleinemeier, Leonardo Rodrigues, Leonie van Vliet, Leslie Ruiz, Leslie Tolentino, Lev Gribakin, Leyna Bajaj, Li Gu, Liam Buki, Liana Greyber, Liana Pavlovica, Lianne Louise Manalili, Liew En Quan, Lily Mastin, Lincoln Hamilton, Linda Saffer, Linh Nguyen, Lino Thomas, Lioba Elisabeth Happel, Lisa Daxer, Logann Ibarra-Ortiz, Lohit Basavaraju, loic Divay, Lois Holmes, Lorenzo Garza, Lorin Monck, Lottiee Haslam-Simmons, Louis Howarth, Louisa Kupke, Loïc Baert, Luan Bermudez, Luca Andreula, Lucas Babuk, Lucas Eilola, Lucas Lévêque, Lucia Hecht, Lucia Maria Ferreira Lima, Luciana Navarrete Gualteros, Lucie Golding, Ludovic Toucheau, Luigi Guzzo, Luka Layton, Luke Carreras, Luke Hanson, Luke Scanlan, Luminita Georgescu, Lynnette Kiluwe, Lyra Glass, Lysandros Wasilopoulos, M. Anderson, M. Zizlsperger, M.S. Batukov, Maciej Woloszka, Madeleine Tuck, Madeline Oxley, Maeve O'Brien, Magdalena Maszewska, Maggie Jardine, Mahmoud Elmagrby, Mai Tanaka, Maia G. Mohsen, Maisie Pattison, Maja Bergmann, Maksym Savenko, Malek Zeidieh, Malia Bouvard Jourde, Mallory Thomsen, Manahil, Mannat Lamba, Manuel Leister, Mara Ioana Șneapotă, Marc Hanset, Marc Raff, Marcela Sepulveda, Marcelino Alvarez, Marcelo Gomes Viana Lopes, Marco Bacchi, Marco Callegari, Marcus Schulz, Margaret Weeks, Maria Eduarda Alves de Araújo, Maria Fragoso, Maria Isabella Mesina Mendoza, Maria Martin, Maria Smirnova, Maria Tubol, Maria Wicher, Maria-Olivia Torcea, Marialena Pouliou, Mariasole Melara, Mariella Gozzi, Marika Jandrey, Marikaitlynn VanZandt, Mario Henrique, Mario Keller, Marion Moffatt, Marius Agafitei, Marius Theissen, Mark A. Fortuna, Mark S Fletcher, Mark Stypczynski, Marlene Schulz, Marsha Speevak, Martha Murray Zinn, Martin Belles, Martin Bergmann, Martin Hall, Martin Miller, Martin Soto, Martin Ungless, Martina Cucchi, Martín Orlando Gil Cardona, Martín Zamora Rodríguez, Mary Vance, Maryceline Espanol, Masi Steele, Mason Gay, Mason Shawe, Massimo Tonioli, Matayah G, Mathias Bergmann, Mathis Noizet, Matilda Wagner, Matt Romriell, Matteo Cianfaglione, Matteo Pedani, Matthew Alexander White, Matthew Allis, Matthew Given, Matthew Helms, Matthew Loomis, Matthew Moon, Matthew Quiroz, Matthew Rhys George, Matthew Seth Marcus, Matthew Thomas Bates, Matthias Kohl-Himmelseher, Matti Sironen, Mauri Simon, Maurizio Mandarino, Maxence Chaminade, Maxence Delmotte, Maxime Fouris, Maxime Guarino, Maximus LaBrunda, Maxine Lamont, Maya Sandberg-Smith, McKenzie Hall, Meera Desai, Meera Zaki, Megan Cooper, Mehmed Emin, Melanie A. Downes, Melanie Whitehouse, Melisa Díaz Rivero, Melissa Arsaut, Melissa Reed, Melissa Wyrostek, Meredith Navin, Meriel McManemy, Merle Mia Drews, Meyra Sert, Mia Hanert, Mia Holcombe, Michael A. Capraro, Michael B Williams, Michael Beckham, Michael J. Brodnicki, Michael Krol, Michael Laughton, Michael Mulhern, Michael Redfield, Michael Tepen, Michal Sivak, Michaël Crevoisier, Michał Biesiada, Michał Smółka, Michel Cavayé, Michela Pallaver, Michele T. Mazzucato, Michelle Zolina, Migachev Dmitry, Miguel Buzzega Mira, Mikaela Birdsong, Mike Barrett, Mike Beer, Milhann Chodorowski, Milot Sélène, Mina Harker, Mina Nedeva, Minal Bhosale, Mirza Adnan Mohtashim, Mitchell Rogers, Mittapelly Krishna Reddy, Mohamed Ahmed, Mohamed Altawel, Molly Scott, Molly Zhang, Monica Backes, Monika Seniut, Moran Thebault, Morijah Stutz, Mostafa Arafat, Mrunalini Prasad, Muditmayank Mishra, Muhammad Hadi Jawad, Muhammad Muzammil Anwar, Muhammad Ridha Akhsanu Sujiono, Munira Azzout, Murat Calikoglu, Murray Leslie, Mustafa Fayyaz, Myat Thura, Myra Gbesso, Myranda Keightley, Nabhya Gandhi, Nadia Buza, Nadir Chettah, Nadya Neklioudova, Nainpreet Uppal, Nash Gilbert, Nasser Lavaud, Natalie Freund, Natalie Lines, Natalie Savelyeva, Natalie Sebastian, Nathan Corey Bell, Nathan Dales, Nathaniel Hammon, Nathália Pichinin de Oliveira Martins, Nazar Silko, Neeti N Patil, Neil Winterbauer, Nerea Ruiz Martín, Nevena Niagolova, Nicholas Taylor, Nicola Denti, Nicolas Belhoste, Nicolas Denand, Nicolas Prevot, Nicolas Royer, Nicole Humphries, Nicolò Spigariol, Nik Cain, Nikita Kuzminykh, Nikolas Kontaxakis, Nikolaus Hermanspahn, Niladri S Kar, Nimal Kumar, Nina Grkovic, Nina Trubanova, Ninad Muley, Nino Otto, Nithyasri Senthilkumar, Nitika Verma, Noah Hansler, Noah R, Nolan Reket, Nolann Barbeau, Norah Simdon, Notarpietro Sabrina, Océane Rege Nero, Oleksandr Fil, Oliver Rawding, Olivia Lindahl, Olivia Park, Olivia Steger, Olivier Gaytán, Olivier Lecrenier, Olivier Marty, Oussama Guarraz, Ovalle Pérez Paloma, Owen Butler, Oxana Schmidt, P A Gwatkin, P Gopalan, P Tilbury, P. Seal, P.A. Labud, Paige Charman, Pamela Allen, Paola Bungaro, Paola Coves Puelma, Paramita Mohanty, Parizel--Montardi Mickaël, Parker Stasak, Pascal Dicop, Patrick Dorp, Patrick Mahaffey, Patrick Ngo, Patrick Veyssière, Paul Drye, Paul Henry, Paul Niermaréchal, Paula Pascua García, Pavel Atanasov, Pavel Fedyaev, Pedro Augusto Schimidt Leao, Pedro Bergevoet, Peggy Lehmann, Peter Ansorge, Peter Carr, Peter Csitkovics, Peter Dzwig, Peter Everett, Peter Haveland, Peter O'Toole, Phil Andriole, Philip Downey, Philip Holloway, Philippe Laferriere, Pierre Astor, Pietro Guillén Ferreira, Pietro Maiorana, Pilar Fife, Piotr Konecki, Piper Stafford, Prasad Vengurlekar, Preetha Pushkarini B N, Pritam Raikar, Przemysław Kowalski, Pásztor Benedek, Qi Liu, Quentin Guelle, Rachel Alice Ford, Rachel Binnie, Rachel Matt, Raeanne Nichol, Ragushin Aleksandr, Raimo Michaelis, Rainer Gorissen, Rainer Töpler, Rajshekhar Gupta, Rajwa Khansa Dzakirah, Ralf Lengers, Ralf Stephan, Ralitza Radoukanova, Ramita Kemavit, Random Howard, Rapaka Raja Ratnam, Raphael Eckersley, Raphaël Jimenez, Raul Torres, Ravshan Asatov, Raymond Gardocki, Raymond Wilkinson, Razvan Tanase, Rebecca Brown, Rebecca Denae, Rebecca Eid, Rebecca S. Warren, Rebekah F, Renat Saifullin, Renata Pataki, Renato Raduka, Rene-Michel Paquet, Renos Anastasi, René Eichelbaum, Reta Nissan, Retaj Atef Elsobky, Rhianna Cohen, Rhiannon Ford, Rhys Morris, Richard Allen, Richard Goodwin, Richard Huziak, Richard J. Acton, Richard Knijnenburg, Richard Renner, Richard S Blasi, Richard Scott, Richard Trubshaw, Rick Barnwell, Rida Islam, Riley Keppel, Rishaan Akula, Rishi Tangirala, Rob Garrison, Robert Desmond Maher, Robert E. Wheat, Robert J. Adamson, Robert J Heyden, Robert Neil Haeuser, Robert Schwartz, Roberto Aretxaga-Burgos, Roberto Bonamico, Roberto Gorelli, Roberto Haver, Robyn Keats, Robyne Major, Rodolphe DI MAIO, Roger Bowden, Roger Long, Roger Marrotte, Rohan Krishna, Roland Heringhaus, Roman Rakityanskiy, Ron Usry, Ronan Badel, Ronan Hughes, Rosa Bermejo, Rosemary Billington, Roshini Gade, Rowan Tekampe, Ruby Cline, Rudra Protap Nandi, Ruslan Migranov, Russell Parman, Russell Reed, Ruth Kevess-Cohen, Ruth Tucker, Ryan Chen, Ryan Craft, Ryan Fick, Ryan Li, Ryan Mavis, Ryan Ross, Rylee Funseth, S THILAK, S. Ominami, Saadet Polat, Saber Williamson, Sadewo Abdi Panuntun, Sai Likhith Poreddy, Sajeev T Prabhakaran, Sakari Vedenpää, Sallyann Chesson, Sam A. Fazelinia, Sam Deen, Sam Dulys, Sam Lee, Samanta Holdaway, Samantha Henry, Samantha Schenk, Samuele Meschini, Sana Farhat Abdul Qayyum, Sandra Lee Harris, Sandro José de Souza, Sandro Montagner, Sandupa Sanudaya, Sanjit Kolluru, Santi Ranjan Atta, Santiago Clements-Rodriguez, Santino Giacometti, Sara Capacchione, Sara I. Estrada, Sara Kurdy, Sara Vardanyan, Sarah Grissett, Sarah Hünermund, Sarah Musumeci, Sarah Sayer, Sarah Szita, Sarah White, Sarp Akar, Satanay Makhmud, Sauro Gaudenzi, Scott Grant, Scott Joseph Sokalski, Scott Richards, Sean Miller, Sebastian Lefebvre, Sebastian Oppermann, Sehr Almas Noordin, Serge Robert, Sergio Borrás Pardo, Seth Smithey, Shannen Chovanec, Sharlotte Phifer, Sharon Dinh, Shaunie Beswick, Shay Shalit, Sherie Kuek, Shirley Xie, Shivani Hancock, Shneor Gubits, Shogo Noma, Shrestha Sarkar, Shrihaan Sindhushri, Shrujal H. Nandurkar, Shuming Wang, Shuning Bian, Siddharth Mehrotra, Silke Jonda, Silke Wachholz, Simon Basler, Simon Nicholson, Simon Tarin, Simon Vincent Götz, Simone Glau, Siqi Yao, Siya Singh, Siyona Das, Skye Grimm, Sofia Yaroshenko, Sofiia Gudkova, Sophia Berinski, Sophia Ottner, Sophie Markman, Sovan Acharya, Soveros Arctos, Spriha Pudasaini, Sreeja Pal, Srujanika Rudra, Stacey Cornell, Stacey Cornell, Stacey E. Farr, Stacey Lewis, Stan Smallman, Stefanie Seidel, Stefanos Drakoulis, Stephana Bregovich, Stephanie Ferguson, Stephanie Garcia, Stephanie Potter, Stephen Halston, Stephen Lowcock, Stephen Morris, Steve Mealy, Steve Saunders, Steven Tan, Sudhakar Subramaniam, Sue Sivonen, Sung Kei Li, Susan Feder, Susan Magyarody, Susannah Stubbs, Sushant Pandit, Svetoslav Alexandrov, Swapnasish Ghorui, Sydney Saha, Sydney Saltzman, Syeda Sukaina, Sylvain Martineau, Sylvain Paszkowski, Sylwia Plenzner, Tada Wang, Takshil Doshi, Tammy Checketts, Tamás Gyurka, Tanja Kuusela, Tanusha Subramaniam, Taos Medjoub, Taryn Allie, Tashfin Shifan, Taylor Thornton, Teemu Oinasvirta, Teresa Israels, Terrance P Felegie, Sr, Terry Berryhill, Thanvi Gorla, Thibault Adam, Thijn Kriesels, Thomas Brindle, Thomas E Collett, Thomas Gravekamp, Thomas Rosa, Thomas Wigbels, Thorben Ruthke, Thorsten Eschweiler, Théo Janvier, Théo Neveu, Tiffany Lloyd, Tim Albrecht, Tim Pilz, Timur Podornicov, Tina Mathews, Tirthankar De, Titiksha Pandit, Tobias Elsner, Tom Bickle, Tom Gilde, Tom Udding, Tommy Reynolds, Tone Iversen, Tony Frost, Tora Frankic, Tora Frankic, Trace Jeffries-Ortiz, Tristan Marchand, Tyler Atkinson, Udaiveer Singh, Ushakov Rostislav Konstantinovich, Uswa Fatima, Vaishnavi Shinde, Vajda János, Valasia Anna Valavosiki, Valentino Dalla Verde, Valeria Bondarenko, Valerie Sousa, Vanisha Thiyyagura, Vansh Bahety, Veda Vishal, Veronica Couñago Vidal, Veronica Motta, Veronika Sivayeva, Victor Alteirac, Victor Gabriel Bibé, Victor Juan Garcia Porcel, Victor Teodor Cozea, Victor Yu, Victor-Francois Ros, Victoria Alicja Machnik, Viera Holik, Vika Manjgaladze, Ville Mänttäri, Ville Tissari, Vincent Decker, Vincent Parmentier, Vincenzo Manto, Virgilio Gonano, Vishakha Jingar, Vivian Jin, Vlad Andrei Manolescu, Vladimir Ezhelyy, Vladimir Komarov, Vladimir Tishkin, Vladimír Šintaj, Walter Feichtinger, Walter MacDonald, Wania Latif, Warren Kinninger, Wasif Amin, Wayne Stripe III, Wendy Hawkins, Wesley Webb, Wiesław Feduniewicz, Wihan van Heerden, William Balmer, William Beaudoin, William Hawkesworth, William MacLennan, William Russell, Winfried Schoberth, Wolfgang G. Kurtz, Xi Liao, Xiangyu Huang, Xincheng Zhu, Xingping Lin, Xuanlin Jiang, Yamil A. Montañez Díaz, Yang Feichi, Yann Lelievre, Yannick Facolle, Yaroslava Brilli, Yassen Mahmoud, Yevheniia Furutina, Yi Ying Mok, Yi-Wen Lasure, Yurii Andryshin, Zachary Benfield, Zachary Kronlage, Zachary Swift, Zachary Wood, Zain R.-Schill, Zainab Ahmad, Zainah Imran Waheed, Zakariya El Jilali, Zhao Long Dong, Zharov Dmitriy, Zipora Morgan, Ziyi Zhou, Zoe Posnette, Zoe Spasaro, Zoltán Huszár, Zoran Von Orel Bronte, Đoàn Khánh Hưng, Łukasz Hyrchel-Sakowicz, Šimon Urbaník, \foreignlanguage{greek}{ΒΑΣΙΛΙΚΗ ΠΕΤΡΟΠΟΥΛΟΥ}, \foreignlanguage{kazakh}{Банияз Сағи}, \foreignlanguage{russian}{Дмитрий Сергеевич Фёдоров, Дмитриков Федор, Евгений Юрченко, Исаков Евгений Дмитриевич, Костя Гоман, Мясников Иван Артемович, Панаиотова Наталья Александровна, Сапожникова София, Цыро Снежана, Юлия Емельянова}, 余映逸(Ying Yat,Yu), 片山洋臣 Yoshifumi Katayama}
\end{CJK}
\end{spacing}

\end{appendix}

\label{LastPage}
\end{document}